\documentclass[a4paper,fleqn]{cas-dc}

\usepackage[authoryear]{natbib}
\usepackage{xcolor}
\usepackage{lineno}

\begin{document}
\let\WriteBookmarks\relax
\def\floatpagepagefraction{1}
\def\textpagefraction{.001}

\newcommand{\apj}{ApJ}
\newcommand{\apjl}{ApJ}
\newcommand{\mnras}{MNRAS}
\newcommand{\aap}{A\&A}
\newcommand{\aapr}{A\&A Reviews}
\newcommand{\aj}{AJ}
\newcommand{\nat}{Nature}
\newcommand{\planss}{P\&SS}
\newcommand{\pre}{Phys.~Rev.~E}
\newcommand{\psj}{PSJ}
\newcommand{\araa}{ARA\&A}
\newcommand{\icarus}{Icarus}
\newcommand{\dlink}{d_\mathrm{link}}
\newcommand{\phieff}{\phi_\mathrm{eff}}
\newcommand{\kcm}{\ \mathrm{kg/m}^{3}}
\newcommand{\flag}[1]{{\color{red} #1}}  
\newcommand{\editone}[1]{\textbf{#1}}  

\shortauthors{Wimarsson et~al.}
\shorttitle{Rapid formation of binary asteroid systems post rotational failure}
\title[mode=title]{Rapid formation of binary asteroid systems post rotational failure: a recipe for making atypically shaped satellites}

\affiliation[1]{
    organization = {Space Research \& Planetary Sciences, Physics Institute, University of Bern}, 
    addressline={Gesellschaftsstrasse 6},
    city = {Bern},
    postcode = {3012},
    country = {Switzerland}
}

\affiliation[2]{
    organization = {Department of Aerospace Science and Technology, Politecnico di Milano}, 
    city = {Milano},
    postcode= {20156},
    country = {Italy}
}

\affiliation[3]{
    organization = {Université Paris Cité, Institut de Physique du Globe de Paris, CNRS},
    city = {Paris},
    postcode = {F-75005},
    country = {France}
}

\affiliation[4]{
    organization = {Colorado Center for Astrodynamics Research, University of Colorado Boulder},
    addressline={3775 Discovery Drive},
    city = {Boulder},
    postcode = {CO 80303},
    country = {USA}
}

\author[1]{John Wimarsson}[type=editor,orcid = 0009-0004-6736-309X]
\ead{john.wimarsson@unibe.ch}
\cormark[1]
\cortext[cor1]{Corresponding author}
\credit{Conceptualization, Methodology, Software, Formal analysis, Visualization, Writing - Original Draft, Writing - Review \& Editing}
\author[1]{Zhen Xiang}[orcid = 0009-0004-5467-872X]
\credit{Methodology, Software, Writing - Original Draft}
\author[2,1]{Fabio Ferrari}[orcid = 0000-0001-7537-4996]
\credit{Supervision, Conceptualization, Writing - Review \& Editing}
\author[1]{Martin Jutzi}[orcid = 0000-0002-1800-2974]
\credit{Supervision, Conceptualization, Writing - Review \& Editing}
\author[3]{Gustavo Madeira}[orcid = 0000-0001-5138-230X]
\credit{Writing - Review \& Editing}
\author[1]{Sabina D. Raducan}[orcid = 0000-0002-7478-0148]
\credit{Writing - Review \& Editing}
\author[4]{Paul S{\'a}nchez}[orcid = 0000-0003-3610-5480]
\credit{Writing - Review \& Editing}

\begin{abstract}
Binary asteroid formation is a highly complex process, which has been highlighted with recent observations of satellites with unexpected shapes, such as the oblate Dimorphos by the NASA DART mission and the contact binary Selam by NASA's Lucy mission. There is no clear consensus on which dynamical mechanisms determine the final shape of these objects. In turn, we explore a formation pathway where spin-up and rotational failure of a rubble pile asteroid lead to mass-shedding and a wide circumasteroidal debris disk in which the satellite forms. Using a combination of smooth-particle hydrodynamical and \textit{N}-body simulations, we study the dynamical evolution in detail. We find that a debris disk containing matter corresponding to a few percent of the primary asteroid mass extending beyond the fluid Roche limit can consistently form both oblate and bilobate satellites via a series of tidal encounters with the primary body and mergers with other gravitational aggregates. Principally, satellites end up prolate (elongated) and on synchronous orbits, accreting mainly in a radial direction while tides from the primary asteroid keep the shape intact. However, close encounters and mergers can break the orbital state, leading to orbital migration and deformation. Satellite--satellite impacts occurring in this regime have lower impact velocities than merger-driven moon formation in e.g.\ planetary rings, leading to soft impacts between differently sized, non-spherical bodies. The resulting post-merger shape of the satellite is highly dependent on the impact geometry. Only moons having experienced a prior mild or catastrophic tidal disruption during a close encounter with the primary asteroid can become oblate spheroids, which is consistent with the predominantly prolate observed population of binary asteroid satellites.
\end{abstract}

\begin{keywords}
 Asteroid dynamics (2210) \sep Asteroid satellites (2207) \sep Asteroid rotation (2211) \sep Debris disks (363) \sep Small Solar System bodies (1469)
\end{keywords}

\maketitle

\section{Introduction} \label{section:intro}

The presence of atypically shaped satellites in binary asteroid systems has been an trending topic as of late \citep{Madeira_&_Charnoz_2024,Agrusa_et_al_2024}. First, close-up observations were made of the spherically oblate Dimorphos orbiting its main body Didymos by NASA's DART (Double Asteroid Redirection Test) spacecraft, which then proceeded to impact the body on September 26th 2022, altering its orbit and likely its shape in a historical event \citep{Daly_et_al_2023a,Chabot_et_al_2024,Rivkin_et_al_2021,Raducan_et_al_2024a}. More recently, images of a contact binary satellite, now named Selam, were provided by NASA's Lucy mission during a flyby of the asteroid Dinkinesh \citep{Levison_et_al_2021,Levison_et_al_2024}. Observations report that $\sim 15\%$ of the near-Earth asteroid (NEA) population have satellites, a number that increases to $\sim 65\%$ for rapidly rotating bodies with a diameter of at least $\sim 300$ m \citep{Margot_et_al_2002,Pravec_et_al_2006}. A large majority of the observed satellites are prolate (elongated) in form \citep{Pravec_et_al_2016}, making the shapes of both the oblate Dimorphos and bilobate Selam surprising. Notably, while significant parts of these bodies have yet to be imaged, which adds a level of uncertainty to their inferred shapes, lightcurve measurements of asteroid systems are biased against detecting satellites that are more oblate \citep{Pravec_et_al_2016,Agrusa_et_al_2024}. Consequently, oblate spheroidal satellites could be more prominent than expected from current observational data.

The primary bodies in both systems are small, with Didymos having a volume-equivalent diameter of $\sim 760$ m \citep{Barnouin_et_al_2023} and Dinkinesh a similar effective diameter of $\sim 720$ m \citep{Levison_et_al_2024}. Moreover, the two display the typical spinning-top-shape of rapidly rotating asteroids \citep{Pravec_et_al_2006,Pravec_&_Harris_2007,Benner_et_al_2015}. The high rotation rate of these objects might be primarily caused by the Yarkovsky-O'Keefe-Radzievskii-Paddack (YORP) effect, where a weak net torque acts on asteroids due to anisotropic re-emission of absorbed sunlight over long periods of time caused by their irregular shape \citep{Rubincam_2000}. Many of these asteroids are so-called `rubble piles' and are gravitational aggregates consisting of a large number of smaller constituent particles with little-to-no bulk tensile strength and moderate porosity. These objects have observed spin-rates exceeding the theoretical disruption limit for fluid materials, indicating that they have shear strengths consistent with angles of friction around $40^\circ$, which yields behaviours similar to terrestrial granular materials \citep{Scheeres_et_al_2010,Sanchez_&_Scheeres_2014,Sanchez_&_Scheeres_2016,Walsh_2018}.

The observed spin-rates of asteroids with diameters larger than 200 m and smaller than 40 km show that very few bodies are rotating with periods shorter than $\sim 2.2$ h, indicative of a critical spin limit \citep{Pravec_&_Harris_2000,Pravec_Harris_&_Michalowski_2002}. In turn, there must exist some mechanism to prevent these bodies from spinning faster. It was postulated early on after the identification of this critical rotational period that YORP could be spinning rubble pile asteroids up beyond this limit, leading to rotational failure and disruption, which would in turn reduce the rotational angular momentum of the body and decrease its rotation rate \citep{Rubincam_2000}. This mechanism was then further tied to the possible formation of asteroid binaries \citep{Vokrouhlicky_&_Capek_2002,Bottke_et_al_2002,Bottke_et_al_2006,Walsh_et_al_2008,Harris_Fahnestock_&_Pravec_2009}.

The top-shapes of small, rapidly rotating asteroids such as Didymos, Dinkinesh, Ryugu \citep{Watanabe_et_al_2019} and Bennu \citep{Barnouin_et_al_2019} further reinforce this idea, hinting at a ubiquitous dynamical history for top-shaped rubble piles involving even shorter rotational periods, causing deformation through global landslides and mass-shedding by their equators \citep[e.g.][]{Walsh_et_al_2008,Walsh_et_al_2012,Hirabayashi_2015,Scheeres_2015}. Their ability to retain their axis ratios despite the strain caused by high rotation rates shows that they have the mechanical strength to resist total disruption and internal deformation. A homogeneous, cohesionless body spinning at these rates would undergo internal deformation before any surface shedding could occur, which has been shown through both theoretical analysis \citep{Holsapple_2010,Hirabayashi_2015}, soft-sphere discrete element method (SSDEM) simulations \citep{Sanchez_&_Scheeres_2012,Sanchez_&_Scheeres_2016,Hirabayashi_et_al_2015,Zhang_et_al_2017,Zhang_et_al_2018,Leisner_et_al_2020} and a model based on a plastic finite element method \citep{Hirabayashi_&_Scheeres_2015,Hirabayashi_2015}. Even homogeneous bodies with some level of cohesion will still be susceptible to disruption before surface failure \citep{Sanchez_&_Scheeres_2016,Zhang_et_al_2018}, meaning there is reason to believe these rubble piles are heterogeneous, either because of polydisperse populations of particles \citep{Zhang_et_al_2017,Zhang_et_al_2021} or large, rigid cores \citep{Walsh_et_al_2008,Walsh_et_al_2012,Hirabayashi_et_al_2015,Sanchez_&_Scheeres_2018,Ferrari_&_Tanga_2022} that prevent internal deformation. 

In this paper, we explore a formation scenario for binary asteroid systems where spin-up of an initially spheroidal rubble pile asteroid leads to a chaotic, instantaneous mass-shedding event, generating a wide debris disk and explore how the dynamical evolution of the disk post rotational failure can produce asteroid satellites. In particular, we focus on the various dynamical mechanisms that govern the shapes of moons in this regime and constrain the different formation pathways, both qualitatively and quantitatively, that can produce the atypical shapes of Dimorphos and Selam. In order to do so, we employ a combination of spin-up simulations of rapidly rotating rubble pile asteroids performed with a smooth-particle hydrodynamics code that provides initial conditions for a gravitational \textit{N}-body code with non-spherically shaped particles that evolves the disk until we obtain a satellite of significant mass. In section \ref{section:previous_work}, we introduce and summarise the previous work on binary asteroid formation. In section \ref{section:method}, we go through the two different numerical codes used in our simulations and the interface we have created to connect them. In section \ref{section:results} we show the results from our main model, based on a Ryugu-like body with similar properties and in section \ref{section:didymos}, we apply our model to the Didymos system and compare how the outcome changes with a smaller primary body. We then discuss the effects of disk geometry, moonlet merger properties and particle size frequency distribution in section \ref{section:discussion}, which is followed by conclusions in section \ref{section:conclusion}.

\subsection{Binary asteroid formation}\label{section:previous_work}

While the narrative of spin-up via YORP leading to structural failure of rubble pile surfaces and global landslides is heavily supported by both theoretical and numerical work along with observations, there is yet no consensus in the community regarding just how these mass-shedding events lead to the formation of binary asteroid systems, especially when considering the existence of atypically shaped satellites such as Dimorphos or Selam. The timescales for the formation process are poorly constrained, where e.g.\ surface analyses of Didymos and Dimorphos indicate that the main body is between 40 and 130 times older than its satellite \citep{Barnouin_et_al_2023}. In turn, it is difficult to determine how much mass needs to be shed over how long a period to successfully form a stable satellite. 

\subsubsection{Continuous mass-shedding}

A first attempt to explicitly tie equatorial mass-shedding to binary formation was made by \cite{Walsh_et_al_2008}, who used a numerical hard-sphere discrete element model (HSDEM) to perform \textit{N}-body simulations of the slow shedding of mass from a cohesionless, rapidly rotating primary spun up by YORP. The primary itself was modelled using one thousand rigid, similarly sized spheres arranged into a spheroidal aggregate using hexagonal closest packing (HCP). This particle configuration yielded an angle of friction of around 40$^{\circ}$, providing the shear strength needed to withstand internal deformation until surface shedding could occur. As the rotation rate of the main body was gradually increased to imitate the process of slow spin-up via the weak torques from YORP and approached the critical spin limit, \citeauthor{Walsh_et_al_2008} observed how particles began moving from the poles towards the equator which came to serve as the origin of the subsequent mass loss. In turn, the primary became oblate and obtained the characteristic top-shape observed for so many rapidly rotating asteroids. Mass was then ejected continuously through shedding events from the equator of the primary and eventually, with enough mass in orbit, the debris began quickly accumulating into a satellite. A follow-up study investigating a broader set of initial conditions with higher resolution was done in \cite{Walsh_et_al_2012}, reaffirming the correlation between this formation pathway and the observed population of binary asteroid systems. 

While the formed satellites have orbits and masses that align with the observed population, the two studies did not explore their final shapes. Due to the nature of tidal forces for granular aggregates, most of bodies created in this manner must first begin accumulating near the theoretical fluid Roche limit, as most bodies formed interior of this distance would be tidally disrupted before reaching a high enough mass \citep{Holsapple_&_Michel_2006,Sharma_2009}. Initially having an orbit near the Roche limit, the tidal forces acting on the aggregate should render it prolate \citep{Porco_et_al_2007,Tiscareno_et_al_2013} and in synchronous rotation, which aligns with observations of binary satellites on close orbits \citep{Pravec_et_al_2016}. As we will show in this work, a rubble pile moon formed without undergoing any events of deformation does indeed exhibit this pattern, which was also indicated by other recent studies from \cite{Madeira_&_Charnoz_2024} and \cite{Agrusa_et_al_2024}.

Hence, the long-term formation process explored in \cite{Walsh_et_al_2008,Walsh_et_al_2012} can account for the predominantly prolate population, but offers no satisfying explanation for the oblate shape of Dimorphos, which has estimated semi-major axes of $\approx 87.90\times 86.96\times 57.16$ m \citep{Daly_et_al_2023b}. The body further exhibits geological properties of a rubble pile \citep{Barnouin_et_al_2023}, which aligns with results from impact simulations \citep{Raducan_et_al_2024a,Raducan_et_al_2024b}, meaning it cannot simply be a large monolithic boulder from a past, chaotic dynamical event. A detailed review of the dynamical state of Dimorphos before and after the DART impact can be found in \cite{Richardson_et_al_2024}. To obtain further data on the post-impact state and the geological properties of Dimorphos, the European Space Agency's Hera mission is sending a spacecraft which will arrive in 2027 to perform a close-up analysis of both Didymos and Dimorphos, as well as in situ measurements of the latter with CubeSats \citep{Michel_et_al_2022}. This will greatly aid the endeavour to constrain the internal structure of Dimorphos and in turn its dynamical history.

As previously mentioned, polydisperse aggregates react differently to external forces and strain than those consisting of similarly sized particles. Moreover, the packing of material and their effective shapes can greatly affect the dynamics of granular aggregates, which was shown by e.g\ \cite{Zhang_&_Michel_2020}, who investigated tidal disruption of rubble piles with different packing, particle shapes and numerical treatments for particle--particle contacts. More specifically, they found that models based on HSDEM and HCP packing produce different dynamical results than the more complex SSDEM treatment with random closest packing. This comes from HSDEM being unfit to simulate granular systems where the particles sustain long-lasting contacts \citep[e.g.][]{Sanchez_&_Scheeres_2016}. Furthermore, as pointed out by \cite{Agrusa_et_al_2024}, new inhomogeneities caused by global deformation of the surface of spinning asteroids can highly affect the net torques caused by YORP throughout the satellite formation process, causing large variations in the amount of mass that is being shed over long periods of time \citep{Statler_2009,CattoFigueroa_et_al_2015,Zhou_et_al_2022}.

\subsubsection{Contact binary fission}

Another formation scenario was explored by \cite{Jacobson_&_Scheeres_2011a}, based on the `rotational fission' of a contact binary asteroid due to long-term spin-up via YORP, where the bilobate object is disrupted and separated into two components and the larger one is the primary while the smaller is the secondary \citep{Scheeres_2007,Scheeres_2009}. This yields a chaotic and instantaneous shedding event where mass ejected into the system is less prone to escape before being accumulated into a stable satellite, which is another issue that long-term YORP-driven formation scenarios must take into account \citep{Jacobson_&_Scheeres_2011a}. Under the assumption that the contact binary is initially arranged to satisfy its global minimum energy configuration, they generated each contact binary body as two fused ellipsoids with their longest axes aligned. They also used a second type of configuration, where the primary is an ellipsoid while the `secondary' consists of two spheroids, allowing for a `secondary fission' event where the created satellite also can undergo fission due to spin-orbit coupling, leading to a ternary (triple) system. The introduction of this secondary event allowed for the fissioned body to dissipate energy in an alternative way for systems with a low mass ratio between the secondary and primary body (<0.2), resulting in positive free energy. When the secondary body was kept monolithic for these initial conditions, it would then ultimately always escape the system. 

Because of the implementation approximating the granular aggregates as consisting of two or three components, they were able to perform very long-term dynamical evolution (up to 1000 years) of the systems post rotational fission using a high-order integration scheme, accounting for inelastic collisions, the non-spherical gravitational potentials of each component and energy dissipation due to mutual body tides using a semi-analytical implementation. Nevertheless, despite the advanced treatment of the system dynamics capturing influential long-term dynamical effects, the choice of using rigid bodies as representations for each component may not capture the full complexity of granular mechanics. As we have already established, tidal forces highly affect rubble piles forming within the theoretical fluid Roche limit, causing deformation, mass-shedding or total disruption and the resilience of a rubble pile to these effects depends strongly on internal packing, cohesion, as well as the size frequency distribution and shape of its particles \citep{Zhang_&_Michel_2020}. For a granular aggregate to survive within the theoretical tidal disruption limit, it would be necessary for its material to have a friction angle larger than zero and particle interlocking effects \citep{Movshovitz_et_al_2012}, significantly higher bulk density than the resulting primary or some level of cohesion (or perhaps a combination of all these effects), providing the gravitational aggregate with higher internal mechanical strength such that it can withstand disruption \citep[e.g.][]{Holsapple_&_Michel_2008,Sanchez_&_Scheeres_2016}. For example, as shown by \cite{Agrusa_et_al_2022}, Dimorphos would be tidally disrupted if it formed from a rotational fission event with Didymos, which makes such a formation scenario implausible given its seemingly low level of cohesion \citep{Raducan_et_al_2024a,Raducan_et_al_2024b} and a bulk density similar to that of Didymos \citep{Daly_et_al_2023b}. 

The contact binary fission scenario has since its introduction by \cite{Jacobson_&_Scheeres_2011a} been further explored in several studies \citep{Jacobson_et_al_2016,Boldrin_et_al_2016,Davis_&_Scheeres_2020,Ho_et_al_2022}. Moreover, a similar mechanism for binary asteroid system formation was tied to the existence of equatorial cavities on rapidly rotating asteroids, where \cite{Tardivel_et_al_2018} suggested that the ejection of a large chunk of material could lead to the formation of a satellite, assuming the material would have some level of cohesion which would prevent it from being tidally disrupted immediately after the fission event.

\subsubsection{Debris disks}

More recently, a third possible formation scenario has been suggested that combines the two previous ideas. As mentioned, several studies on spin-up of rubble piles have found that the presence of cohesion and/or friction caused by the angle of friction of the granular material can lead to higher resistance to rotational failure \citep{Hirabayashi_et_al_2015,Sanchez_&_Scheeres_2016,Sanchez_&_Scheeres_2020,Zhang_et_al_2017,Zhang_et_al_2018,Zhang_et_al_2021,Tardivel_et_al_2018,Sugiura_et_al_2021}. In turn, as aggregates with higher mechanical strength can spin up to more rapid rates prior to mass shedding, the ejected material will have higher momentum and travel farther out into the system onto wider orbits, generating a debris disk rather than just an equatorial ridge where particles can accumulate into a satellite over time. In turn, the slow mass-shedding from movement of surface particles towards the equator during rotational failure observed in \cite{Walsh_et_al_2008,Walsh_et_al_2012} is linked with the instantaneous mass-shedding of \cite{Jacobson_&_Scheeres_2011a}, overcoming the main issues with these theories, namely the unpredictable nature of YORP and escaping ejecta, along with the tidal disruption for the contact binary fission scenario, preventing early formation of satellites.

The instantaneous mass-shedding was first explicitly tied to the formation of asteroid satellites in \cite{Hyodo_&_Sugiura_2022}, who dynamically evolved a debris disk around a top-shaped primary using three-dimensional smoothed-particle hydrodynamics (SPH) simulations. The initial conditions for their simulations were taken from the study of rotational failure by \cite{Sugiura_et_al_2021}, who also used SPH to study the deformation of rapidly rotating rubble piles. This type of numerical hydrodynamical method has been used extensively to investigate disruption and deformation of rubble pile bodies due to high-energy impacts \citep[e.g.][]{Benz_&_Asphaug_1999,Jutzi2008,Jutzi2022,Jutzi2014,Jutzi2015,Raducan_et_al_2022,Sugiura_et_al_2018}. In their simulations, \citeauthor{Sugiura_et_al_2021} applied a slow angular acceleration ($10^{-10}$ rad s$^{-2}$) to a spherical, Ryugu-like body consisting of $\sim 25,000$ similarly sized particles with random packing. Using a new treatment where they combine cohesion and friction angle into one single parameter in SPH (introduced in section \ref{section:method_sph}), they found that rotational failure for bodies with some level of cohesive strength in their model leads to debris disks with masses around 10$\%$ of the primary mass after $\sim30$ h of simulation, with their outer edge inside the theoretical fluid Roche limit. \cite{Hyodo_&_Sugiura_2022} then used the same model and numerical implementation as in \cite{Sugiura_et_al_2021} and proceeded to propagate the system with debris disks with masses of $10-20\%$ of the primary for $\sim 700$ h, studying the formation of moons in the debris disk. By the end of their simulations, they found that several moons had formed with a combined mass of $\sim 4\%$ of the primary. This is significantly more massive than Dimorphos and satellites of similar systems which are $\sim 1\%$ of the primary mass \citep{Pravec_et_al_2016,Pravec_et_al_2019}. In this work, we use a similar SPH-based model (see section \ref{section:method_sph}) to investigate the spin-up and subsequent mass-shedding of rubble pile asteroid, adding a more detailed investigation of the subsequent disk evolution, tracking the shape and deformation of formed satellites over time.

In an accompanying paper to \cite{Hyodo_&_Sugiura_2022}, \cite{Madeira_et_al_2023} used 1D hydrodynamical simulations with the code \texttt{HYDRORINGS} \citep[created to investigate the formation of satellites in the rings of Saturn in][]{Charnoz_et_al_2010,Salmon_et_al_2010} to study the viscous evolution of a debris ring around Didymos and how it can be tied to the formation of Dimorphos. In this study, the authors started their simulations directly when a disk has formed, or at the onset of mass shedding with an analytic prescription for depositing material into the disk. At the start of their simulations, the disk was always confined within the fluid Roche limit and its material then migrated over time due to viscous spreading, taking at least a year to reach this distance from the primary's surface. Any mass moving beyond the fluid Roche limit was assumed to be a satellite and turned into a spheroidal body that was subsequently tracked using an analytical treatment for its orbit, incorporating tidal effects and ring-satellite torques. The disk population was set to be monodisperse with its constituents having diameters of 1 m and equal density. 

For the case with an instantaneously formed disk, assuming an initial disk mass equal to 25\% of $M_\mathrm{Didymos}$, they found that a satellite with 0.93$M_\mathrm{Dimorphos}$ can grow within just a few days. It then migrates outwards, away from the disk due to tidal effects and dynamical resonance with its material and slowly grows via accretion of migrating material, reaching an equal mass to that of Dimorphos after more than 40 years. When material is deposited into the disk over time, they found that long deposition timescales lead to the formation of many similarly sized moonlets that migrate outwards and merge, ultimately forming one massive satellite. This behaviour, where similar mass bodies migrate through a debris disk and merge is referred to as the growth in the \textit{pyramidal regime} \citep{Charnoz_et_al_2010,Crida_&_Charnoz_2012}.

They revisit their previous work and expand on their model in the follow-up article \cite{Madeira_&_Charnoz_2024}. In order to avoid issues where ejected material escapes the system before it can be accreted by a satellite \citep{Jacobson_&_Scheeres_2011a}, they decreased the timescale for the formation of Dimorphos in the pyramidal regime by adjusting the duration for shed mass to spread to the Roche limit, while also altering the geometry of the resulting debris disk such that it extended to 1.5 primary radii from Didymos surface. With this alteration, they showed that their model can form a satellite with a mass of Dimorphos from a ring of only $0.04 M_\mathrm{Didymos}$, significantly smaller than what was indicated in the initial study. Forming in the pyramidal regime, the largest satellite undergoes a series of mergers with impacts of $\sim 2-3$ mutual escape velocities that boost its mass over time. Using the results from \cite{Leleu_et_al_2018} in a qualitative comparison, who investigated mergers between similar-sized moonlets in the pyramidal regime, the authors suggest that their model can reproduce the shape of Dimorphos, as well as the contact binary satellite Dinkinesh.

The mechanism of Dimorphos forming in a lower mass, wide disk that is not narrowly confined within a region near the primary was initially suggested by \cite{Agrusa_et_al_2024}. Basing their model on an instantaneous, single mass-shedding event detailed in \cite{Zhang_et_al_2017}, they dynamically evolve debris disks with a few percent (2-4\%) of the primary mass in a region within 1.5 primary radii from the main body surface, near the theoretical effective Roche limit for a granular aggregate with a friction angle of $35^\circ$ \citep{Holsapple_&_Michel_2006}. They evolve each system using the gravitational \textit{N}-body code \texttt{pkdgrav} \citep{Richardson_et_al_2000,Stadel_2001}. While the code is based on spherical particles it uses SSDEM to handle contacts and allows for interparticle friction, along with plastic twisting and rolling friction, approximating the behaviour of granular aggregates consisting of material with a given friction angle \citep{Schwartz_et_al_2012,Zhang_et_al_2017}. Providing a proof of concept simulation, detailing the evolution of a rapidly spinning aggregate accelerated by YORP with a friction angle of 40$^\circ$ consisting of $\sim 90\ 000$ particles with diameters between 4 and 16 m (consistent with observations of boulders on Dimorphos surface \citep{Pajola_et_al_2023}), they demonstrate the full process of mass-shedding near the equator of the object leading to the formation of Dimorphos-like satellite in about five days. For the remainder of the simulations, they instead let the primary be a rigid, hollow and oblate object consisting of many bound spheres with properties similar to Didymos (as a solid body). Limiting the minimum particle size to 6 m in diameter, they could perform over 100 simulations spanning 100 days as the total particle numbers varied between $\sim4\ 200$ and $\sim 8\ 400$. Also varying the friction angle of the material along with the total mass of the disk, they obtain a large volume of simulations with varying parameters.

The results of the study indicate that the formation process of binary asteroid satellites is highly chaotic and unpredictable. Nevertheless, they conclude that a disk mass of $0.02M_\mathrm{Didymos}$ can produce moons with masses and diameters similar to that of Dimorphos, with the process becoming more consistent for a disk mass of $0.03 M_\mathrm{Didymos}$. Assessing the size and orbital properties of their satellites, they find that they most often end up in a state of synchronous rotation with prolate shapes, likely due to the outer boundary of their disk being near the theoretical Roche limit. Yet, a handful of the simulations produce the more atypical oblate shapes and some of the satellites end up with very prolate shapes.

In this work, we aim to further explore the details of the dynamical mechanisms that determine the shape of asteroid satellites by performing state-of-the-art \textit{N}-body simulations using irregularly shaped particles and studying the mechanics that deform granular aggregates in this formation regime, focusing on analysing disk geometry, tidal encounters and impact properties for mergers.

\section{Method}\label{section:method}

In an attempt to accurately model the complex dynamical mechanics related to the deformation of rubble pile asteroids and subsequent satellite formation, we have opted to divide the process into two stages. The first is characterised by a deformation event caused by spin-up via e.g.\ YORP. Numerical hydrodynamical codes, such as SPH, are a good choice for modelling deformation as they use very high resolution, and excel at tracking structural reshaping \citep[e.g.][]{Raducan_et_al_2024a}. Moreover, as the implementation evaluates contact mechanics over a smoothed continuum rather than discrete particle--particle interactions, it becomes computationally cheaper than \textit{N}-body simulations. The next stage is represented by the dynamical evolution of the ejected mass settled into a debris disk. This regime is modelled using a gravitational \textit{N}-body code. A number of different studies have previously combined SPH and \textit{N}-body in this manner to investigate rubble pile dynamics, mainly related to asteroid collision, fragmentation and disruption \citep{Michel_et_al_2001,Michel_et_al_2002,Michel_Benz_&_Richardson_2003,Michel_Benz_&_Richardson_2004,Durda_et_al_2004,Durda_et_al_2007,Benavidez_et_al_2012,Ferrari_et_al_2022}.

In order to transition between the different codes a hand-off is required \citep[e.g.][]{Ballouz_et_al_2019}. Said operation is not trivial due to two main issues: (i) Since SPH particles do not represent discrete particles, but constituents of a smoothed continuum, they tend to overlap in the physical space. In turn, it is not possible to translate them one-to-one as physical particles in a gravitational \textit{N}-body setting; (ii) SPH simulations require a high resolution, i.e.\ large number of particles (on the order of $\sim10^6$), to properly resolve e.g.\ mass-shedding or deformation. However, using that many particles offers little improvement in capturing key physical effects in an \textit{N}-body simulation and simply leads to a large increase in elapsed real time until it is resolved. To facilitate the hand-off between SPH and \textit{N}-body, we have developed an interface that identifies clusters of SPH particles and groups them as single, angular fragments with physical properties based on their constituents.

We now go through the details of the two different codes and proceed to introduce the developed interface that performs the hand-off. Finally, we summarise the simulation setups.

\subsection{Smooth particle hydrodynamics code}\label{section:method_sph}

In this study, we use the Bern SPH code to simulate the rotation and breakup of our asteroid. It is a shock-based grid-free numerical hydrodynamics code created to simulate various geological materials. Originally developed by Benz \& Asphaug \citep{Benz_&_Asphaug_1994,Benz_&_Asphaug_1995} to model collisional fragmentation of rocky bodies, the code has been expanded with additional physics for a more accurate representation of materials. Some features of this code include various equations of state, modelling of porous \citep{Jutzi2008} and granular materials, pressure-dependent strength \citep{Jutzi2015}, and a tensile fracture model making the code highly suitable to simulate and analyse asteroids with diverse material properties. The simulations produced have been tested extensively and are in close agreement with experimental results \citep[e.g.][]{Jutzi2015}. A recently developed fast integration scheme allows for the modelling of low-velocity granular flow \citep{Jutzi2022,Raducan_et_al_2024a}. 


For the modelling of friction and cohesion, we follow the approach by \cite{Sugiura_et_al_2021} who showed that interparticle cohesion has a significant effect on the rotational failure modes of rubble-pile asteroids. To implement this effect into their SPH simulations, they opted to use a parameter referred to as the \textit{effective friction angle}, $\phieff$. The parameter was introduced as a simplification to reduce the number of variables that the deformation modes could depend on when representing shear strength of a cohesive granular material as $Y = \tan(\phi) p + C$, where $\phi$ is the true friction angle, $p$ is the confining pressure and $C$ is the cohesive strength. Assuming that the effective friction angle can account for cohesion between the granular particles in their simulations, the expression was rewritten as $Y = \tan(\phieff)p$. For reference, a value of $\phieff=0^{\circ}$ means the granular media will behave like a fluid. \citeauthor{Sugiura_et_al_2021} found that for cases where $\phieff\leq60^{\circ}$, the rotation spheroid will undergo internal deformation, while surface landslides begin to occur for values $\phieff\geq70^{\circ}$. They also note that the landslides are axisymmetric when spinning up the initial, spheroidal rubble pile over the course of a few days, meaning the body deforms into an axisymmetric top-shaped object. For longer timescales, leading to slower spinups, above one month, the landslides are localised and result in a nonaxisymmetric shape for the body. We note that this treatment for cohesion is still under some scrutiny \citep[see][]{Agrusa_et_al_2024} and will need further study.

\subsection{\texttt{GRAINS} - an \textit{N}-body code for angular particles}\label{section:method_nbody}

To investigate post rotational failure dynamics of the asteroid system, we employ the gravitational \textit{N}-body DEM (Discrete Element Method) code \texttt{GRAINS} \citep{Ferrari_et_al_2017,Ferrari_et_al_2020}. The software has been directly developed for the purpose of studying granular mechanics in rubble pile asteroids and allows for the use of non-spherical, angular particles. Using the \texttt{c++} libraries of \texttt{Chrono::Engine} \citep[\texttt{Chrono},][]{Chrono2016} in combination with a GPU acceleration scheme, \texttt{GRAINS} can perform numerical integrations of complex systems with thousands of particles. The use of angular particles allows us to replicate physical effects that are more difficult to capture when using spheres to represent the shape of each body, such as interlocking, off-centre collisions and particle spin orientation \citep{Korycansky_&_Asphaug_2006,Korycansky_&_Asphaug_2009}. Physical effects characteristic for granular media can be approximated using sphere-based codes such as \texttt{pkdgrav} with the inclusion of a linear spring-dashpot method with multi-directional friction and a shape parameter that gives a statistical measure of a non-spherical shape \citep{Schwartz_et_al_2012,Zhang_et_al_2017}. It was shown by \cite{Ferrari_&_Tanga_2020} using \texttt{GRAINS} that the use of non-spherical particles increases our ability to capture key dynamical effects in simulations of granular mechanics. A similar study for \texttt{pkdgrav} has been performed by \cite{Marohnic_et_al_2023} who also concluded that the use of non-spherical particles, here in the form of aggregates of bounded spheres, affects the dynamics of rubble piles. As for resolving contacts between particles in \texttt{GRAINS}, there exist capabilities for using both non-smooth (impulse-based) and smooth (force-based) implementations. In this work we have opted to use only the Smooth-Contact Method (SMC) of \texttt{Chrono}, which behaves similarly to soft-sphere contact methods implemented in other \textit{N}-body DEM codes \citep{Sanchez_&_Scheeres_2011,Schwartz_Michel_&_Richardson_2013} and is optimal for resolving low-velocity and long-lasting contacts.

Regarding the SMC method for computing contacts (i.e.\ force-based), it is based on a two-way normal-tangent spring-dashpot system \citep{Ferrari_et_al_2020}. For this problem we use a non-linear Herzian model which is described in detail in \cite{Fleischmann_et_al_2016}. There are several parameters that govern surface interactions including static and sliding friction; restitution; adhesion; stiffness; and damping. We kept these parameters at their default values which have been used to verify \texttt{Chrono} against the behaviour of granular material in laboratory studies \citep{Chrono2016}. Energy dissipation during contact between the particles is mainly determined by the coefficient of restitution set to make collisions inelastic. Moreover, the adhesion coefficient is kept at zero, meaning that aggregates formed in our simulations are kept together purely via self-gravity. Note that the way these coefficients relate to granular behaviour in space is an ongoing field of study and will be further constrained in the future.

\subsection{SPH to \textit{N}-body interface}\label{section:method_handoff}

In our interface, each aggregate is created using a friends--of--friends algorithm \citep[e.g.][]{Huchra_&_Geller_1982} with a dynamic linking length based on the combined SPH smoothing length of a particle pair, $i,\ j$, given by $\dlink = 0.5(h_i+h_j)$. The smoothing length is the characteristic radius of the kernel function and defines the range of influence from neighbours interacting with a given particle $i$. For our spin-up model, this value varies between 20 m for the disk of a Didymos-like primary and 25 m for the Ryugu-like primary case. To avoid outlier values in the smoothing lengths, which can occur for isolated single SPH particles and lead to unrealistic fragments, we require that each smoothing length is within 5\% of the median smoothing length value in the system. While there are other prevalent clumping algorithms, such as DBScan \citep{Ester_et_al_1996} or K-means, which has been used in combination with machine learning to identify clusters in SPH simulations \citep{Sakong_et_al_2019}, friends--of--friends is better suited for this problem due to the fact that it does not rely on a pre-defined number of fragments (K-means), static search distance for the fragment or struggle with large variations in density (DBscan). Hence, our implementation is completely dynamic and only relies on the smoothing length of the SPH particles without the need for additional input that could significantly affect the outcome of the hand-off. To briefly summarise the algorithm, it superimposes a cubic lattice on a three-dimensional computational box and generates a linked list by mapping the sorted values of the SPH particle coordinates along the axes of the box to the cells of the lattice. Afterwards, we choose a particle, $i$, and check its distance to all potential neighbours inside the lattice which are within one smoothing length, $h_i$ in every direction. If the distance between two particles fulfils $|\Vec{r}_i-\Vec{r}_j|\leq \dlink$, they are part of the same fragment.

After it has been generated, each fragment gets assigned physical properties based on its constituents. Summing up the mass, we determine the body's barycenter and compute the corresponding position and velocity relative to the origin. We also determine the resulting orbital angular momentum of each fragment and compare it to the total corresponding sum for each SPH particle that has been used to generate it. If the total change in angular momentum in the system is greater than 10\%, we reject the solution and attempt to tweak the input parameters to generate a more physically accurate output. 

The ability of \texttt{GRAINS} to perform integrations with angular particles allows us to further improve the hand-off by capturing complex granular dynamics during the evolution of the disk. Hence, instead of simply merging the fragment constituents into a sphere, we can assign convex or concave shapes for each fragment based on the distribution of their SPH particles. This is accomplished using a geometrical algorithm based on Delaunay triangulation, known as $\alpha$-shape \citep{Edelsbrunner_&_Mucke_1994}, which has been used in previous studies of asteroid dynamics such as \cite{Ferrari_&_Tanga_2020}. The method was also used in the SPH to \textit{N}-body interface by \cite{Ballouz_et_al_2019} in the scope of asteroid family formation, but just for the largest remnant in the system while we have expanded the use of $\alpha$-shapes to all granular aggregates in the system. 

\subsubsection{$\alpha$-shape implementation}

To elaborate, the geometric algorithm allows us to determine the concave shapes of fragments, with their individual SPH particles acting as vertices. We require that the $\alpha$-shape has a density that falls inside a pre-defined range based on a target density, $\rho_\mathrm{target}$, given by $ (1-\kappa)\rho_\mathrm{target}\leq\rho_\mathrm{frag}\leq (1+\kappa)\rho_\mathrm{target}$, where $\kappa\in(0,1)$. For our purposes, we found that a tolerance of $\kappa=0.05$ worked well. If the density of a fragment is too small, the most distant vertex of the shape is removed and a new $\alpha$-shape is created. Note that the fragment still retains the same physical quantities, such as mass and velocity, which are calculated relative to the centre-of-mass of its SPH particle members. Shapes with less than 20 constituents are instead given a randomised convex hull by filling a box with a volume based on the target density value with 30 randomly placed vertices.

Due to this approach, the high resolution of the SPH simulation is preserved while removing any overlap between particles and improving performance of the \textit{N}-body code. The main drawback of $\alpha$-shape when implemented in interfaces such as ours has been that it needs input based on a characteristic size for individual SPH particles, which makes little sense physically as they represent infinitesimal fluid elements. Moreover, an additional input referred to as $\alpha$ is required, which sets the concavity of the final three-dimensional shape. It can be thought of as the squared radius of a ball rolled over the vertices of a point set which yields the final surface of the shape. A small value for $\alpha$ increases the concavity, providing a higher resolution of the vertex distribution while the shape becomes convex when $\alpha\to\infty$. To avoid choosing the values of these parameters arbitrarily, we use an improved approach where $\alpha$ is computed individually for each fragment based on the minimum value needed for a set of vertices to form a single coherent shape. This method is available in the \texttt{Alpha\_shapes\_3} library \citep{cgal:alpha_shapes_3} of the Computational Geometry Algorithms Library \citep[\texttt{CGAL},][]{cgal:eb-23a}. Furthermore, \texttt{CGAL} has two modes for its $\alpha$-shape computation: \textit{generalised} and \textit{regularised}. While we leave the details of these implementations to the library documentation, the regularised version will exclude any singular facets from the shape, meaning the algorithm automatically identifies and excludes outlier members of the fragment from its final geometrical form. In turn, using the regularised mode ensures that we do not accidentally combine two separate fragments as one. The formal definition for the optimal value, $\alpha_\mathrm{opt}$, is that it must satisfy the following two conditions: (i) Each data point in the set is contained within or on the boundary of the regularised alpha shape; (ii) The alpha shape has a number of solid constituents that is equal to or smaller than the specified number of components in the final output \citep{cgal:alpha_shapes_3}. Since this provides us with the highest possible resolution for the resulting shape, the final product is not necessarily representative of real asteroid surfaces. To end up with a smoother output, we chose to multiply $\alpha_\mathrm{opt}$ with a factor of 10 to serve as the initial guess for each fragment shape when performing the hand-off. Artificially inflating the $\alpha$ value like this only marginally affects the density and volume, but removes artifical crevices and craters on the surfaces which can produce undesired interactions between fragments when resolving contacts in \texttt{GRAINS}.

All-in-all, the use of \texttt{CGAL} in our interface greatly facilitates our modelling of granular mechanics in rubble pile asteroids.

\subsection{SPH simulation setup}\label{section:metod_sph_simulation_setup}

The SPH spin-up simulations were done for two separate cases, the first being a Ryugu-like body and the second Didymos-like. The two bodies differ in size and estimated bulk density, where Ryugu has a value of $\rho_0 = 1190\kcm$ \citep{Watanabe_et_al_2019} and Didymos $\rho_0 = 2400\kcm$ \citep{Daly_et_al_2023a}, allowing us to investigate whether the bulk density affects the mass and shape of the secondary asteroid in the system post rotational failure. We note that the density value for Didymos was chosen before the latest values of $2760\pm130\kcm$ from \cite{Naidu_et_al_2023} were published. For both scenarios, we generate spheres consisting of 50 165 particles. The sizes of the spheres correspond to 500 m and 400 m for Ryugu and Didymos, respectively. 
We used the fast integration scheme \citep{Jutzi2022,Raducan_et_al_2024a} which allows us to carry out the SPH simulations for sufficiently long times. The spheres were initially rotating with a period of $\approx$ 3.6 h. The final spin-up was then performed by using an angular acceleration of $ d \omega/dt \sim 10^{-9} s^{-2}$, until rotational failure occurred.

For each main body case, we used two different values for the effective friction value. Given that the goal was to investigate cases where a distinct debris disk is formed post rotational failure, we opted for values such that $\phieff> 50^{\circ}$, allowing us to avoid scenarios where the spin-up only results in internal deformation.

The SPH simulations were run until $3\times10^5$ s to ensure that significant mass had been lost from the primary such that it could settle in a resulting debris disk.

\subsection{\textit{N}-body simulation setup}\label{section:method_nbody_setup}

\begin{figure*}
    \resizebox{\hsize}{!}{\includegraphics[trim=0 0 0 0, clip]{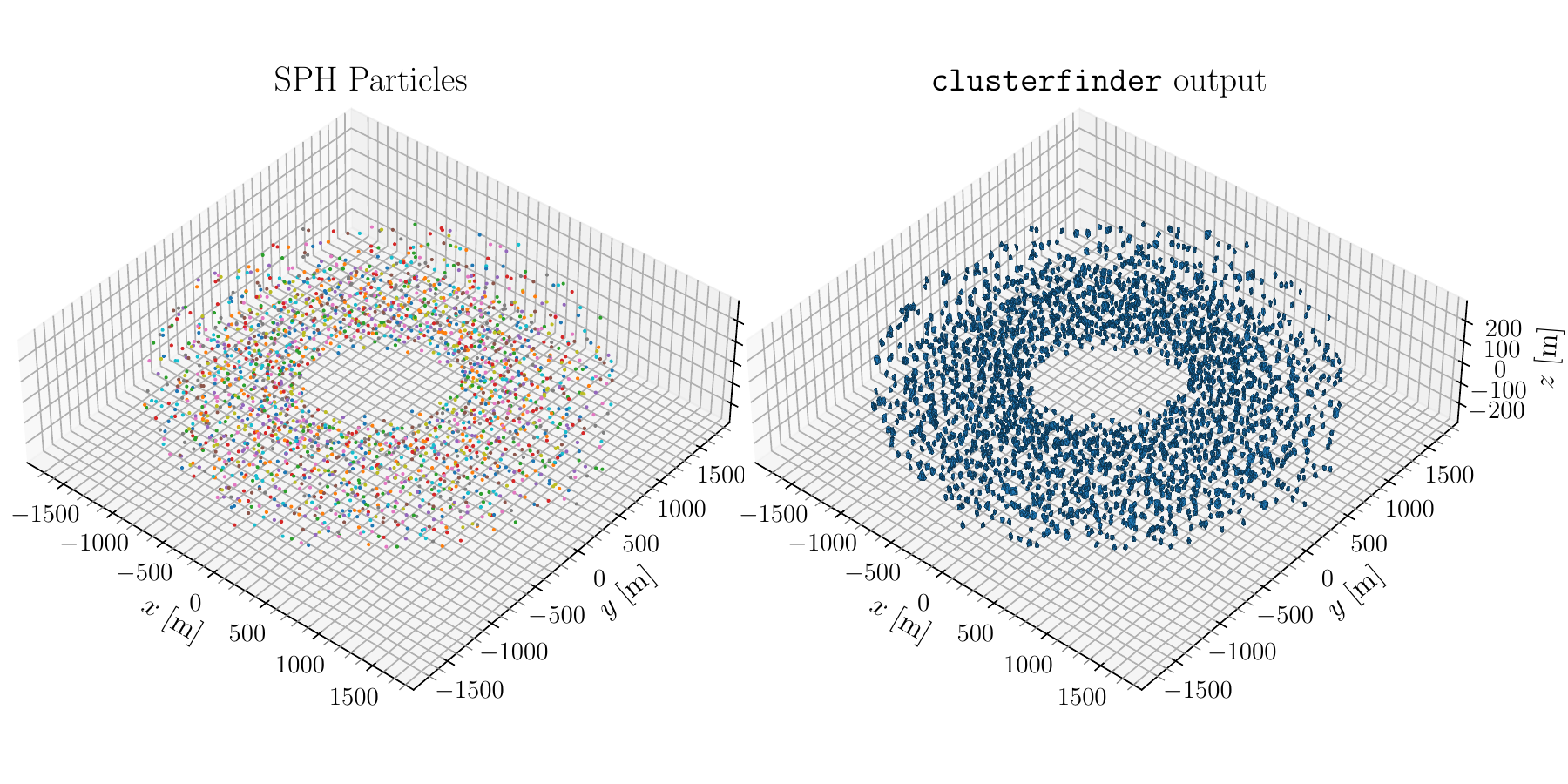}}
    \caption{\textbf{Left:} the position of the SPH particles at the time of the hand-off for a symmetric debris disk formed from a Ryugu-like primary. \textbf{Right:} the final output of \texttt{clusterfinder}. The algorithm has generated $\alpha$-shapes and convex hulls with volumes matching the densities and masses of SPH particle clusters. Zooming in will show the resulting 3D shapes with black grid lines marking the edges that separate the faces in blue.}
    \label{fig:ryugu_handoff}
\end{figure*}

\begin{figure*}
    \resizebox{\hsize}{!}{\includegraphics[trim=0 0 0 0, clip]{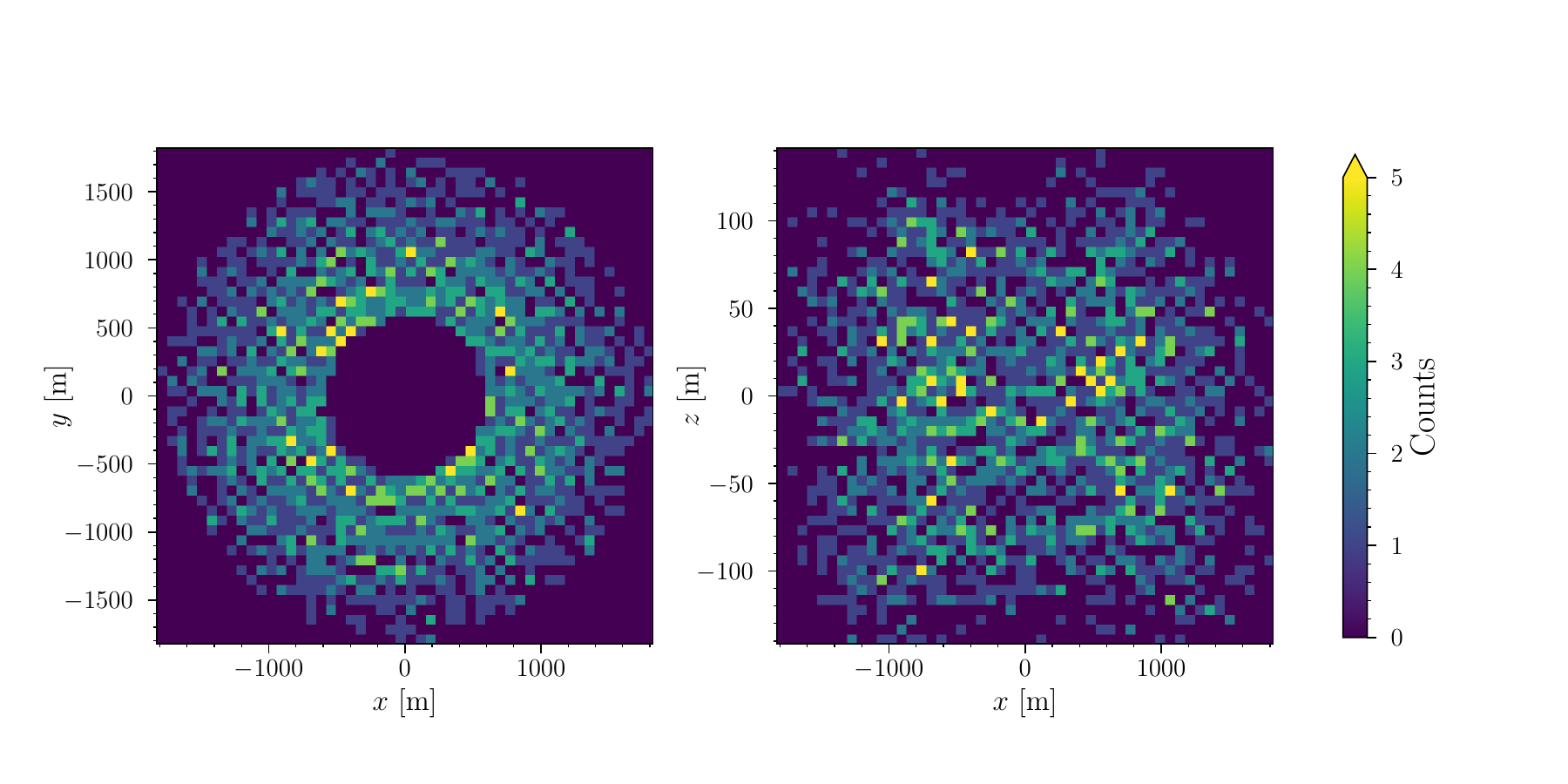}}
    \caption{\textbf{Left:} a histogram with 50 bins along each axis showing the number density of particles the disk obtained after performing the hand-off for the example simulation \texttt{f60\_R2} in the $xy$-plane. Corresponds to the same distribution seen in figure \ref{fig:ryugu_handoff}. \textbf{Right:} the same metric but for the $xz$-plane. Note the differing scales between the $x$- and $y$-axis of this plot.}
    \label{fig:ryugu_handoff_density}
\end{figure*}

Analysing several debris disks produced in the SPH simulations, we could narrow down the spread of properties to generate initial conditions for the \textit{N}-body simulations. Sticking to the case of $\phieff = 60^{\circ}$, the average mass of the debris disk ended up being around $\sim5\%$ of the initial primary mass, radially extending about 1 km ($2 R_\mathrm{main}$) from the inner edge with a thickness of $\sim200$ m ($0.4R_\mathrm{main}$). Assuming that particles within the range $R_\mathrm{main}<r_\mathrm{p}<R_\mathrm{main}+100$ m would get reaccreted onto the main body, we set the inner edge of the debris disk as $R_\mathrm{main}+100$ m. An example of a disk containing SPH particles compared with the angular particles post hand-off is shown in figure \ref{fig:ryugu_handoff}. The number density of the disk post hand-off can be found in figure \ref{fig:ryugu_handoff_density}.

At the point of the hand-off for each SPH simulation, the expansion of the disk had mostly halted from dissipation due to gravitational interactions with the primary and the rest of the shed matter, along with inelastic particle--particle contacts. As a result, using the velocities for each particle computed by \texttt{clusterfinder}, we found that their orbits were effectively circularised when passing the initial conditions to \texttt{GRAINS}. Hence, when generating disks for the standalone \textit{N}-body simulations we simply gave each particle a Keplerian velocity for a circular orbit and noticed no apparent difference in the behaviour of the disk during its evolution compared to the hand-off simulations. As for the individual spin of particles, there is no way of inferring any angular velocity from the SPH data so we opted to not add any artificial spin at the onset of each \textit{N}-body simulation.

For the resulting particle sizes from ten different spin-up simulations, we found that the minimum boulder diameter was $\sim 20$ m, which corresponds to single, isolated SPH particles. The mean diameter varied between 25.3 m and 25.9 m for the different disks. The lower limit is inconsistent with observed sizes of rubble on asteroids, where the minimum diameters on for example Dimorphos surface appear closer to $\sim1$ m \citep{Pajola_et_al_2023,Robin_et_al_2023}. This is a direct result of the resolution of the SPH simulation determined by the number of particles in the primary, which sets a minimum mass and size for any isolated grains produced during the hand-off procedure. The exponential SFD we used to generate random particle distributions for the standalone \textit{N}-body simulations satisfies 

\begin{equation}\label{equation:particle_sfd}
    f_{D_\mathrm{p}}(D_\mathrm{p},D_\mathrm{min},D_\mathrm{mean}) = e^{-(D_\mathrm{p}-D_\mathrm{min})\lambda_D},
\end{equation}

where $\lambda_D = 1/(D_\mathrm{mean}-D_\mathrm{min})$. While using a distribution based on the mean and minimum diameter values from the SPH simulations could replicate the corresponding, steep SFD, this would diverge from SFDs derived using observational data. In turn, we opted for a flatter distribution with a smaller minimum value for the diameter. Noting that an average diameter of 25 m also significantly increased the number of particles up towards $\sim10^4$, which would severely slow down the simulations, we also increased the $D_\mathrm{mean}$ value. To keep the lapsed real time down for resolving the simulations, the final SFD was set to have $D_\mathrm{min} = 5$ m and $D_\mathrm{mean}=30$ m. This distribution also sometimes generates particles with diameters close to 100 m, which is much larger than the maximum size of particles observed on the surface of Dimorphos. However, the maximum boulder size observed on the surface of Didymos is 93 m with a lower limit of $\sim$16.5 m set by the resolution of images from the DART mission \citep{Pajola_et_al_2023}. The particles generated using \texttt{clusterfinder} can also reach these values when there has been significant clumping during the mass shedding phase with maximum values of 54.5 m for the Ryugu-like disk (see section \ref{section:ryugu_spinup}) and 125.8 m for the Didymos-like systems (see section \ref{section:didymos}). Nevertheless, given that the SFD of particles on Didymos likely has a much lower minimum size than indicated by the low spatial resolution images of its surface along with the sizes observed on Dimorphos, it is clear that the resolution of our SPH simulations introduces a possible limitation in our \textit{N}-body modelling. Consequently, we emphasize that not each body used in the simulations corresponds to an individual grain, but also clusters of particles that are represented as one monolithic body. Given that our SPH model includes some level of cohesion built into the effective friction angle implementation, clustering occurs even early on during the mass-shedding phase in accordance with \cite{Tardivel_et_al_2018}. The level of clustering does not change significantly if we choose an earlier or later point to perform the hand-off. The fact that these clusters survive within the fluid Roche limit could be explained if they represent large chunks of matter, or small aggregates kept together by e.g.\ particle interlocking or the presence of fine-grained material $\lesssim 10 \mu$m providing cohesive strength to the body \citep[e.g.][]{Sanchez_&_Scheeres_2014}. The overall potential presence of finer grain material in the system after the mass-shedding event would have little consequence for our simulations as they would be lost from aggregates during mergers or tidal disruptions and subsequently ejected from the system due to solar radiation pressure \citep{Ferrari_et_al_2022}. In turn, we do not include any cohesion for the \textit{N}-body simulations. Moreover, our implementation does not yet include any fragmentation for these larger boulders in our system, causing them to stay intact no matter the stress they endure during the course of the simulation due to e.g.\ spin, tides or collisions, which might introduce unphysical behaviour. The implications of our SFD choice are further discussed in section \ref{section:discussion_particle_sfd}. 

\begin{figure}
    \resizebox{\hsize}{!}{\includegraphics[trim=0 0 0 0, clip]{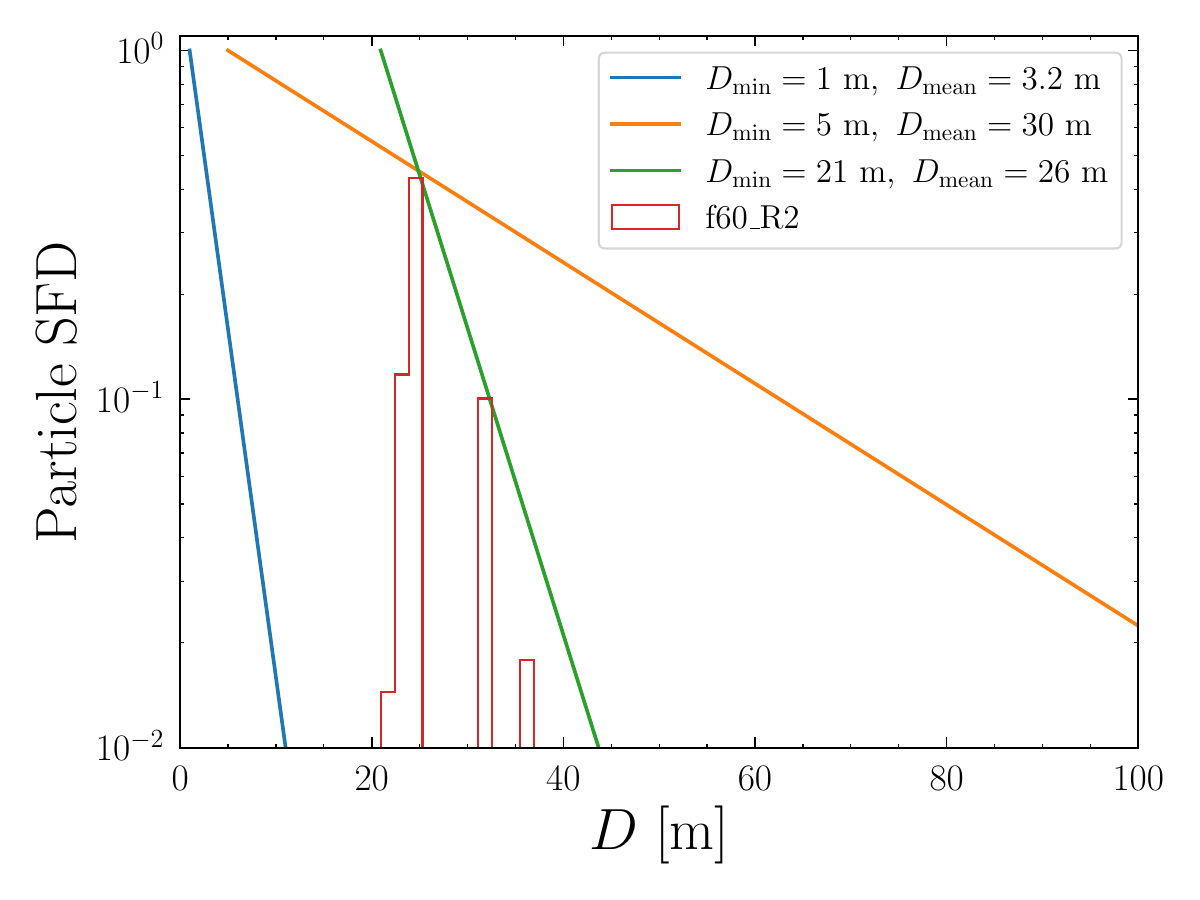}}
    \caption{This figure depicts three different SFDs mentioned in the text, as well as the probability density of the particle SFD obtained from the hand-off \texttt{f60\_R2} as a histogram with 20 bins, which is our example simulation discussed in section \ref{section:ryugu_hand_off_results}. The SFDs have been generated using the previously defined exponential function in equation \ref{equation:particle_sfd}. The blue curve is our closest approximation to observed sizes on Dimorphos surface, while the green matches the particles from the \texttt{f60\_R2} hand-off. The orange curve shows the much flatter distribution we use to randomly generate particles for our standalone \textit{N}-body simulations.}
    \label{fig:particle_sfds}
\end{figure}

Three different SFDs generated with our exponential distribution, as well as a probability density of particle diameters obtained from an example hand-off discussed in section \ref{section:ryugu_hand_off_results}, have been plotted in figure \ref{fig:particle_sfds}. The steepest function (blue) uses a minimum and mean that yields distributions that have particles with diameters between 1 and $\sim16$ m, which are the minimum and maximum diameters observed on Dimorphos surface \citep{Pajola_et_al_2023}. The green curve with $D_\mathrm{min}=21$ m and $D_\mathrm{mean}=26$ m approximates the SFD of particles generated from the SPH to \textit{N}-body hand-off. Finally, the orange curve shows the chosen SFD for randomly generating debris disks for our standalone \textit{N}-body simulations.

Furthermore, the shape generation scheme involves randomly generating $N_\mathrm{vert}$ vertices in a square box with a side of $D_\mathrm{p}$ and then computing the resulting convex hull. This often creates particles that are coarser than observed grains on the surface of Dimorphos and other rubble pile asteroids \citep{Robin_et_al_2023}. This trend and its consequences for our simulations is also discussed in section \ref{section:discussion_particle_sfd}.

As for the primary body, we opted to use solid, monolithic bodies in order to reduce the total number of particles to track during the simulations\footnote{We note that we detected an issue in \texttt{GRAINS} during the analysis of the simulation data that caused disk particles to sometimes pass through the boundary of the primary body, not properly colliding with the generated mesh. Re-running two of our simulations with a corrected version of the code, we concluded that the observed dynamical mechanisms that govern the shape of satellites during the formation phase were still valid. In turn, the conclusions drawn from the results obtained within this work still hold.}, significantly increasing performance. For Ryugu, we used a simple sphere of radius $R_\mathrm{Ryugu}=500$ m, a bulk density of $1190\kcm$ \citep{Watanabe_et_al_2019} and corresponding moments of inertia for a spheroid. In the second case where we tried to recreate the Didymos-Dimorphos system, we opted to use a derived pre-impact shape from the DART mission for the main body with a radius of around $R_\mathrm{Didy}=390$ m. The corresponding shape also has pre-defined estimates for the moments of inertia, which we fed into the simulation while also giving it a density of $2400\kcm$ \citep{Barnouin_et_al_2023}. For the primary rotation rate, we opted to use the pre-impact value of 2.26 h measured by \cite{Pravec_et_al_2006}.

Our initial conditions post rotational failure show disks that are more radially and vertically extended than those dynamically evolved in \cite{Agrusa_et_al_2024}. In their work, the disks extended radially for 1.5$R_\mathrm{main}$ and $0.05R_\mathrm{main}$ (20 m) in each direction along the $z$-axis, making them narrower by half a primary radius and much flatter. This can be attributed to the nature of the rotational failure of the spinning primary asteroid. The mass-shedding in \cite{Agrusa_et_al_2024} occurs right by the equator, while the primary in our model ejects mass in a wider region that extends $\sim 200$ m above and below the equator. The two models differ mainly in the fact that they use \textit{N}-body simulations to perform the spin-up for a Didymos-like shape while our model is based on SPH and uses spherical aggregates to approximate the shapes of Ryugu and Didymos. Moreover, the resolution of the spin-up simulations in \cite{Agrusa_et_al_2024} is higher at $\sim 90\ 000$ particles while we use $\sim 50\ 000$. With the many differences between the two simulation methods, it is not trivial to pinpoint what could be the major cause of this deviation in disk geometry as it would require an in-depth analysis. In the future, we aim to compare spin-up simulations between our SPH code and \texttt{GRAINS} to better understand any differences in the disks they generate. However, for the purposes of this article, we direct the focus to the disk evolution post rotational fission and we compare our results to \cite{Agrusa_et_al_2024} in more detail in section \ref{section:agrusa_comparison}.

\subsubsection{Numerical setup}

For the \texttt{GRAINS} simulations, we opted for a Barzilai-Borwein solver \citep{Barzilai_&_Borwein_1988}, which is an iterative gradient descent method which offers good convergence, using a tolerance of $10^{-4}$ and a maximum of 300 iterations to reach a solution. The time step was kept below 0.5 s, going as low as 0.01 s for complex debris disks to ensure that contacts between particles could be resolved accurately.

In order to detect the fragments in the \texttt{GRAINS} simulations of the debris disk evolution, we opted for a modified version of the \texttt{clusterfinder} algorithm from section \ref{section:method_handoff} with a linking length of $\dlink = \max(D_i,D_j)$. In this method, we skipped the shape determination scheme and simply saved the $\alpha$-shapes or convex hulls representing each aggregate constituent and used them to generate a composite shape. Moreover, due to the angular nature of each grain, we determined the moment of inertia of the aggregate using the parallel axis theorem. With this value at hand, we proceeded to compute the rotational and orbital angular momenta for each detected cluster at each time step in the simulation. 

\section{Results: Ryugu scenario}\label{section:results}

We begin by introducing the results from the SPH spin-up, going through the disk formation process and proceed by presenting the results from two specific hand-offs and the subsequent dynamical evolution of said disk, which ultimately leads to the formation of a satellite. After that, we show results from five less computationally demanding standalone \textit{N}-body simulations.

\subsection{Spin-up and rotational failure of rubble pile}\label{section:ryugu_spinup}

All of the spin-up cases, where we let $\phi_\mathrm{eff} = 60^{\circ}$, behave similarly as the failure of the rotating primary body occurs at around $8\times10^4$ s (in the final spin-up phase; see section \ref{section:metod_sph_simulation_setup}). The mass shedding is localised, leading to chunks of matter getting ejected near the equator, which is consistent with previous studies of rotational failure \citep{Hirabayashi_et_al_2015,Tardivel_et_al_2018,Ferrari_&_Tanga_2022}. However, due to the fast rotation of the body, the resulting flow of matter quickly becomes symmetric. Moreover, because of the slight increase in strength that the body gets with this effective friction angle (and correspondingly higher rotation rate at failure), the matter flowing into orbit carries significant momentum with velocities greater than the orbital speed and travels several primary radii before it settles into an orbit due to energy dissipation caused by gravitational interactions with the rest of the disk and contacts between particles..

\begin{table}[]
    \centering
    \begin{tabular}{lllll}
        \hline\hline
         Run & $M_\mathrm{disk}\ [M_\mathrm{main}]$ & $N_\mathrm{p}$ & $D_\mathrm{frag}$\ [m] & $D_\mathrm{mean}$\ [m] \\
         f60\_R0 & 0.049 & 2005 & [20.9, 47.3] & 25.8 \\
         f60\_R1 & 0.033 & 1485 & [19.9, 45.9] & 25.3\\
         f60\_R2 & 0.055 & 2233 & [20.9, 50.0] & 25.9 \\ 
         f60\_R3 & 0.057 & 2301 & [20.4, 54.5] & 25.8\\
         f60\_R4 & 0.052 & 2112 & [20.5, 48.8] & 25.9\\
         f60\_R5 & 0.049 & 2054 & [20.0, 43.5] & 25.6\\
         f60\_R6 & 0.043 & 1847 & [19.5, 43.4] & 25.4\\
         f60\_R7 & 0.035 & 1478 & [20.2, 48.8] & 25.5\\
         f60\_R8 & 0.042 & 1814 & [20.2, 43.0] & 25.5\\
         f60\_R9 & 0.039 & 1643 & [21.3, 43.6] & 25.5\\
         \hline
    \end{tabular}
    \caption{The distribution of disk masses, number of particles and fragment sizes for the post rotational failure debris disks created by spinning up a Ryugu-like primary using SPH.}
    \label{tab:ryugu_spin_up}
\end{table}

Figure \ref{fig:ryugu_handoff} shows an example of the hand-off result for a Ryugu-like case simulation with $\phi_\mathrm{eff} = 60^{\circ}$ that takes place 38.8 h ($1.4\times10^5$ s) after the onset of the spin-up simulation. The resulting disk has a width of $\sim1$ km (2$R_\mathrm{main}$) measured from its inner edge at 600 m from the centre of the primary and a thickness of 200 m. The disk has a mass of 0.055$M_\mathrm{main}$ containing 2233 particles and the fragment size distribution computed from the volume equivalent diameter fulfils $D_\mathrm{frag} \in [20.9,50.0]$ m with an average size of $D_\mathrm{mean} = 25.9$ m. The corresponding parameters for the remaining spin-up cases with the same initial properties for the primary can be found in table \ref{tab:ryugu_spin_up}. Due to the chaotic nature of the rotational failure, each resulting disk has slightly different properties.  

\subsection{Dynamical evolution post SPH to \textit{N}-body hand-off}\label{section:ryugu_hand_off_results}

\begin{figure*}
    \resizebox{\hsize}{!}{\includegraphics[trim=65 0 65 0, clip]{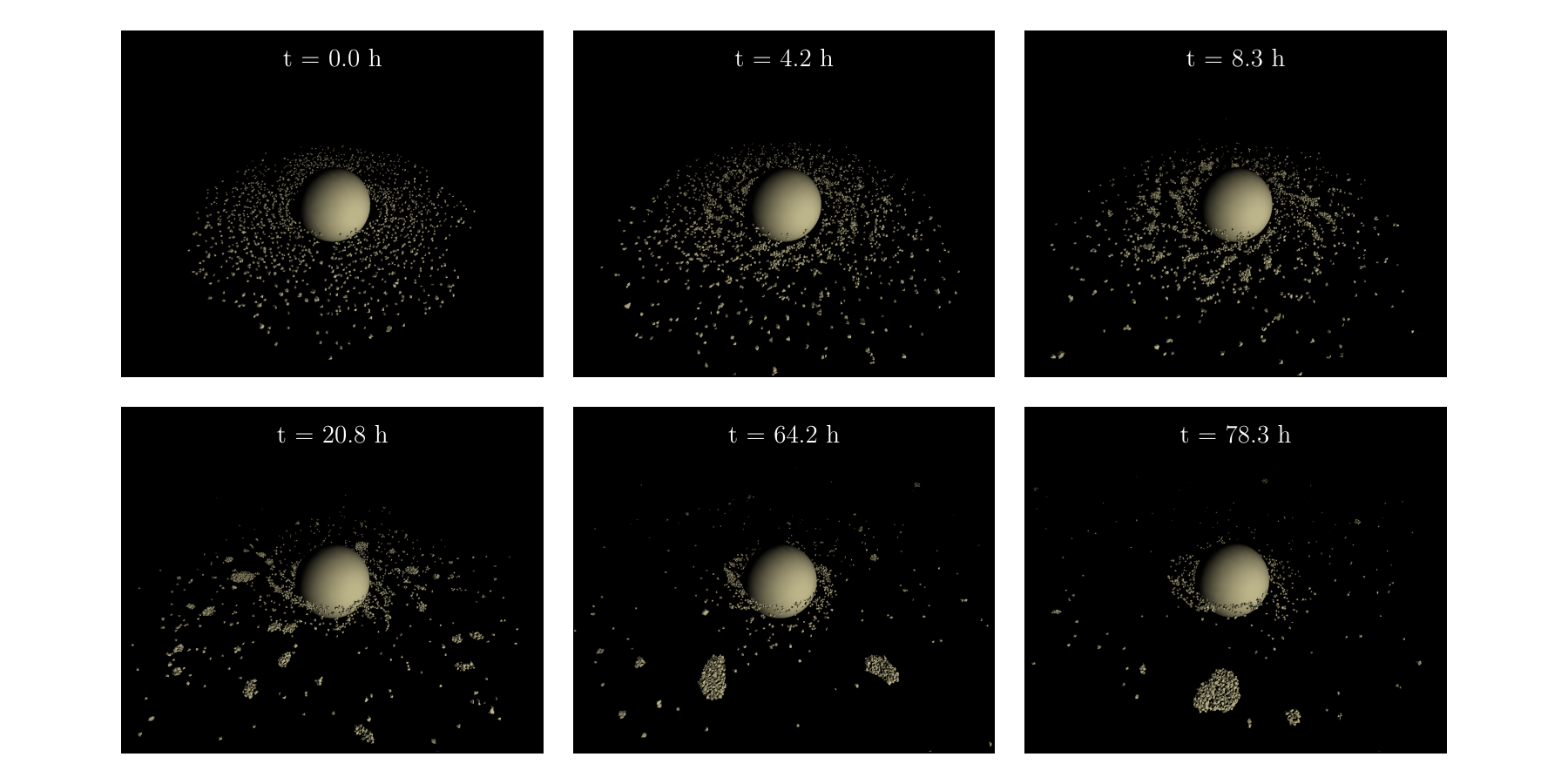}}
    \caption{Snapshots from the hand-off simulation \texttt{f60\_R2} where a debris disk around a Ryugu-like primary quickly forms a satellite over the course of 73 hours. The final shape of the moon is largely determined by the merger between the two largest remnants in the system, occurring during the period separating the two final snapshots at 64.2 and 78.3 h. The two prolate aggregates collide side-to-side and form a more oblate shape.}
    \label{fig:ryugu_handoff_sim_snaps}
\end{figure*}

\begin{figure}
    \resizebox{\hsize}{!}{\includegraphics{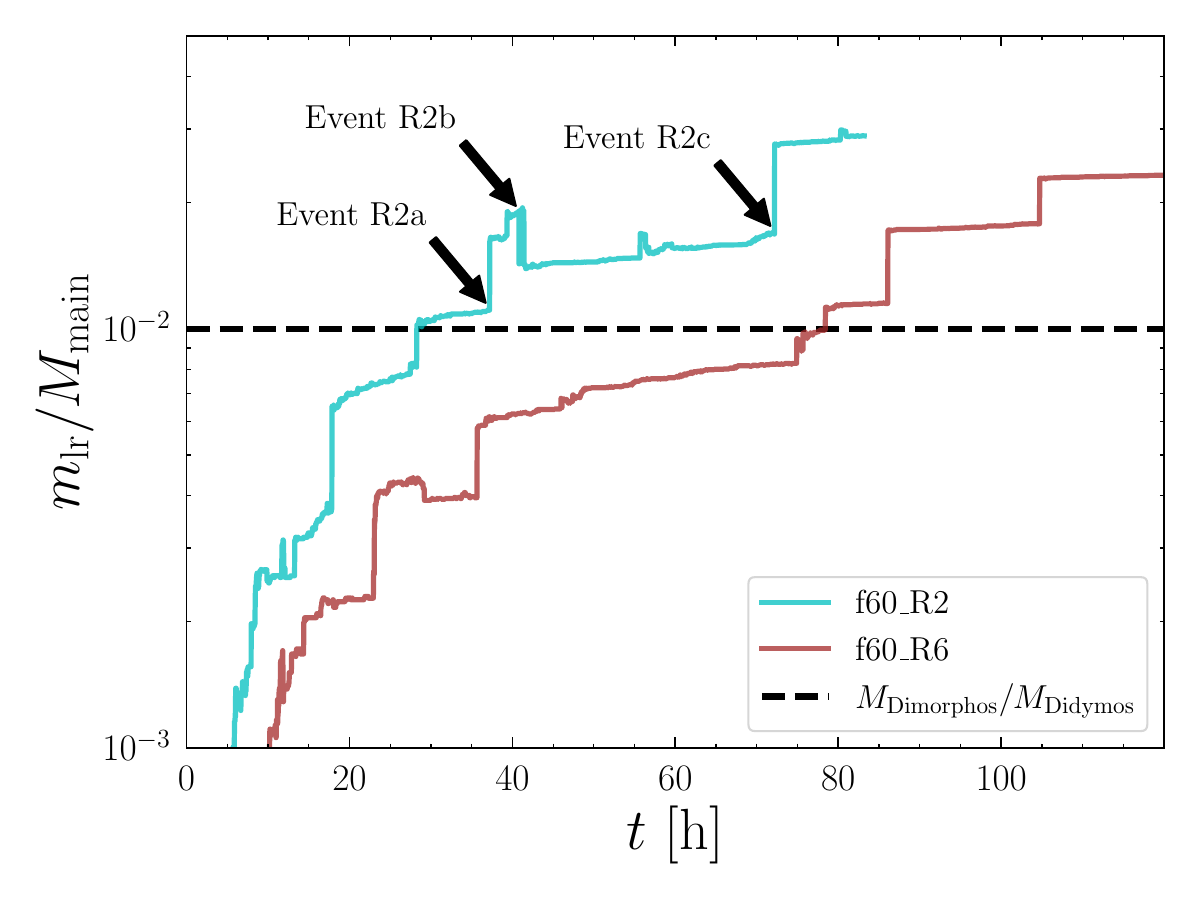}}
    \caption{Growth over time after the formation of the debris disk, $t$, for the largest remnant in the system with mass $m_\mathrm{lr}$, scaled by the mass of the primary body $M_\mathrm{main}$. The dashed black line indicates the target mass ratio between Dimorphos and Didymos, assuming equal densities for the two bodies.}
    \label{fig:ryugu_handoff_growth}
\end{figure}

We performed two hand-off simulations for two different disk mass cases, namely runs \texttt{f60\_R2} and \texttt{f60\_R6} from table \ref{tab:ryugu_spin_up}. We use the former as an example and go through it step by step to highlight the different physical mechanisms involved. We have provided a movie of this simulation in the supplementary media repository\footnote{Found at \url{https://doi.org/10.5281/zenodo.12592633}. The repository also contains movies of a few relevant standalone \textit{N}-body simulations mentioned later in the article.}, which we encourage the reader to view while going through this example. The formation process for a satellite body in this setting is largely hierarchical. As the disk settles around the primary and we perform the hand-off, the distribution of azimuthal velocities causes particles farther out from the primary to move at lower velocities. Along with gravitational collapse, this quickly leads to the formation of denser filaments in the disk over the course of a few orbits, creating a spiral pattern in the disk due to Keplerian shear. This aligns with the behaviour of pure particle circumplanetary disks generated by giant impacts \citep[e.g.][]{Ida_et_al_1997,Kokubo_et_al_2000}. The converging trajectories of the particles in these dense regions lead to rapid formation of smaller gravitational aggregates with more than 10 constituents, reaching masses around 0.5\% of the primary mass after only about 24 hours. In turn, the spiral patterns in the disk quickly diminish, meaning angular momentum transfer is thereafter largely dominated by particle--particle collisions \citep{Takeda_&_Ida_2001}. Snapshots from the example simulation can be found in figure \ref{fig:ryugu_handoff_sim_snaps}, while the growth over time for the largest remnant in the system, with mass $m_\mathrm{lr}$, is shown in figure \ref{fig:ryugu_handoff_growth} (cyan line). As the trajectory of growth indicates, there is significant fluctuation of the mass of the largest remnant, which is not necessarily always the same object. This is due to the accretion or loss of individual grains from the aggregate or because of another body becoming the most massive remnant in the disk. While the initial growth is slow and largely dominated by the steady and slow process of grain accretion, the subsequent growth after 18 h is driven by mergers of aggregates and examples of such events can be seen at 18, 29, 37, 40, 57 and 73 h where there are significant increases in the mass. The two more significant mergers occur at 37 and 73 h, events that we refer to as R2a and R2c, respectively. 

\begin{figure*}
    \resizebox{\hsize}{!}{\includegraphics{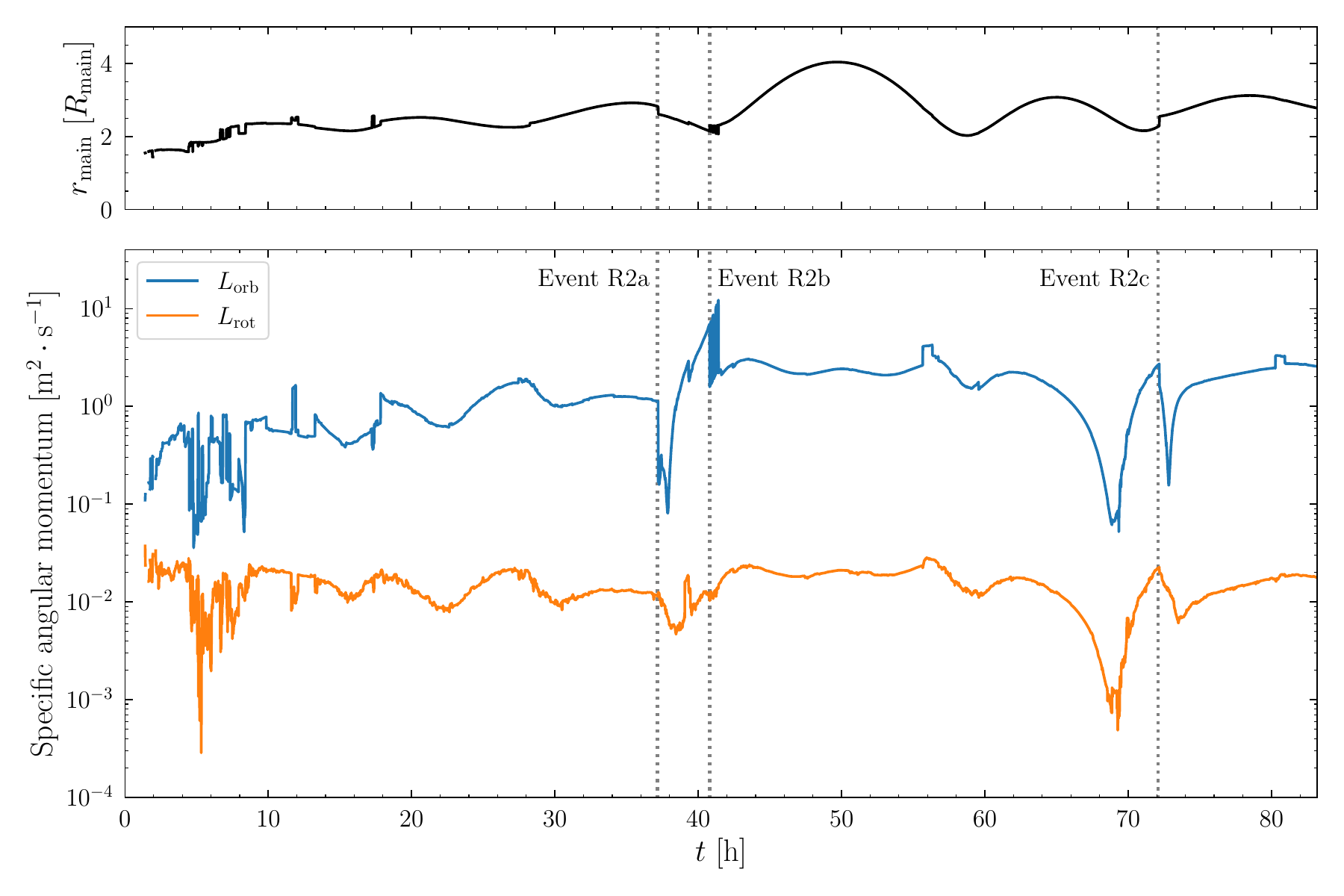}}
    \caption{Correlation between distance to the primary and the angular momentum of the satellite in \texttt{f60\_R2}. In all plots, the specific events of interest referred to in the text as R2a, R2b and R2c have been marked with dotted lines. \textbf{Top}: the distance to the barycenter of the primary for the largest remnant in the debris disk over time. \textbf{Bottom}: specific orbital (blue) and rotational (orange) angular momentum over time for the largest remnant. Not that the value displayed does not necessarily always correspond to the same object, e.g.\ for time steps before 18 h.}
    \label{fig:ryugu_handoff_spec_angmom}
\end{figure*}

\begin{figure}
    \resizebox{\hsize}{!}{\includegraphics[trim=0 0 0 0, clip]{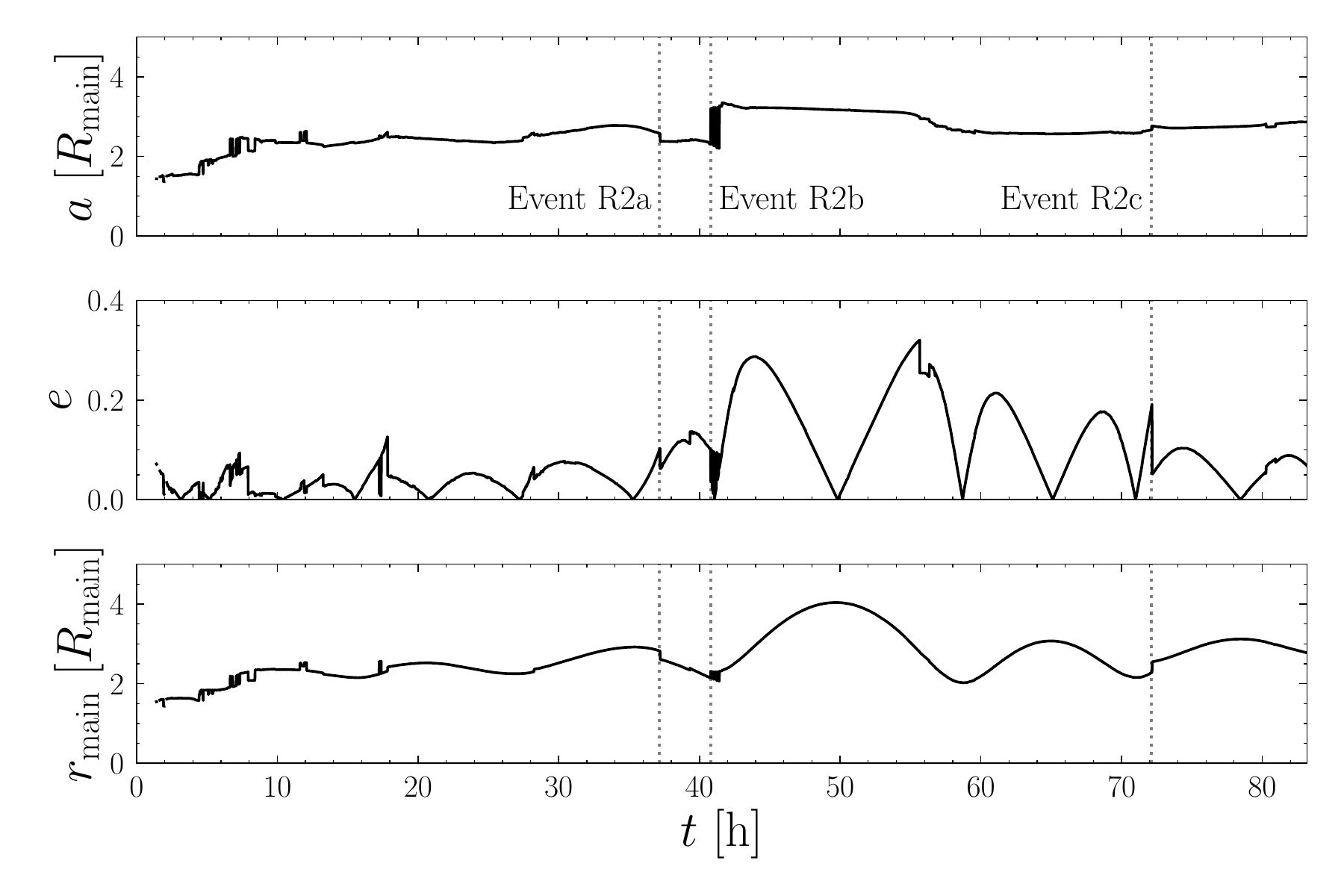}}
    \caption{The changes in orbital elements for the largest remnant in the binary system during its formation in simulation \texttt{f60\_R2}. In both plots, the specific events of interest referred to in the text as R2a, R2b and R2c have been marked with dotted lines. \textbf{Top}: the change in semi-major axis over time for the largest remnant in the system. \textbf{Middle}: the corresponding change in eccentricity over time. \textbf{Bottom}: the distance to the barycenter of the primary.}
    \label{fig:ryugu_handoff_orbital_elements}
\end{figure}

Notably, there is also a sharp decrease in mass occurring shortly after the 40 h mark, an event we will call R2b. This is due to the largest remnant in the system getting too close to the fluid tidal disruption limit and losing much of the mass it had previously gained during event R2a and a smaller subsequent merger. Furthermore, said remnant was also on a tidally locked synchronous orbit, with respect to the primary, before this mass shedding event took place. The tidal forces stripping away a chunk of mass effectively altered the trajectory due to loss of angular momentum, putting the aggregate on a more eccentric orbit and sending it farther out into the system. The specific orbital and rotational angular momentum for the largest remnant in the system have been plotted against time in the bottom graph of figure \ref{fig:ryugu_handoff_spec_angmom}. The corresponding changes in orbital elements can be seen in figure \ref{fig:ryugu_handoff_orbital_elements}. In addition, the figure displays the distance from the primary's barycenter over time. The close encounter with the primary occurring during event R2b functions as a scattering event, inducing a spin for the aggregate and placing it on a seemingly stable circular orbit for the next 15 hours. At that point, around the 55 h mark, the gravitational pull from the primary once more brings the aggregate into a close encounter, leading to another merger at 57 h and a subsequent scattering event, albeit a weaker one. This effectively put it on a collision course with the second-largest remnant in the disk. These two aggregates finally merged at the 73 h mark during the R2c event, creating the final satellite consisting of 869 grains with a total mass of $M_\mathrm{moon} = 0.028M_\mathrm{main}$. To evaluate the shape of the body, we used the dynamically equivalent equal-volume ellipsoid (DEEVE) measure, determining the ellipsoidal axes from the principal moments of inertia of the body. As shown by \cite{Agrusa_et_al_2024}, this measure provides a good estimate for the true extents of an ellipsoidal body that is an aggregate of many smaller constituents but leads to systematically larger values than the true physical extents. First, we compute the total inertia tensor for the composite object using the Huygens-Steiner theorem and then proceed to evaluate the tensor's eigenvalues to obtain the principal moments of inertia $A,\ B,\ C$. The correlation between $A,\ B,\ C$ and the DEEVE axes $a,\ b,\ c$ are as follows:

\begin{equation}
    A = \frac{m}{5}(b^2+c^2),\quad B = \frac{m}{5}(a^2+c^2),\quad C = \frac{m}{5}(a^2+b^2).
\end{equation}

Given that the axes $a,\ b,\ c$ are oriented along some orthogonal coordinate system $x',\ y',\ z'$ determined by the specific rotation of an object that aligns the angular momentum and angular velocity vectors, an oblate shape in the orbital plane corresponds to $a/b$ values close to unity. This assumption holds as long as there is little pitch and roll for the asteroid. In turn, we define the orbital plane axes such that $a>b$ while we assume that $c$ is pointing out of the plane. The $b/c$ value then determines how flat the body is, i.e.\ how elongated it is perpendicularly to the orbital plane. We found that  $a/b=1.27$ and $b/c=1.58$, meaning the resulting satellite is less oblate and flatter than Dimorphos, which has a shape fitting the values $a/b=1.06\pm 0.03$ and $b/c=1.47\pm 0.04$ \citep{Daly_et_al_2023b}.

\begin{figure*}
    \resizebox{\hsize}{!}{\includegraphics[trim=0 0 0 0, clip]{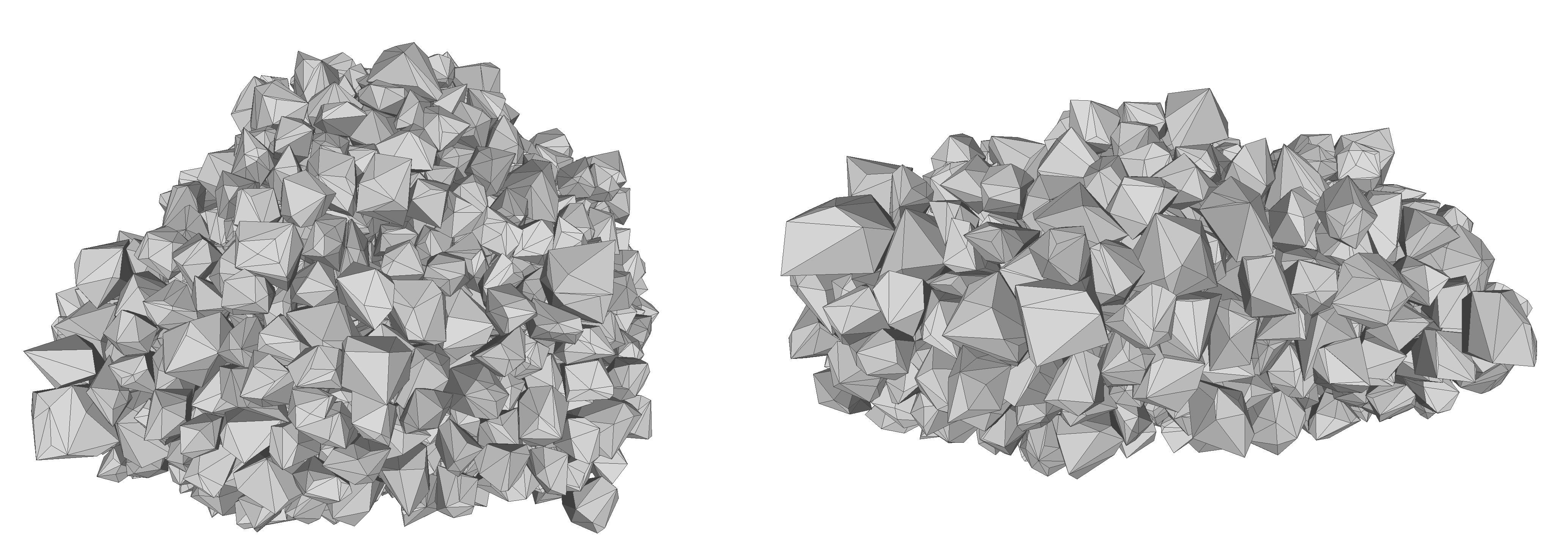}}
    \caption{The final shape for the $0.028M_\mathrm{main}$ aggregate consisting of 869 grains that formed during the dynamical evolution after mass shedding by rotational failure for a Ryugu-like body in simulation \texttt{f60\_R2}. \textbf{Left:} the aggregate as seen from a top-down perspective. \textbf{Right:} an edge-on view of the final satellite.}
    \label{fig:ryugu_handoff_final_aggregate}
\end{figure*}

The final shape from a top-down and edge-on perspective can be seen in figure \ref{fig:ryugu_handoff_final_aggregate}. We note that the sharp features and the size of the individual grain constituents do not align with the observed particles on Dimorphos surface and may generate unrealistic details \citep{Robin_et_al_2023,Pajola_et_al_2023}.

Studying the simulation in detail, we found that the last merger of the largest remnant occurred in a state of rotation where the aggregate was spinning with a period of around 9 h, whereas its orbital period was 13 h. Due to the spinning nature of the target in the merger, i.e.\ the most massive of the two bodies, the final shape of the resulting body was highly dependent on the orientation of the longest axis of the target, along with the location and impact angle of the collision. In this scenario, the prolate impactor merges with the prolate target such that their major principal axes are parallel and side-to-side, creating a more oblate final shape, changing the DEEVE axis ratio from $a/b \sim 1.8$ to a value closer to 1.2. Since the minor axes in the orbital plane align and become longer in this scenario, we will continue to refer to this type of merger as \textit{short-axis}, while using the name \textit{long-axis} for the opposite type that occurs when the two bodies merge at the edges of their longest axis, generating more prolate shapes.

Taking another look at figure \ref{fig:ryugu_handoff_spec_angmom}, there is another significant dip in specific angular momentum right before the final merger which is due to a third scattering by the primary. As can be seen, the orbit of the largest remnant stabilises after this close encounter and the fluctuation in angular momentum is dampened, ultimately placing it on an orbit with a semi-major axis of $2.52R_\mathrm{main}$ with an eccentricity expected to land somewhere in between 0 and 0.09.

The second simulation, \texttt{f60\_R6}, went through a slightly different growth trajectory than the largest remnant of the example simulation, as can be seen when comparing the two tracks in figure \ref{fig:ryugu_handoff_growth}. While the formation process of the largest remnant is also driven by mergers, it did not undergo any disruptions and takes more than 20 h longer to reach its mass at the final time step of $0.023 M_\mathrm{main}$. This can be attributed to the lower mass and fewer particles of the initial disk post hand-off. Furthermore, it ended up with a more prolate shape than our example simulation, having DEEVE axes ratios of $a/b=1.34$ and $b/c = 1.42$. 

Here, the shape was also largely affected by the final impact. As a smaller prolate cluster spiralled in towards the primary, it was catastrophically disrupted and an oblate remnant was scattered outwards, ending up on a trajectory that converged with that of the largest remnant. The oblate impactor then underwent a head-on short-axis merger with the spinning target close to the centre of its major principal axis with a small impact angle. This resulted in a less prolate target, with its DEEVE axis ratio going from $a/b\sim 1.4$ to 1.34. Moreover, the nature of the impact resulted in a significant slowdown in the rotation rate of the target, placing it on a tidally locked and synchronous orbit with a period of 15 h.

\subsection{Standalone \textit{N}-body simulations for Ryugu-like systems}

\begin{figure}
    \resizebox{\hsize}{!}{\includegraphics{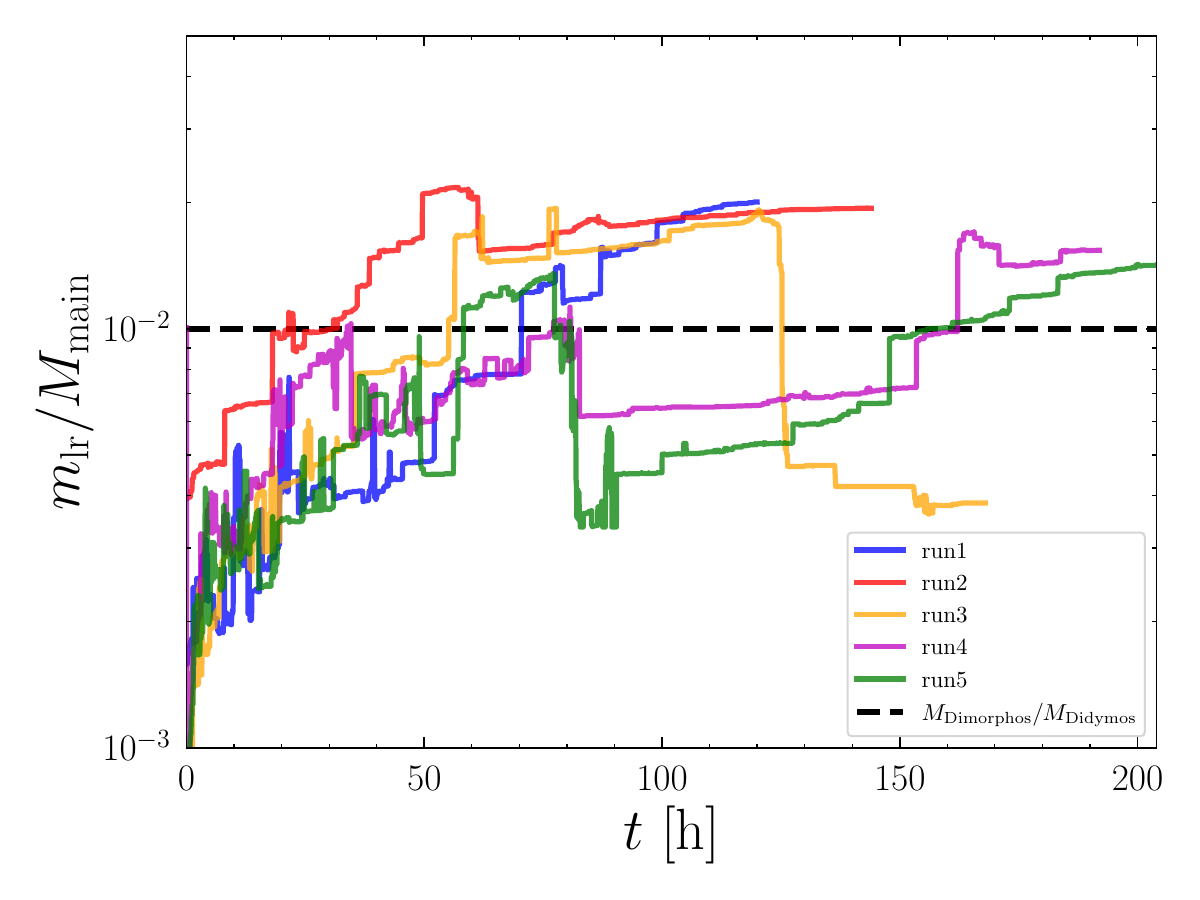}}
    \caption{The same plot as in figure \ref{fig:ryugu_handoff_growth} but for the case of five separate standalone \textit{N}-body simulations of a Ryugu-like system.}
    \label{fig:ryugu_Nbody_growth}
\end{figure}

To compare with the hand-off simulations, we also performed a series of pure \textit{N}-body simulations where we generated a system with a debris disk orbiting an asteroid and let it propagate dynamically. The total mass was set to be equal to the estimated mass of Ryugu \citep{Watanabe_et_al_2019}. Using the SPH simulation results as a basis, we distributed the mass such that 5\% of the total mass was put in the debris disk and the rest in the primary.

We ran five simulations for at least 120 h and propagated them further if there were still many aggregates in range for mergers or if the orbit of the largest remnant was deemed unstable and likely to decay or become hyperbolic. Due to the random nature of the debris disk population synthesis, each disk ended up with a different number of particles from the same grain size distribution. From runs 1 through 5, the disk consists of 5458, 4926, 5530, 5465 and 5301 particles. Hence, allowing for a lower minimum particle diameter effectively more than doubled the number density of particles despite the disks having similar masses to the hand-off cases. 

\begin{figure}
    \resizebox{\hsize}{!}{\includegraphics{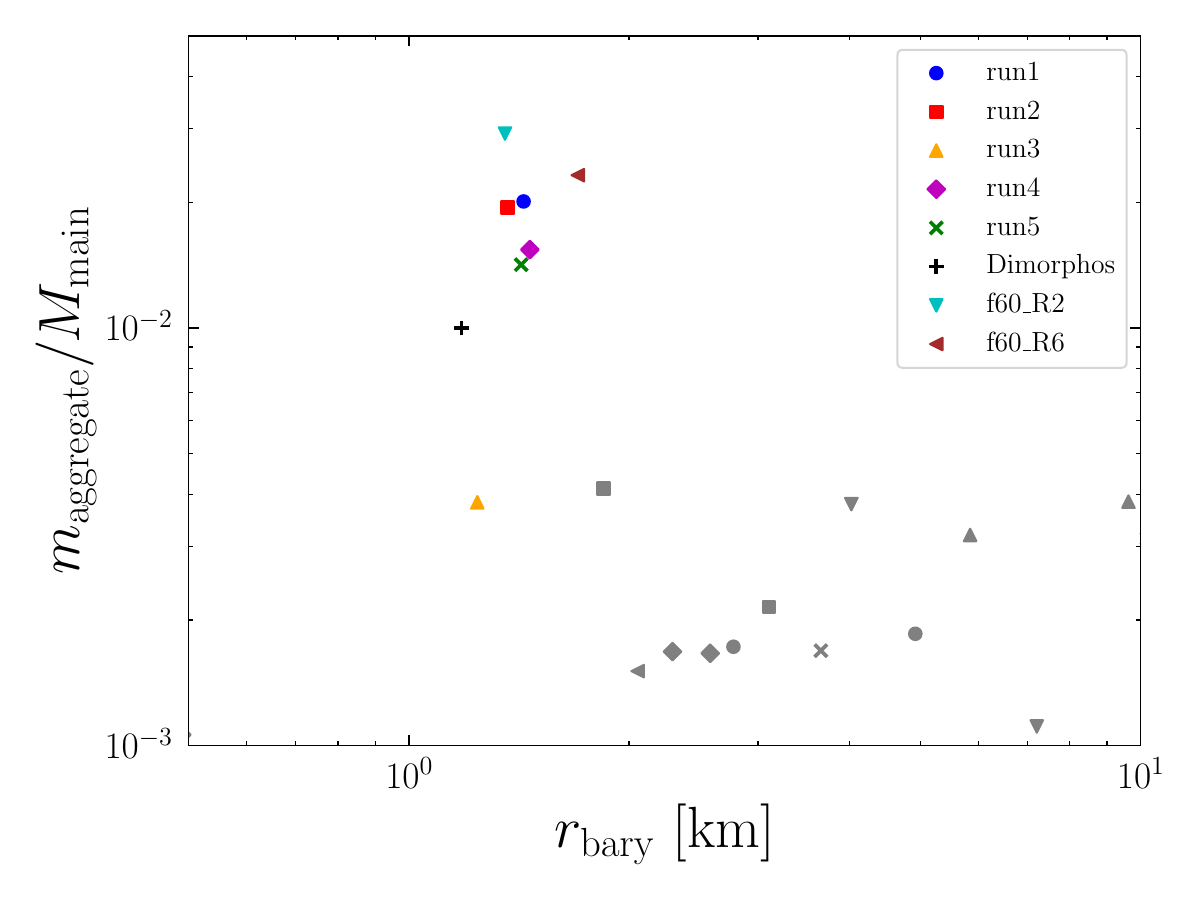}}
    \caption{The distribution of masses and distance from the barycenter of the primary body at the final time step of each standalone \textit{N}-body simulation. The largest remnant in each system has colour, while the grey markers represent the remaining aggregates that either are too low mass or too far out in the system (i.e.\ outside 5 km from the primary) to count as the most stable and massive satellite in the timescales we consider. The cyan and brown points marked with upside down and sideway triangles represents the result from the hand-off simulations described in section \ref{section:ryugu_hand_off_results}.}
    \label{fig:ryugu_Nbody_final_masses}
\end{figure}

The mass growth for the five different cases is displayed in figure \ref{fig:ryugu_Nbody_growth}. From the trajectories, it is clear that within our simulation framework, an aggregate can reach a relative mass similar to that of Dimorphos on average $\sim50$ hours after the formation of the debris disk and in as little as 24 hours. This is made possible by early mergers that drastically boost the mass of each largest remnant, in turn increasing their ability to accrete both other clusters and single grains, promoting mass growth. A similar mechanism can be seen in planet formation, where collisions between migrating planetesimals in protoplanetary disks can greatly promote their ability to accrete pebbles, increasing their growth rate \citep[e.g.][]{Liu_et_al_2015,Wimarsson_et_al_2020}. However, just like for the hand-off case (see section \ref{section:ryugu_hand_off_results}) there were many situations where an aggregate would shed particles during close encounters with the primary body in the system. In all of the five cases, there were significant disruption events that either prolonged the growth process or halted it. Looking at the magnitude of the most critical mass shedding event in each case, it is clear that the systems with the smallest relative changes in mass reached the largest final masses, i.e.\ runs 1 and 2 which both have satellites two times our target mass. In runs 1 and 2 there were only a few mild disruption events where a smaller fraction ($<0.5 m_\mathrm{lr}$) of the body's total mass was stripped off from tidal interactions with the primary, which slowed down the growth rate. While the largest remnant in run 1 still went through another merger, significantly increasing its mass, that of run 2 only grew via accretion of single grain or small clusters making it take an additional 60 hours until it had regained a substantial fraction of the mass lost in its disruption event, occurring at the 60 h mark. The accretion rate still being efficient for the secondaries of runs 1 and 2 can largely be attributed to them being more massive post tidal disruption. In turn, they more easily reaccreted single grains and clusters surrounding them in the disk. 

In figure \ref{fig:ryugu_Nbody_final_masses}, we have plotted the relative masses of the aggregates in each system against their distance from the system barycenter at the final time step. The coloured markers represent the bodies that are massive enough and close enough to be considered the most stable and massive satellite within the scope of this article, while the grey markers show the remaining aggregates with masses greater than $0.001 M_\mathrm{main}$. Looking at the system configuration for these three cases, we can see an explanation as to why the growth rate slowed down. For both runs 1 and 2, the minor satellites are on much wider orbits and will likely not undergo any more mergers with the largest remnant, meaning all the smaller aggregates in the vicinity of the largest remnant have already been absorbed by the final time step and growth mainly occurs through single-grain accretion. 

In simulations 3, 4 and 5, the largest remnants reached relative masses above the target of $0.01 M_\mathrm{main}$ but ended up going through catastrophic disruptions during close encounters with the primary and lost more than half their mass. The aftermaths of these tidal events differ significantly however, as the largest remnants in runs 4 and 5 retained their orbit close to the primary within the debris disk and reaccreted much of the shed mass before they each ultimately encountered another remnant formed in the tidal event and underwent a critical merger, once more putting their relative mass at a level above $0.01 M_\mathrm{main}$. On the contrary, the corresponding body in system 3 did not regain any of its lost particles and most of the mass ended up in orbits separated from the main debris disk. Moreover, there is again a correlation between the amount of mass held in additional aggregates still present in the system and where they are positioned with respect to the largest remnant. Looking at the final configuration for the aggregates in system 3 (see figure \ref{fig:ryugu_Nbody_final_masses}), all of the remaining clusters have similar masses between $0.003 M_\mathrm{main}$ and $0.004 M_\mathrm{main}$ and the largest remnant is separated from the other two, meaning it is unlikely to go through subsequent mergers and further growth as the system continues to evolve. 


\subsubsection{Satellite shapes}

\begin{figure}
    \resizebox{\hsize}{!}{\includegraphics{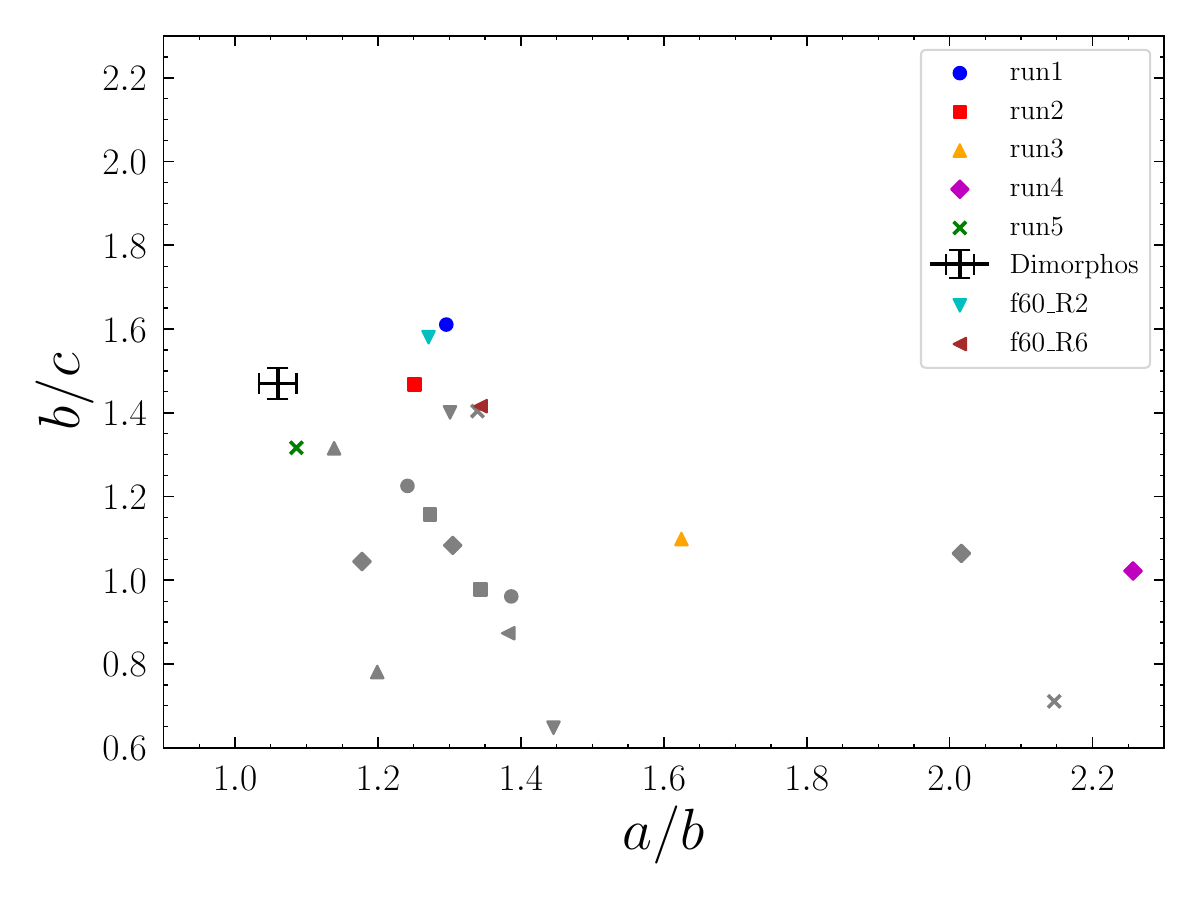}}
    \caption{The final shapes at the end of each simulation for the Ryugu-like scenario. Each ellipsoidal axis has been evaluated from the corresponding dynamically equivalent equal-volume ellipsoid to the moment of inertia of each aggregate. Note that only the coloured markers represent bodies that have a large enough mass and are situated within 5 km of the primary body. They can be compared to the final masses and positions in the different systems from figure \ref{fig:ryugu_Nbody_final_masses}. We note that these values are likely slightly larger than the those obtained from the principal axes corresponding to the actual physical extents of the bodies \citep{Agrusa_et_al_2024}.}
    \label{fig:ryugu_Nbody_final_shapes}
\end{figure}

The distribution of the $a/b$ and $b/c$ values for the DEEVE axes of each aggregate in the pure \textit{N}-body simulations and the hand-off case can be found in figure \ref{fig:ryugu_Nbody_final_shapes}. The marker and colour of each data point follow the same scheme as in figure \ref{fig:ryugu_Nbody_final_masses} and the corresponding values for Dimorphos from \cite{Daly_et_al_2023b} have also been added. Only the largest remnant from run 5 with an axis ratio of $a/b = 1.08$ displays a similar type of oblate shape as Dimorphos when comparing them visually. Three of the outcomes, two of them being hand-off cases, end up with similar $b/c$ values but are more elongated than Dimorphos and the rest have prolate bodies more similar to the majority of observed asteroid satellites. The final body of run 4 stands out as it has a value of $b/c = 1.02$ and a much longer $a$-axis yielding $a/b = 2.26$. 

\begin{figure*}
    \resizebox{\hsize}{!}{\includegraphics{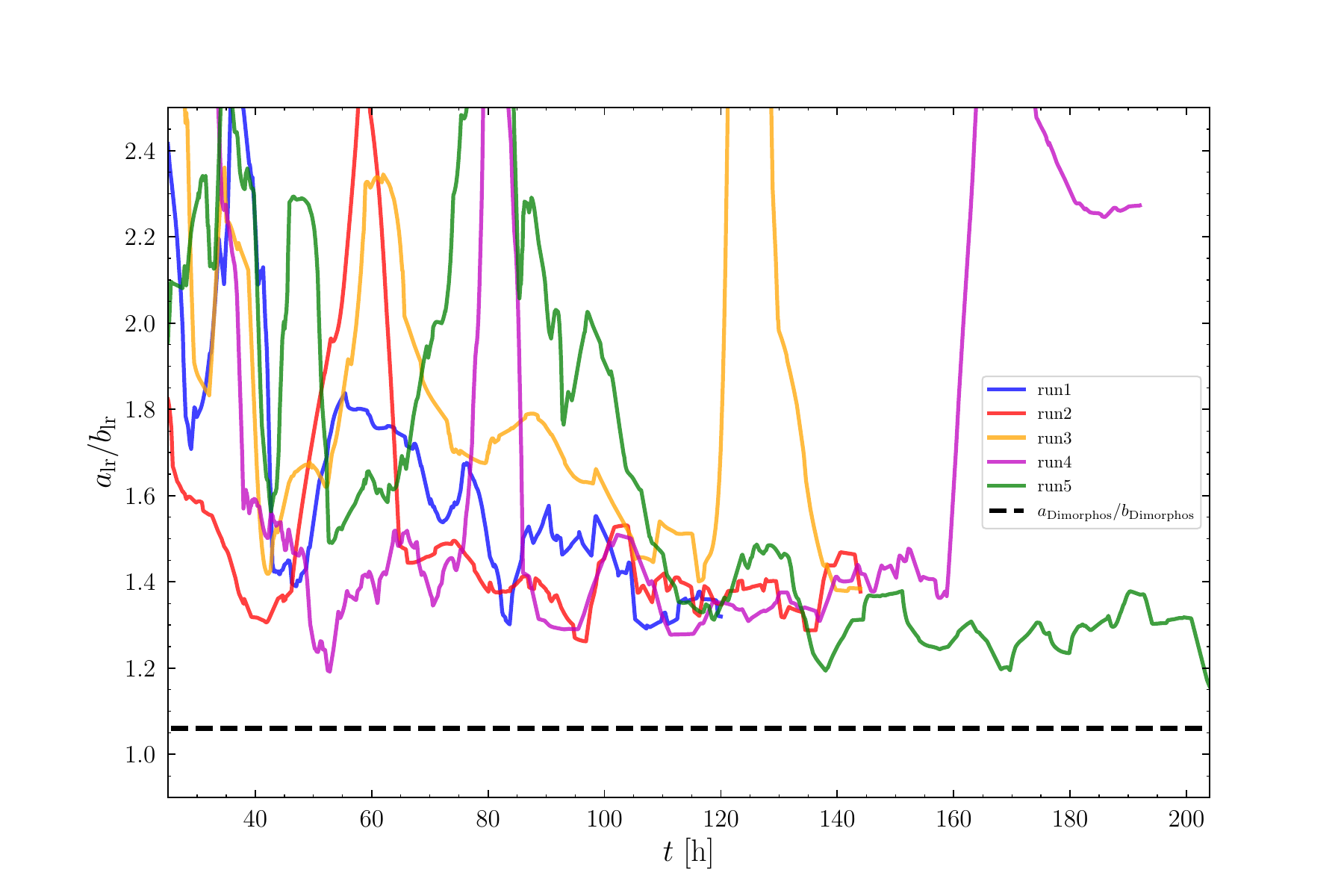}}
    \caption{The change in principal axis ratios $a/b$ for the corresponding DEEVE shape of each largest remnant over time for the five Ryugu pure \textit{N}-body simulations. Due to excessive noise, the data has been smoothed using a uniform one-dimensional filter to facilitate interpretation. Note that this may skew the individual data points from their actual value. The evolution prior to 25 h has been omitted due to the naturally chaotic behaviour of the formation process. Comparing the tracks to those representing the largest aggregate mass over time in figure \ref{fig:ryugu_Nbody_growth}, there are clear overlaps between large mass shedding or accretion events and spikes in $a/b$ values.}
    \label{fig:ryugu_Nbody_axes_vs_time}
\end{figure*}

\begin{figure*}
    \resizebox{\hsize}{!}{\includegraphics{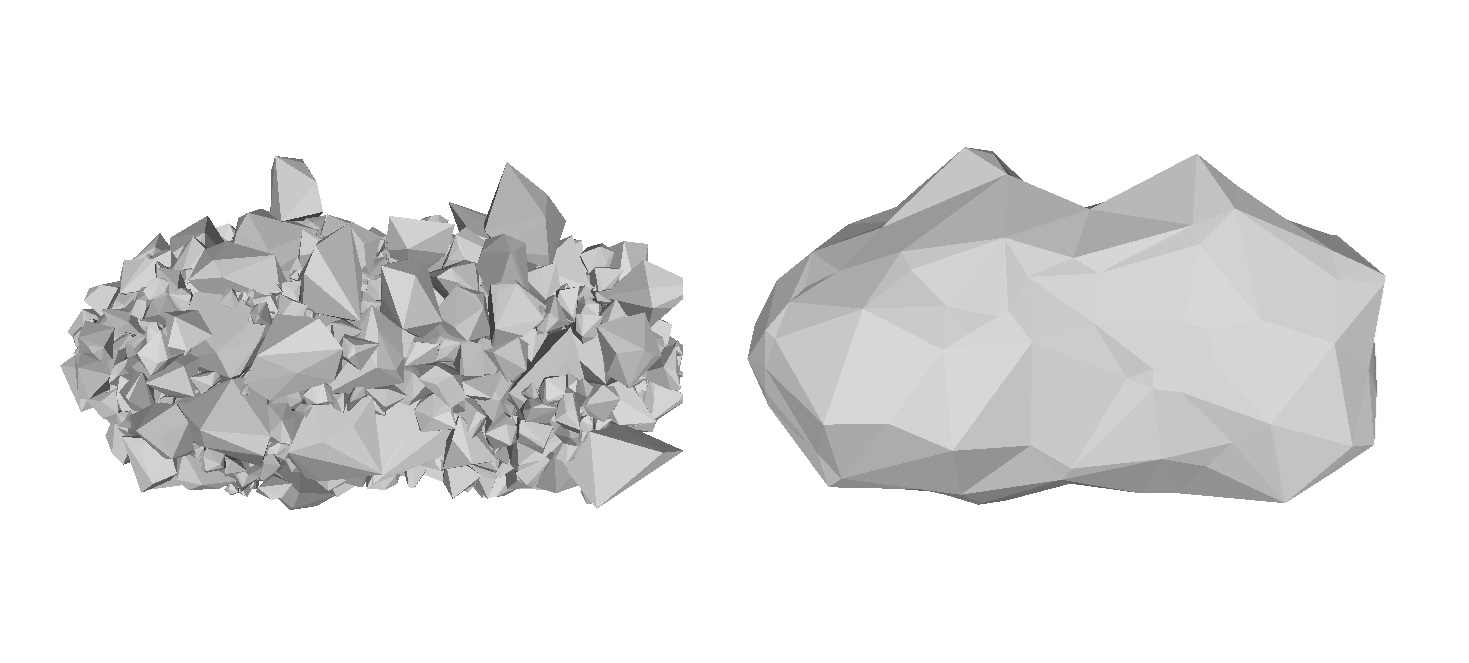}}
    \caption{The final shape of the largest satellite in run 4 of the standalone Ryugu \textit{N}-body simulations. \textbf{Left:} the configuration of individual grains in the aggregate at the final time step of the simulation. \textbf{Right:} the corresponding $\alpha$-wrap shape for the same configuration of particles.}
    \label{fig:ryugu_Nbody_run4_shape}
\end{figure*}

To be able to better understand the formation process and evolution of the axis ratio $a/b$ over time, we have plotted them for each pure \textit{N}-body case in figure \ref{fig:ryugu_Nbody_axes_vs_time}. Note that the many changes in mass and deformation through tidal interactions, along with fluctuation between which principal axis is larger have made the data noisy. In turn, we applied a one-dimensional uniform filter to smooth it out and make it easier to interpret. Hence, some of the values might not be correctly displayed but the overall trends remain the same. One of the main features of the evolution tracks is large spikes, where the $a/b$ values increase drastically. Comparing the behaviour with the mass growth patterns in figure \ref{fig:ryugu_Nbody_growth}, these features can be compared with significant changes in mass, such as disruption events or mergers. While the large deformations in runs 1, 2, 3 and 5 are due to close encounters with the primary, leading to tidal disruptions, this is not the case for run 4. Instead of its main satellite being reshaped due to mass shedding at 160 h, the change into a highly prolate shape came from a merger with another aggregate, indicating that the largest remnant is, in fact, a bilobate satellite. We show the shape of the final aggregate in figure \ref{fig:ryugu_Nbody_run4_shape} along with the corresponding $\alpha$-wrap\footnote{An $\alpha$-wrap is a geometrical algorithm available in \texttt{CGAL} \citep{cgal:eb-23a} which is also based on Delaunay triangulation that uses iterative refinement. It is more appropriate for capturing detailed features of a coherent body and can be imagined as the shape generated when shrink-wrapping a sheet on top of a collection of facets or points.} for the configuration of grains. Due to a significant relative velocity between the bodies at the time of the merger (equal to $0.84$ of the mutual espace velocity), the long-axis impact resulted in a subtle bilobate shape that is only evident because of narrow concavities at the centre of the body. These results agree with \cite{Leleu_et_al_2018}, who showed that the relative velocity and impact angle in a merger of aggregate bodies greatly affect the final shape. We discuss this further in section \ref{section:discussion_impact_properties}.

The evolution of the largest satellite shape in run 5 also provides key insight into what determines the final form. In the first 70 h of the simulation, it followed the same behaviour as the rest of the simulations where accretion of grains and small aggregates, along with tidal deformations, generated the characteristic prolate shape of asteroid satellites. However, the previously discussed catastrophic disruption that began around 75 h into the simulation (see figure \ref{fig:ryugu_Nbody_growth}), led to a drastic change in shape that differs in nature from the other cases. Instead of the tides deforming the body such that some mass was lost from both long-axis edges, leaving a more oblate shape behind, here the disruption was so strong that only the outermost part of the body survived. The close encounter and strong disruption event resulted in the single remnant getting ejected from the system, ending up on an orbit 10 km from the primary, as can be seen in figure \ref{fig:ryugu_Nbody_final_masses}. As a result, due to the main satellite being practically disintegrated, there was a hierarchical shift in which aggregate was the most massive. As the disruption ejected a barrage of grains farther out into the system, some of the mass got accreted by the new largest aggregate, which then steadily grew and reached a relative mass of $>0.01 M_\mathrm{main}$ after another 75 h. There are two more significant events of deformation, as it first experienced a short-axis merger right before the 150 h mark and then continued to accrete aggregates that have $\sim 10\%$ of its mass. The final deformation that can be observed at around 180 h of figure \ref{fig:ryugu_Nbody_axes_vs_time} was induced by a close encounter with another aggregate which was subsequently ejected farther out into the system, which then allowed the largest remnant to relax and settle at its final shape, being an oblate but a less flat version of Dimorphos with axis ratios of $a/b = 1.08$ and $b/c = 1.32$ and a slightly higher relative mass of $0.014 M_\mathrm{main}$.

\subsubsection{Formation patterns for satellites in a radially extended disk}\label{section:formation_patterns}

Looking at the formation histories and final shapes of the largest satellites, it is clear that the process is highly chaotic, which agrees with the findings of \cite{Agrusa_et_al_2024} and can lead to different shapes and number of satellites just from a small change in the initial conditions of the simulation. In order to make more sense of the formation mechanisms involved and find conditions that favour the formation of oblate satellites, we formulate three general patterns for our type of debris disk that will aid us in the analysis:

\begin{itemize}
    \item \textbf{Insignificant disruption history (IDH):} The main satellite forms farther out in the system, which results in little-to-no disruption caused by tidal interactions with the primary and its growth is steady throughout the simulation, where it accretes single grains and merges with other, smaller aggregates. Given its distance from the primary, the body is less likely to become tidally locked, meaning that its angular rotation rate is mainly decided by impacts and mergers that either stop or induce rotation, depending on the impact angle and whether they are short- or long-axis. The mass ratio, and in turn diameter ratio, between target and impactor, $m_\mathrm{impactor}/m_\mathrm{target}$ is likely small. An example of this formation pathway is \texttt{f60\_R6} in figure \ref{fig:ryugu_handoff_growth}.
    \item \textbf{Mild disruption history (MDH):} Being closer to the primary, the largest remnant is more susceptible to tidal disruption and undergoes one or a few such events where the tidal forces deform the aggregate, elongating the body until it structurally fails and loses less than half its mass. Post disruption, the body regains some of its lost mass through single-grain accretion and mergers leading to a gradual change in shape. Given the mass shedding of the largest remnant, the ratio $m_\mathrm{impactor}/m_\mathrm{target}$ during mergers is likely greater than for the insignificant disruption history scenario, leading to more significant fractional changes in $m_\mathrm{lr}$, $D_\mathrm{lr}$ and $a/b$ values. The rotation rate of the body is kept low due to tidal locking but impacts and mergers can induce higher rates. Hand-off run \texttt{f60\_R2} along with \textit{N}-body runs 1 and 2 are examples of this type of formation history.
    \item \textbf{Catastrophic disruption history (CDH):} Due to close encounters with the primary, the largest remnant undergoes strong tidal interactions and loses more than half its mass, which significantly slows down the growth process. The remaining body tends to be less prolate, but due to mergers having a $m_\mathrm{impactor}/m_\mathrm{target}$ close to unity, the fractional change in $m_\mathrm{lr}$, $D_\mathrm{lr}$ and $a/b$ can be very large, causing significant deformation. In some cases, the disruption leads to a complete halt in growth after ejecting a significant portion of the disk mass from the system. \textit{N}-body runs 3, 4 and 5 display these types of behaviour.
\end{itemize}

In general, short-axis mergers favour oblate shapes while long-axis mergers often lead to prolate shapes and the creation of contact binaries, which is further explored in section \ref{section:discussion_impact_properties}. We will henceforth refer to the different formation history archetypes as IDH, MDH and CDH, respectively. 

\section{Results: Didymos scenario}\label{section:didymos}

Here, we perform the same type of analysis as we did in the previous section but for a Didymos-like primary, using the newly established formation patterns to describe the evolution of each system. The main difference from the general Ryugu-like setup is that each grain, along with the primary, has a higher density, which leads to on average smaller particles in the system.

\begin{table}[]
    \centering
    \begin{tabular}{lllll}
        \hline\hline
         Run & $M_\mathrm{disk}\ [M_\mathrm{main}]$ & $N_\mathrm{p}$ & $D_\mathrm{frag}$\ [m] & $D_\mathrm{mean}$\ [m] \\
         f60\_D0 & 0.067 & 2767 & [18.6, 41.7] & 22.5 \\
         f60\_D1 & 0.072 & 2406 & [19.1, 125.8] & 23.0\\
         f60\_D2 & 0.065 & 2562 & [18.0, 93.2] & 22.5 \\ 
         f60\_D3 & 0.076 & 2462 & [17.4, 151.1] & 23.0\\
         f60\_D4 & 0.062 & 2601 & [18.7, 43.6] & 22.4\\
         \hline
    \end{tabular}
    \caption{The distribution of disk masses, number of particles and fragment sizes for the post rotational failure debris disks created by spinning up a Didymos-like primary using SPH.}
    \label{tab:didymos_spin_up}
\end{table}

For the Didymos-like primary, we performed five SPH simulations, again with $\phieff = 60^{\circ}$ and the properties of the resulting disks can be found in table \ref{tab:didymos_spin_up}. There is a clear difference in the number of particles and total mass shed from the primary post rotational failure which can be seen in the average disk mass being $0.068 M_\mathrm{main}$ compared to the $0.045 M_\mathrm{main}$ for the Ryugu-like primary. While the topic of rotational failure still needs dedicated research and any analysis related to this mechanism and our SPH simulations in particular are far from conclusive, there are still patterns here worth mentioning. Given that the density of the primary is higher and the radius is smaller, the structural integrity of the body is stronger than for the Ryugu-like primary. This aligns with the density dependency of the critical spin rate for rubble pile asteroids which can be inferred from lightcurve amplitude using $P_\mathrm{crit}/3.3\ \mathrm{h} = \sqrt{(1+\Delta m)/\rho}$, where $\Delta m \sim 2.5 \log(a/b)$ is the lightcurve amplitude from observed magnitude converted to the axis ratio $a/b$ \citep{Walsh_2018}. While the debris disks for the Ryugu-like scenario were principally symmetrical, this is not the case for all debris disks in the Didymos-like case. Only two out of five are symmetrical, namely \texttt{f60\_D0} and \texttt{f60\_D4} which also have smaller maximum particle diameters. We show an example of a nonaxisymmetric disk in figure \ref{fig:didymos_handoff}. From the right-hand plot which displays the output from the \texttt{clusterfinder} algorithm, it is evident that the disk is denser towards the bottom of the plot where it contains two large boulders with large diameters of 83 and 93 m. This shows us that the nature of the mass shedding for this body was more localised, with large chunks of particles being ejected from its equator \citep[as postulated by][]{Tardivel_et_al_2018}, whereas the symmetric disks have a history of global symmetric landslides. Moreover, the distinct extended features on the left side of the debris disk also confirm that the rotational failure of the primary was initially more localised in this scenario, starting at a specific location, allowing the corresponding shed mass to spread farther out in the system than the rest of the debris. 

\subsection{Dynamical evolution post hand-off for Didymos-like systems}\label{section:didymos_handoff}

\begin{figure*}
    \resizebox{\hsize}{!}{\includegraphics[trim=0 0 0 0, clip]{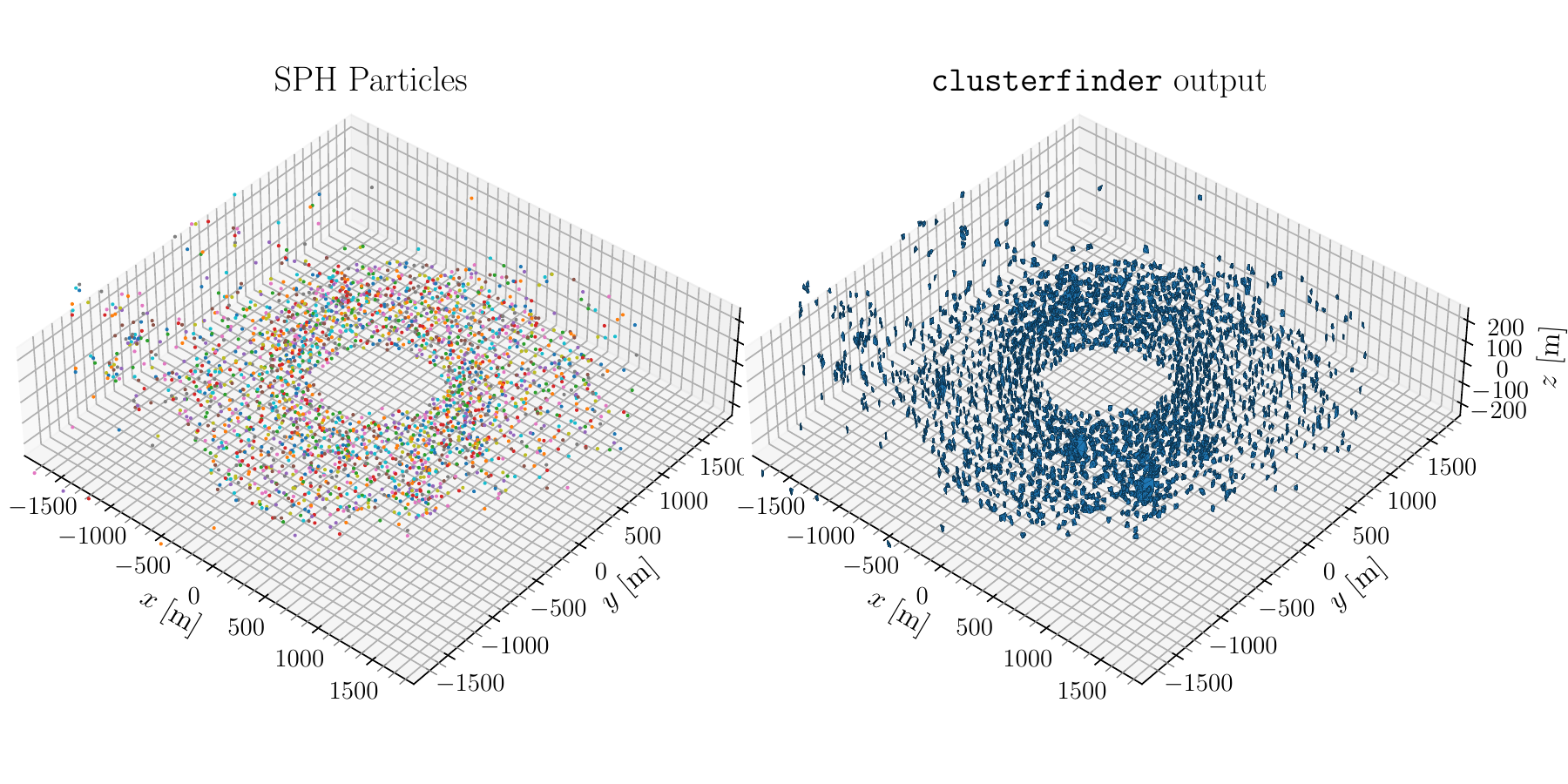}}
    \caption{\textbf{Left:} the position of the SPH particles at the time of the hand-off for an nonaxisymmetric debris disk formed from a Didymos-like primary. \textbf{Right:} the final output of \texttt{clusterfinder}. The algorithm has generated $\alpha$-shapes and convex hulls with volumes matching the densities and masses of SPH particle clusters. Zooming in will show the resulting 3D shapes with black grid lines marking the edges that separate the faces in blue.}
    \label{fig:didymos_handoff}
\end{figure*}

\begin{figure*}
    \resizebox{\hsize}{!}{\includegraphics[trim=0 0 0 0, clip]{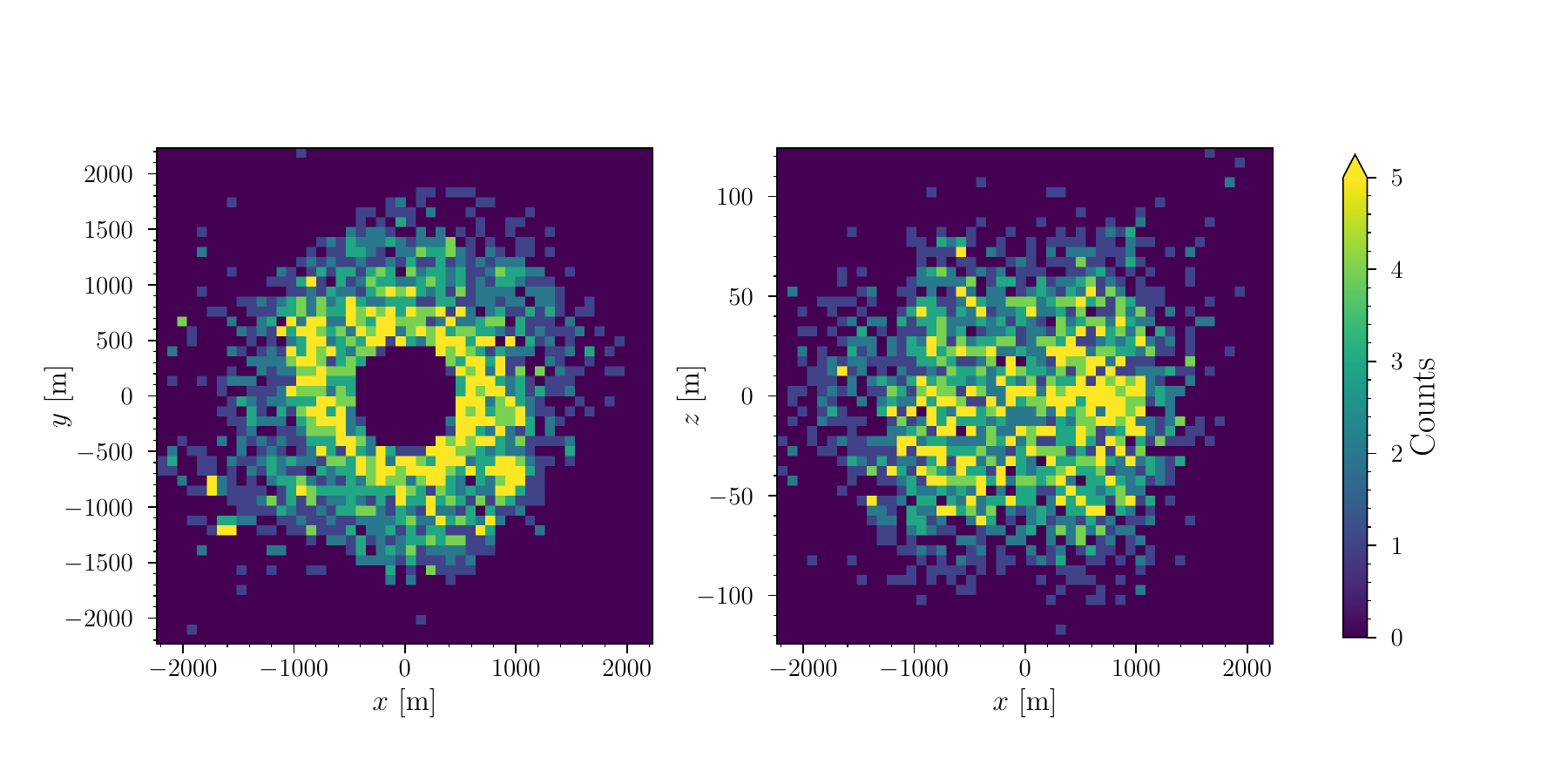}}
    \caption{\textbf{Left:} a histogram showing the number density of the disk obtained after performing the hand-off for the simulation \texttt{f60\_D2} in the $xy$-plane. Corresponds to the same distribution seen in figure \ref{fig:didymos_handoff}. Can be compared to the Ryugu example case with a more symmetric distribution in figure \ref{fig:ryugu_handoff_density}. \textbf{Right:} the same metric but for the $xz$-plane. Note the differing scales between the $x$- and $y$-axis of the figure.}
    \label{fig:didymos_handoff_density}
\end{figure*}

\begin{figure}
    \resizebox{\hsize}{!}{\includegraphics{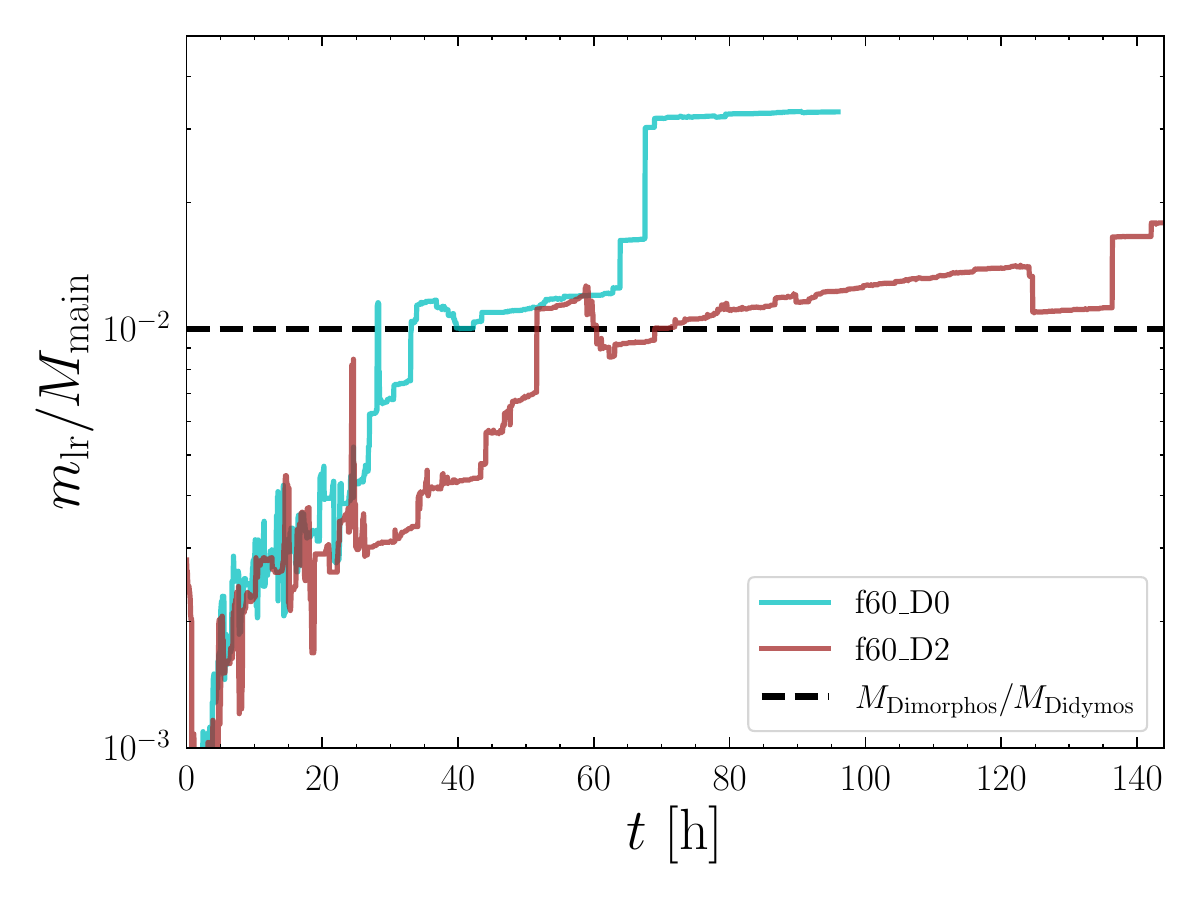}}
    \caption{Growth over time after the formation of the debris disk, $t$, for the largest remnant in the system with mass $m_\mathrm{lr}$, scaled by the mass of the primary body $M_\mathrm{main}$. The dashed black line indicates the target mass ratio between Dimorphos and Didymos, assuming equal densities for the two bodies.}
    \label{fig:didymos_handoff_growth}
\end{figure}

\begin{figure}
    \resizebox{\hsize}{!}{\includegraphics[trim=0 0 0 0, clip]{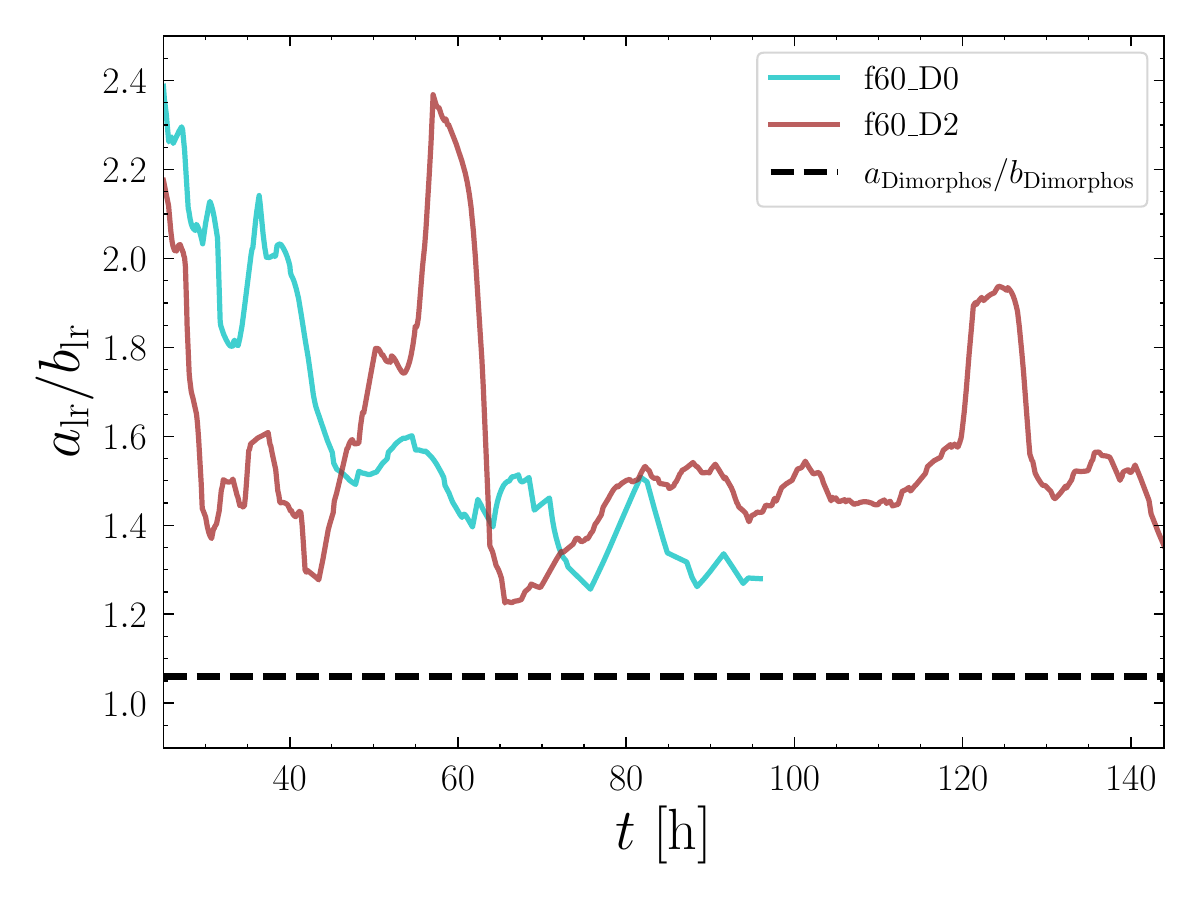}}
    \caption{The change in principal axis ratios $a/b$ for the corresponding DEEVE shape of each largest remnant over time for the hand-off simulations for the Didymos-like system. Due to excessive noise, the data has been smoothed using a uniform one-dimensional filter to facilitate interpretation. Note that this may skew the individual data points from their actual value. The evolution prior to 25 h has been omitted due to the naturally chaotic behaviour of the formation process. Comparing the tracks to those representing the largest aggregate mass over time in figure \ref{fig:didymos_handoff_growth}, there are clear overlaps between large mass shedding or accretion events and spikes in $a/b$ values.}
    \label{fig:didymos_handoff_axes_vs_time}
\end{figure}

We evolved two of the hand-off systems dynamically, choosing one symmetric and one nonaxisymmetric disk in simulations \texttt{f60\_D0} and \texttt{f60\_D2}, respectively. Other than their shapes, the geometry of the disks from each simulation is similar to that of those formed from the rotational failure of the Ryugu-like bodies. They radially extend for about 1 km (2.5$R_\mathrm{main}$) from the surface of the primary and have a thickness of $\sim 200$ m, with a few individual particles being farther out for the nonaxisymmetric case, as can be seen in figure \ref{fig:didymos_handoff}. The resulting growth tracks for the two systems can be found in figure \ref{fig:didymos_handoff_growth}. Just like in the case of the Ryugu-like system, there is one case of IDH (\texttt{f60\_D0}) and one of MDH (\texttt{f60\_D2}). While there was a mild disruption for the case of \texttt{f60\_D0} around 30 h into the simulation, it had little consequence for the growth rate and final shape of the body. Comparatively, the formation process of the largest remnant in \texttt{f60\_D2} got severely slowed down by its disruption events. Moreover, the body ended up on an orbit less than a primary radius from the main body after the first instance of mass shedding. This explains why it mainly grew via accretion of individual grains and small clusters from 60 hours onward until it underwent a second disruption event at 125 h, leading to fragmentation and migration outwards for the largest remnant. Looking at the change in the $a/b$ axis ratio over time for the two cases in figure \ref{fig:didymos_handoff_axes_vs_time}, we observe how large of an effect the final disruption event had on the resulting shape of the body. The tidal forces split the moon into two smaller aggregates along its longest axis that eventually recombine in a short-axis merger, making the final body oblate with axis ratios of $a/b = 1.14$ and $b/c = 1.48$. We note that the largest satellite might still undergo further mergers, as two additional aggregates remain in the system on similar orbits. Yet, as the system is unresolved after 148 h and it illustrates the importance of mergers for obtaining oblate shapes, we opted not to continue the dynamical evolution while recognising that the final shape might be different for this satellite when the debris disk has been cleared of material.

The growth patterns for each largest remnant in the two systems are distinctly different, which could potentially be tied to disk geometry, given that the two disks had similar masses and number of particles. For all the symmetric disk scenarios, the growth trajectories are largely similar up until the 50 h mark of the simulations, the sole exception being \texttt{f60\_R6} which has less mass and fewer particles than the rest which significantly slows down the growth process. However, for the nonaxisymmetric disk in \texttt{f60\_D2} growth slowed down after the 20 h mark. This could also be tied to the amount of mass lost from the system early on, i.e.\ single grains and aggregates that end up with hyperbolic orbits and have semi-major axes larger than 10 km. By the 20 h mark, system \texttt{f60\_D2} had ejected $0.02 M_\mathrm{disk}$ which was two times as much mass as ejected by \texttt{f60\_D0}. By the 40 h mark, this value had increased to $0.04 M_\mathrm{disk}$ which was three times as much. The values intersect once more towards the end of the simulations, but the effect of the early ejection of mass can be significant. Despite the amount seeming small compared to the largest remnant masses at the time, it still corresponds to several small aggregates that could have increased the accretion efficiency of the largest remnants in the system by boosting their mass and diameter. Nevertheless, this is a complex problem where the specific geometry of the overdensities in the disk may also have a large influence on the growth rate. In turn, the effect of disk asymmetries will be a topic for future research as our current understanding is inconclusive and our SPH model is still under development.  

\begin{figure}
    \resizebox{\hsize}{!}{\includegraphics{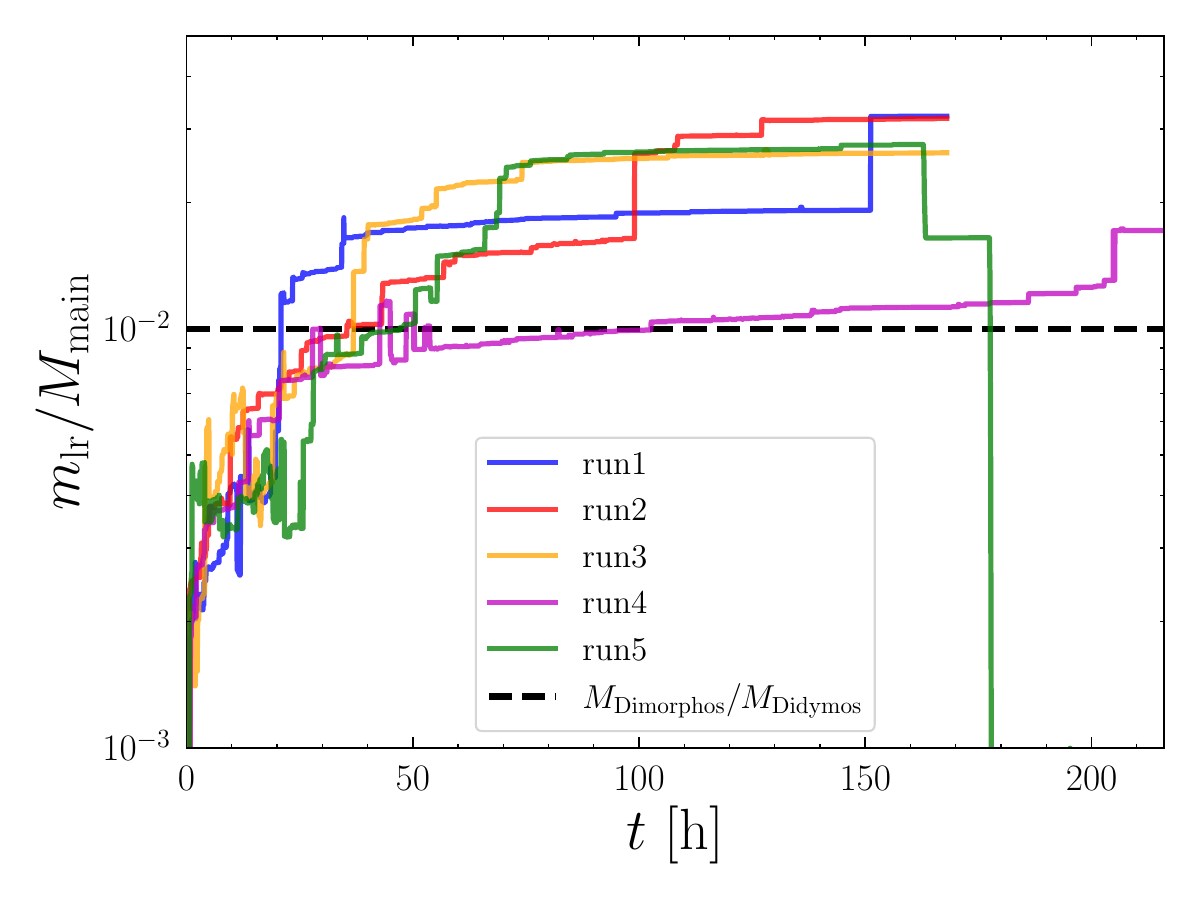}}
    \caption{The same plot as in figure \ref{fig:didymos_handoff_growth} but for the case of five separate standalone \textit{N}-body simulations of a Didymos-like system.}
    \label{fig:didymos_Nbody_growth}
\end{figure}

\begin{figure}
    \resizebox{\hsize}{!}{\includegraphics{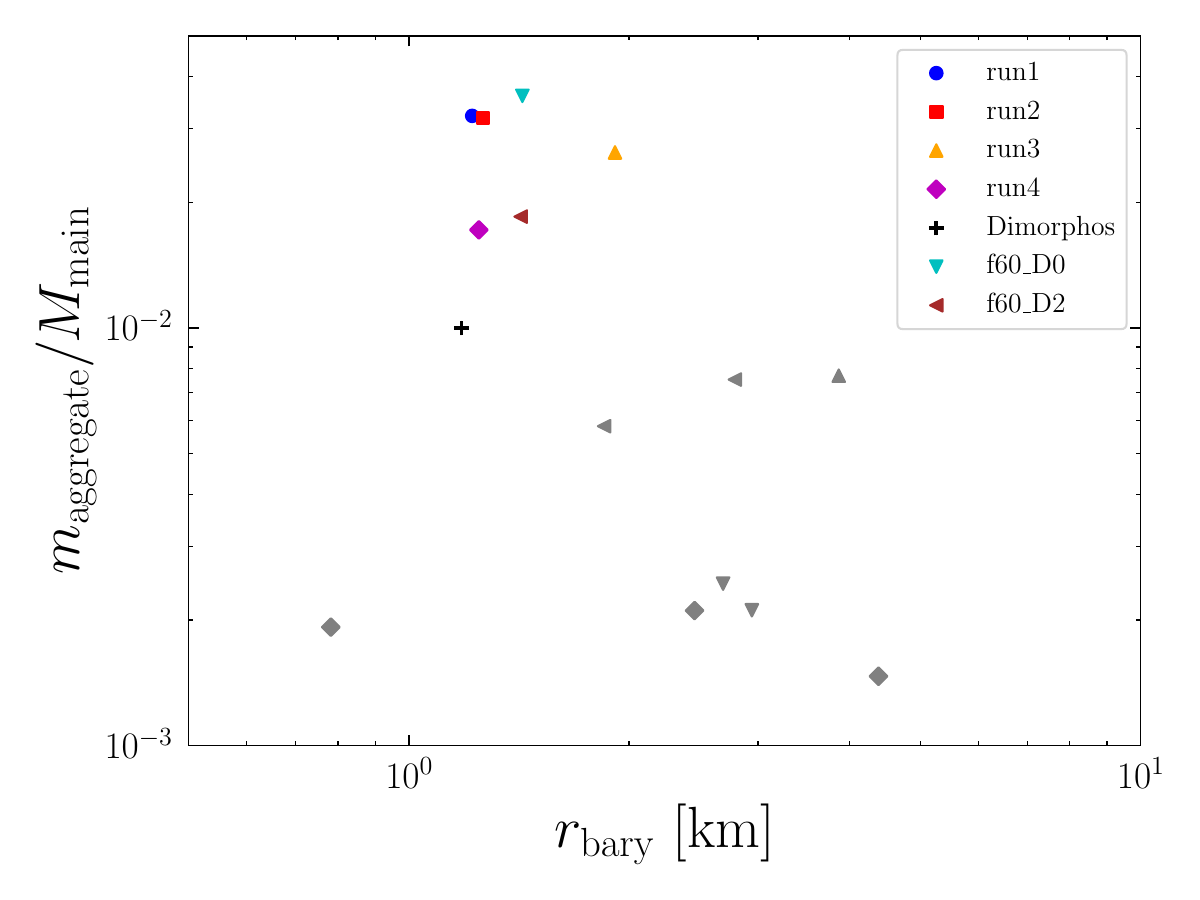}}
    \caption{The distribution of final masses for all major aggregates in the simulations for the Didymos-like case plotted against their distance from the barycenter. The largest remnant in each system has colour, while the grey markers represent the remaining aggregates that either are too low mass or too far out in the system (i.e.\ outside 5 km from the primary) to count as the most stable and massive satellite in the timescales we consider. The cyan and brown points marked with upside down and sideway triangles represents the result from the hand-off simulations.}
    \label{fig:didymos_Nbody_final_masses}
\end{figure}

\begin{figure}
    \resizebox{\hsize}{!}{\includegraphics{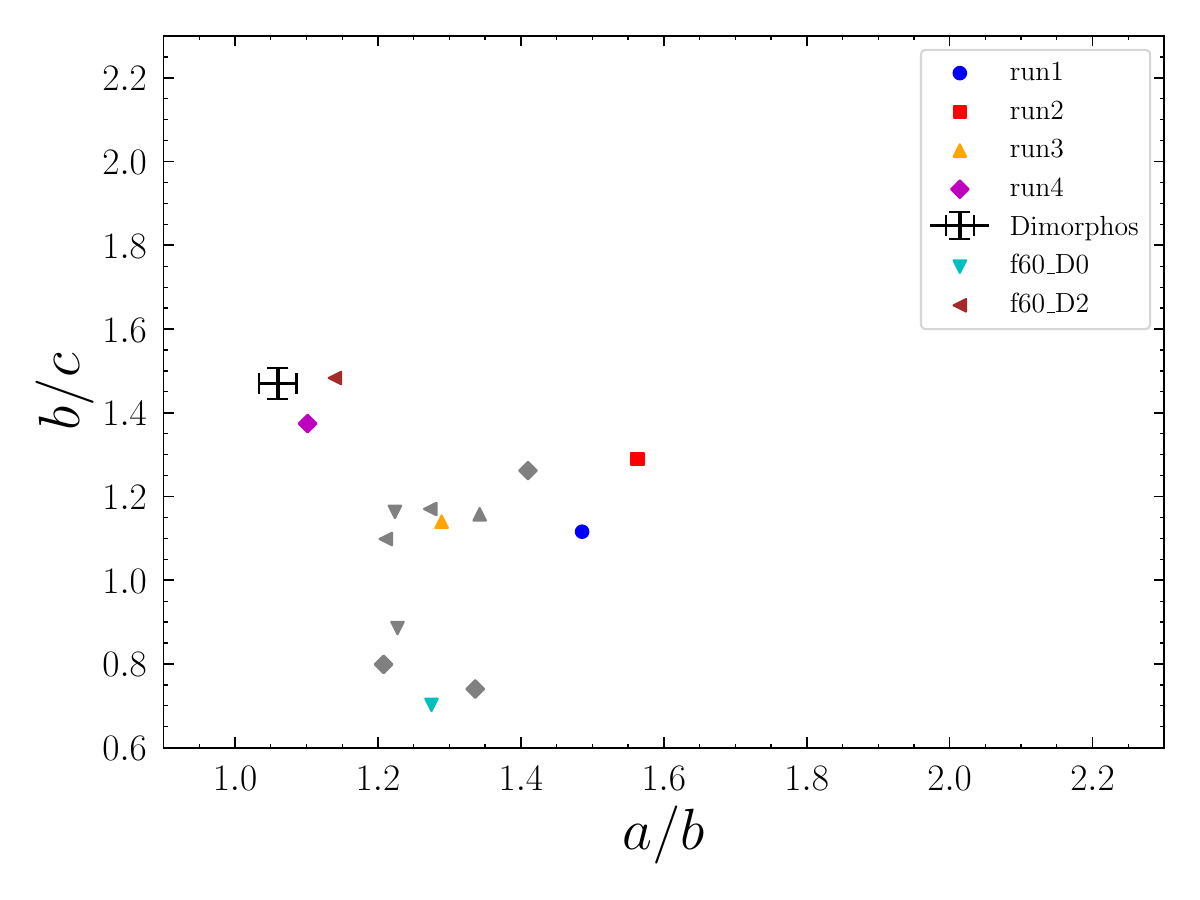}}
    \caption{The final shapes at the end of each simulation for the Didymos-like scenario. Each ellipsoidal axis has been evaluated from the corresponding dynamically equivalent equal-volume ellipsoid to the moment of inertia of each aggregate. Note that only the coloured markers represent bodies that have a large enough mass and are situated within 5 km of the primary body. They can be compared to the final masses and positions in the different systems from figure \ref{fig:didymos_Nbody_final_masses}. We note that these values are likely slightly larger than what we would obtain from the principal axes corresponding to the actual physical extents of the bodies \citep{Agrusa_et_al_2024}.}
    \label{fig:didymos_Nbody_final_shapes}
\end{figure}

\begin{figure*}
    \resizebox{\hsize}{!}{\includegraphics{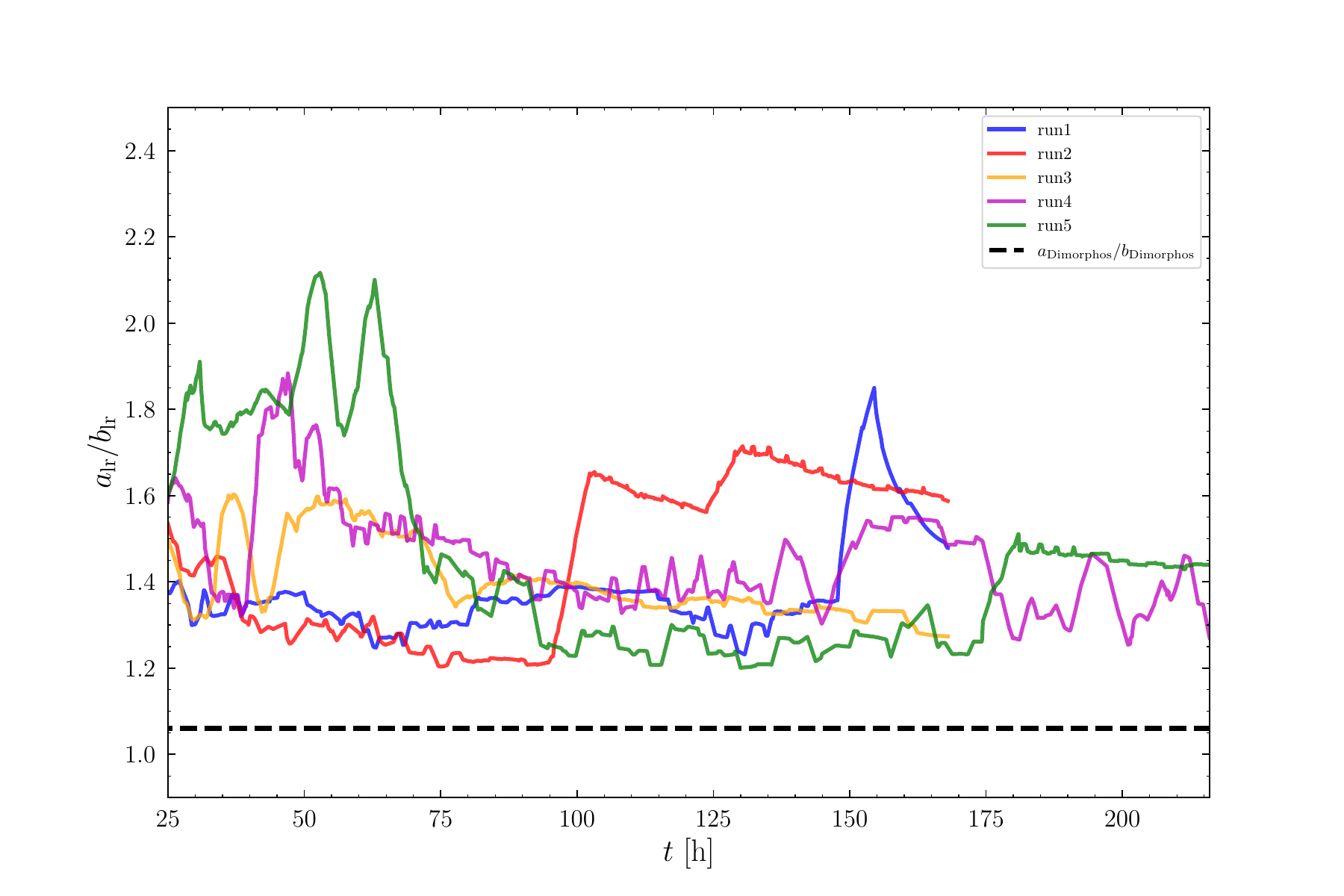}}
    \caption{The same plot as in \ref{fig:didymos_handoff_axes_vs_time} but for the standalone \textit{N}-body simulations for a Didymos-like primary. The graphs can be compared to the change in mass over time in figure \ref{fig:didymos_Nbody_growth} to identify major mass shedding events and mergers.}
    \label{fig:didymos_Nbody_axes_vs_time}
\end{figure*}

\subsection{Pure \textit{N}-body simulations for Didymos-like systems}\label{section:didymos_Nbody}

For the five pure \textit{N}-body simulations, we once more opted to use the same disk mass of $0.05 M_\mathrm{main}$ as in the Ryugu-like case. The SFD for the generated grains was also kept the same for consistency. The higher density along with the lower total disk mass led to a significant decrease in number of particles, averaging 2454 across the five simulations. Examining the growth tracks for each of the simulations, which are presented in figure \ref{fig:didymos_Nbody_growth}, three of the formation histories are IDH (runs 1, 2 and 3), while run 4 is an MDH case and run 5 is a CDH where the satellite gets completely disintegrated. The final masses and positions relative to the barycenter have been plotted in figure \ref{fig:didymos_Nbody_final_masses}. Lastly, the resulting shapes for the aggregates at the end of the simulations are given in figure \ref{fig:didymos_Nbody_final_shapes} while the evolution of the elongation in the orbital plane, i.e.\ the $a/b$ axis ratio, for the largest remnants can be found in figure \ref{fig:didymos_Nbody_axes_vs_time}. Compared to the Ryugu-like system cases, the final bodies generally have higher relative mass, with three out of seven simulations generating secondaries with masses greater than $0.03 M_\mathrm{main}$. Notably, the aggregate most similar to Dimorphos from run 4 is an MDH scenario, growing much more slowly and steadily after its disruption event, primarily through single-grain accretion with one last merger setting its final mass at $0.017 M_\mathrm{main}$. With principal axis ratios of $a/b = 1.10$ and $b/c = 1.38$, it exhibits the same type of flattened oblate shape that we are looking for. Studying the evolution of the $a/b$ values for the secondaries grown in Ryugu-like and Didymos-like cases in figures \ref{fig:ryugu_Nbody_axes_vs_time} and \ref{fig:didymos_Nbody_axes_vs_time}, a key difference between them is that the aggregates are initially less prolate for the latter. Overall, strong tidal interactions with the primary are less common due to its smaller fluid Roche limit since $D_\mathrm{Roche}\approx 2.44 R_\mathrm{main}(\rho_\mathrm{main}/\rho_\mathrm{p})^{1/3}$ \citep{Roche_1847}, where $\rho_\mathrm{p}$ is the density of a particle in orbit, which explains why we have more IDH scenarios with these initial conditions for the geometry of the disk. We note that this expression does not take into account the oblateness of the shape model we use for Didymos and address this further in \ref{section:discussion_geometry}. Looking at the growth tracks for the bodies once more, comparing them to the corresponding evolution of $a/b$, it becomes clear that deformations in this case are primarily driven by mergers rather than tidal forces. With this in mind, we can conclude that both runs 1 and 2 have produced bilobate satellites, albeit significantly less prolate ones than the case of the largest moon in Ryugu-like run 4. We study these merger cases in more detail in section \ref{section:discussion_impact_properties}.

\begin{table*}[]
    \centering
    \begin{tabular}{lllllllll}
        \hline\hline
         System & $M_\mathrm{lr}$ [$M_\mathrm{main}$] & Shape & Formation & $a/b$ & $b/c$ & $D_\mathrm{lr}$\ [$D_\mathrm{main}$] & $a$\ [$R_\mathrm{main}$] & $e$ \\
         \textbf{Didymos} & \textbf{0.010} & \textbf{OS} & - & \textbf{1.06} & \textbf{1.47} & \textbf{0.20} & \textbf{3.03} & \textbf{<0.03} \\
         f60\_R2 & 0.029 & PS & MDH & 1.27 & 1.58 & 0.41 & 2.72 & 0.09 \\
         f60\_R6 & 0.023 & PS & IDH & 1.34 & 1.42 & 0.38 & 3.41 & 0.01 \\
         Ryugu\_run1 & 0.020 & PS & MDH & 1.29 & 1.61 & 0.42 & 2.51 & 0.25 \\
         Ryugu\_run2 & 0.019 & PS & MDH & 1.25 & 1.47 & 0.38 & 2.58 & 0.02 \\
         Ryugu\_run3 & 0.004 & PS & CDH & 1.62 & 1.10 & 0.26 & $\infty$ & >1 \\
         Ryugu\_run4 & 0.015 & BL & CDH & 2.26 & 1.02 & 0.48 & 2.88 & 0.09 \\
         Ryugu\_run5 & 0.014 & OS & CDH & 1.09 & 1.32 & 0.31 & 3.02 & 0.26 \\
         f60\_D0 & 0.032 & PS & IDH & 1.27 & 0.70 & 0.42 & 3.36 & 0.05 \\
         f60\_D2 & 0.018 & OS & MDH & 1.14 & 1.48 & 0.38 & 3.51 & 0.06 \\ 
         Didymos\_run1 & 0.032 & BL & IDH & 1.49 & 1.12 & 0.46 & 2.69 & 0.10 \\
         Didymos\_run2 & 0.032 & BL & IDH & 1.56 & 1.29 & 0.49 & 2.93 & 0.00 \\
         Didymos\_run3 & 0.026 & PS & IDH & 1.29 & 1.16 & 0.39 & 3.93 & 0.01 \\
         Didymos\_run4 & 0.017 & OS & MDH & 1.10 & 1.37 & 0.33 & 2.94 & 0.05 \\
         Didymos\_run5 & - & - & CDH & - & - & - & - & - \\
         \hline
    \end{tabular}
    \caption{The properties of the largest moons at the final time step of each simulation along with the corresponding values for Dimorphos, which are highlighted in bold. The shape types correspond to oblate spheroid (OS), prolate spheroid (PS) and bilobate (BL). As for the formation history archetype definitions, see section \ref{section:formation_patterns}.}
    \label{tab:final_moon_properties}
\end{table*}

Properties such as final mass, orbital elements and geometrical values for each largest remnant in our simulations, along with the corresponding Dimorphos values can be found in table \ref{tab:final_moon_properties}. We signify each type of shape for the aggregates as oblate spheroid (OS), prolate spheroid (PS) and bilobate (BL).

\section{Discussion}\label{section:discussion}

Based on our results and analysis in sections \ref{section:results} and \ref{section:didymos}, we can identify initial disk conditions and dynamical evolution histories that facilitate the formation of oblate spheroids such as Dimorphos. Working backwards from how we obtain our aggregates with $a/b < 1.15$, they all went through at least one critical merger that led to significant deformation. As previously mentioned, a formation path where a body is formed via slow accretion of material periodically shed from the primary body in the system logically favours a prolate shape. This also seems to apply to cases where the body rapidly forms via mergers in a debris disk after a gravitational instability has occurred and then mainly grows through the accretion of smaller aggregates and single grains during the later stages of its formation. This comes from the fact that if the body does not undergo any mergers with larger aggregates and remains close to the fluid Roche limit, it will in most cases end up on a tidally locked, synchronous orbit. This aligns with the results of \cite{Agrusa_et_al_2024}, who investigated the formation of binary asteroid systems using a large number of simulations starting from a more narrow and thinner debris disk with different masses around a Didymos-like primary. Much like in our model, in the cases where they ended up with an oblate Dimorphos-like secondary, the later stages of formation was primarily driven by tidal disruption events and/or mergers with favourable geometry. The same results were found by \cite{Madeira_&_Charnoz_2024}, who expanded on their 1D ring-satellite model study \cite{Madeira_et_al_2023} to consider cases where mass is deposited into the debris disk over different timescales. When the mass shedding is instantaneous, the satellite that forms does not undergo any critical mergers that significantly could alter its shape, rendering it prolate.

Looking at the formation histories for the largest remnants in our simulations and their corresponding shapes, it is clear that a body growing slowly via IDH with few to no critical mergers during the later stages of its formation would favour a prolate shape, such as e.g.\ Ryugu run 2. Most of our simulations go through shape-defining deformation events such as mergers and tidal disruptions, which create the many discontinuities observed in figures \ref{fig:ryugu_handoff_growth}, \ref{fig:ryugu_Nbody_growth}, \ref{fig:didymos_handoff_growth} and \ref{fig:didymos_Nbody_growth} which display the growth tracks for each largest remnant. Again, focusing on e.g.\ Ryugu run 2 after the 75 h mark, the shape of the bodies that reach a high mass early on will be more weakly affected by each subsequent merger as each aggregate it collides with will most often have a lower mass. 

Hence, while the formation method is highly chaotic, a debris disk that favours mergers between bodies of similar mass should have a higher likelihood of forming an oblate spheroid. To further this claim, even though similar mass mergers far from guarantee an oblate shape, as the outcome is highly dependent on impact velocity and angle \citep{Leleu_et_al_2018}, tidal locking in our simulations is usually broken by accretion events that induce a higher spin-rate for the largest aggregate and push it out into a wider orbit. In turn, the subsequent accretion of single particles and small aggregates for the spinning satellite will be isotropic and often further reduce its $a/b$-value, making the body more oblate as time passes and the disk is slowly depleted of material.

An example that reinforces this idea is Didymos run 3. In the same manner as Ryugu run 2, it does not undergo any significant mergers after the 75 h mark. However, when studying the trend of the $a/b$ value for the largest remnant in this system, it tends towards a more oblate shape as it accretes single grains and smaller aggregates. This is due to its final major merger that broke its synchronous rotation, causing a yaw rotation back and forth. As a result, the single grain and small aggregate accretion no longer had a preferred direction along its longest axis, making tides less efficient at retaining the prolate shape. Over the later stages of its evolution, the aggregate went from having an $a/b$ value of $\sim 1.4$ to 1.29. Nevertheless, at the final time step, almost the entire disk was depleted of material showing us how the mechanism where single grain accretion makes a satellite more oblate over time depends on the remaining disk mass and the initial $a/b$ value of the body.

Based on these observations, we will proceed with the discussion to eventually motivate which properties a debris disk should have to increase the probability of forming an oblate spheroid satellite. After that, we go through which developments could improve our model and, as a result, our understanding of this complex problem. 

\subsection{Debris disk properties and tidal influence}\label{section:discussion_geometry}

Judging from our different simulation setups, including hand-off cases and \textit{N}-body simulations for Ryugu and Didymos-like bodies, the initial geometry of the disk seems to highly affect the formation history and the final shape of the main satellite, more so than the debris disk mass, the number of particles and their size distribution. This is also consistent with the outcome for \cite{Agrusa_et_al_2024}, who found that the final secondary shape and formation process for disks with masses of $0.02 M_\mathrm{main},0.03 M_\mathrm{main}$ and $0.04 M_\mathrm{main}$ are highly similar, with the lowest mass disk producing more prolate bodies than the other cases. The main difference for these cases is the slower growth rate for lower disk masses, while a higher disk mass leads to greater final relative masses, semi-major axes and eccentricities of the formed secondaries. A more radially extended disk, with its edge situated beyond the fluid Roche limit for the primary body, allows for a higher survival rate of aggregates which leads to a higher number of bodies in the system, more mergers and more possibilities for deformation. The type of chaotic rotational failure that we have explored also results in higher growth rates than for previous studies such as \cite{Walsh_et_al_2008}, where the mass shedding was slow and continuous rather than instantaneous as a result of their implemented spin-up model.

We have also observed that the initial shape (at $\sim 24$ h into the simulation) of the largest remnant before any tidal events or mergers is less prolate for a Didymos-like disk, which according to our SPH simulations extends farther beyond the fluid Roche limit than the disk created in the Ryugu-like rotational failure event. The edges of the disks in these cases are located at $\sim 3 R_\mathrm{main}$ from the primary center-of-mass in the Ryugu case and $\sim 3.5 R_\mathrm{main}$ in the Didymos case. Moreover, the moons formed in our model have a mean semi-major axis of $\sim 3.03 R_\mathrm{main}$ by the final time step, which aligns with observations of binary asteroid systems \citep{Monteiro_et_al_2023} and migrate moderately over the course of the simulations unless they experience a strong collision or tidal event, as can be seen from an example in figure \ref{fig:ryugu_handoff_orbital_elements}. In turn, since our aggregates primarily grow through mergers with other aggregates, angular momentum and energy transfer mainly occur via soft particle--particle contacts and lead to quick migration outwards beyond the fluid Roche limit for aggregates formed inside of it, which for our cases would be located at $\sim 2.44 R_\mathrm{main}$. Since the debris disk ends up more radially extended for the Didymos case relative to the primary radius, the proto-moon will most often form close to the Roche limit or outside of it. As a result, any deformation due to tides is less likely to occur throughout the growth process, which again explains why a majority of the formation histories for the Didymos-like systems are IDH, while most Ryugu-like systems are MDH or CDH. This provides a different pathway for forming Dimorphos-like bodies since single-grain accretion and mergers with lower mass moonlets for an aggregate that is initially less prolate and faster rotating than a tidally locked body will be isotropic and make the shape more oblate over time. We discuss the effect of the initial shape of the proto-moon for deformation via mergers in the next section. Despite this trend for the Didymos-like systems, an overall pattern for our simulations is that only formation histories that are MDH or CDH can produce oblate spheroids similar to Dimorphos in mass and form, emphasising the importance of close encounters with the primary to deform the body from its initially prolate state before it enters another stage of accretion which leads to the final shape. This also indicates that Dimorphos could form over longer timescales than explored in this work after the debris disk has been more or less cleared of matter. In this scenario, a prolate body undergoes orbital decay through tidal interactions with the primary and then has a catastrophic close encounter which results in mass shedding and scattering, leaving behind a less massive but oblate shape on a wider orbit, a phenomenon we have observed in several of our simulations before the disk is depleted.

While the theoretical fluid Roche limit provides a convenient qualitative measure for where the tidal interactions will occur throughout our simulations, using a static limit for the tidal disruption for Rubble pile asteroids fails to capture the intricacies of aggregate dynamics \citep{Holsapple_&_Michel_2006,Holsapple_&_Michel_2008} and has been explicitly numerically modelled and investigated in several \citep{Richardson_Bottke_&_Love_1998,Walsh_&_Richardson_2006,Movshovitz_et_al_2012,Schunova_et_al_2014,DeMartini_et_al_2019,Zhang_&_Michel_2020}, primarily regarding tidal disruption of rubble piles during close encounters with planets, but also with white dwarfs \citep[e.g.][]{Li_Mustill_&_Davies_2021}. The effect of using irregularly shaped grains on the integrity of the rubble pile during a tidal encounter was the topic for \cite{Movshovitz_et_al_2012}, who in the context of rubble pile asteroids emphasised that particle interlocking, dilatancy\footnote{The expansion of granular material when subjected to shear deformation.} and more complex packing of granular media contributes to the structural strength of the aggregate, making it more difficult to disrupt than a collection of spherical particles. With this in mind, it is unsurprising that the aggregates in our simulations often undergo close encounters with the primary, having orbital pericenters closer than $D_\mathrm{Roche}$ without being disrupted, as can be seen when comparing figures \ref{fig:ryugu_handoff_growth} and \ref{fig:ryugu_handoff_orbital_elements} before the critical merger at 72 h. Furthermore, we also expect that this contributes to the total number of aggregates available for collisions in our debris disks, as they will not be disrupted as easily during each stage of the formation process and that this number is greater than for a disk consisting of spherical particles, which we aim to confirm in a future study. As we have already established, the more aggregates that exist in the system, the likelier it is for the largest remnant to have a geometrically favourable merger that gives it a more oblate shape.

\subsubsection{Asymmetries for the primary body and disk}

The problem of identifying the tidal disruption limit becomes more complicated when introducing asymmetries for the primary shape. The model we use for Didymos is the official pre-derived impact mesh for the DART mission which is an oblate, nonaxisymmetric spheroid \citep{Barnouin_et_al_2023}. Not only does such a shape affect the position of the theoretical fluid Roche limit, but its asymmetric inertia tensor can also lead to extra angular momentum transfer to the disk and affect its evolution \citep{Takeda_&_Ida_2001}. The asymmetries on the surface of an asteroid induce chaotic diffusion of the eccentricity of particles in low orbits \citep{Sicardy_2020,Rollin_et_al_2021,Madeira_et_al_2022,Ribeiro_et_al_2023}, which is expected to affect the geometry and velocity of impacts between particles\footnote{In an environment free of inter-particle collisions, this effect leads to a rapid loss of particles due to ejections or collisions with the asteroid, resulting in a hollow region near the asteroid. In our case, however, the impacts between particles dampen the eccentricities, preventing the formation of the hollow region.}. In order to investigate whether or not the effect is significant to the formation process, we intend to expand upon our hand-off scheme to also provide an $\alpha$-shape with a corresponding inertia tensor for the resulting contour of the primary post rotational failure and compare the evolution of two disks with identical initial conditions but different primary shapes, one that is spherical and one that is based on the aforementioned nonaxisymmetric post fission body. This is a non-trivial problem, as the primary often has formed temporary clumps of SPH particles along its equator at the time step for the hand-off which would not be representative of the final shape when the primary aggregate has settled after its initial deformation. 

As previously discussed in section \ref{section:didymos_handoff}, the evolution of a pure particle debris disk seems to be affected by initial nonaxisymmetries, such as the one in figure \ref{fig:didymos_handoff}. While we lack a statistically significant volume of simulations to confirm this phenomenon, the unique growth track pattern for the corresponding disk (see figure \ref{fig:didymos_handoff_growth}) provides us with a strong indication that asymmetries lead to significant changes in the satellite formation patterns and warrants further investigation, especially given the lack of literature dedicated to this problem for pure particle disks and circumasteroidal disks in general. A critical difference with the asymmetric disks is that they already have large fragments of SPH particles at the hand-off point, which explains why the largest grain diameter is at least two times greater than for the symmetric disks (see table \ref{tab:didymos_spin_up}). As previously mentioned, while the initial sizes of these grains diverge compared to observed boulder sizes on Dimorphos \citep{Pajola_et_al_2023,Robin_et_al_2023}, they represent
aggregates that already have formed during the mass shedding event. The presence of these large grains may have a significant effect on the shape and mass of the final moon as they are monolithic and will not get disrupted by any tidal effects or impact events. In turn, for future hand-off implementations, converting these larger grains into aggregates consisting of smaller constituents with smoother shapes and sizes could prove to be important.  

\subsubsection{Particle number dependency}

As shown by \cite{Kokubo_et_al_2000} and \cite{Takeda_&_Ida_2001}, the number of particles has little effect on the evolution of circumplanetary particle disks confined within the Roche limit for values $10^3 \leq N \leq 10^4$. However, \cite{Sasaki_&_Hosono_2018} found that disk behaviours and moon formation histories can differ for very high resolution \textit{N}-body simulations with $N>10^5$. Nevertheless, we cannot argue with certainty that this would be the case for our model, as these scenarios used spherical particles to investigate a problem on a different scale than ours. Moreover, even when using an SFD more true to observations with e.g.\ $D_\mathrm{min} = 3$ m and $D_\mathrm{mean}=15$ m, this would not yield a number greater than $\sim 2.5 \times 10^4$. Nevertheless, in the future we aim to perform simulations with higher particle numbers, restricted only by computational limitations, to verify whether or not this directly affects the disk evolution and formation processes for the moons in our model.

\subsection{Effect of merger geometry for aggregates}\label{section:discussion_impact_properties}

Studying the $a/b$ value evolution for our simulations in figures \ref{fig:ryugu_Nbody_axes_vs_time}, \ref{fig:didymos_handoff_axes_vs_time} and \ref{fig:didymos_Nbody_axes_vs_time}, it is evident that we have several types of mergers that affect the shape of the largest remnant in different ways. More importantly, we can distinguish between mergers that lead to a more oblate body and those that lead to a more prolate body. To better understand which merger properties govern the final shape of the body, we investigate the $a/b$ value of the largest remnant before and after critical mergers identified by comparing the $a/b$ value evolution over time with the mass growth tracks in figures \ref{fig:ryugu_handoff_growth}, \ref{fig:ryugu_Nbody_growth}, \ref{fig:didymos_handoff_growth} and \ref{fig:didymos_Nbody_growth}. For consistency, we fix which of the axes are $a$ and $b$ prior to the impact, meaning that we do not alter this definition even if there is a change such that $a<b$ post merger.

\begin{figure*}
    \resizebox{\hsize}{!}{\includegraphics[trim=10 0 0 0, clip]{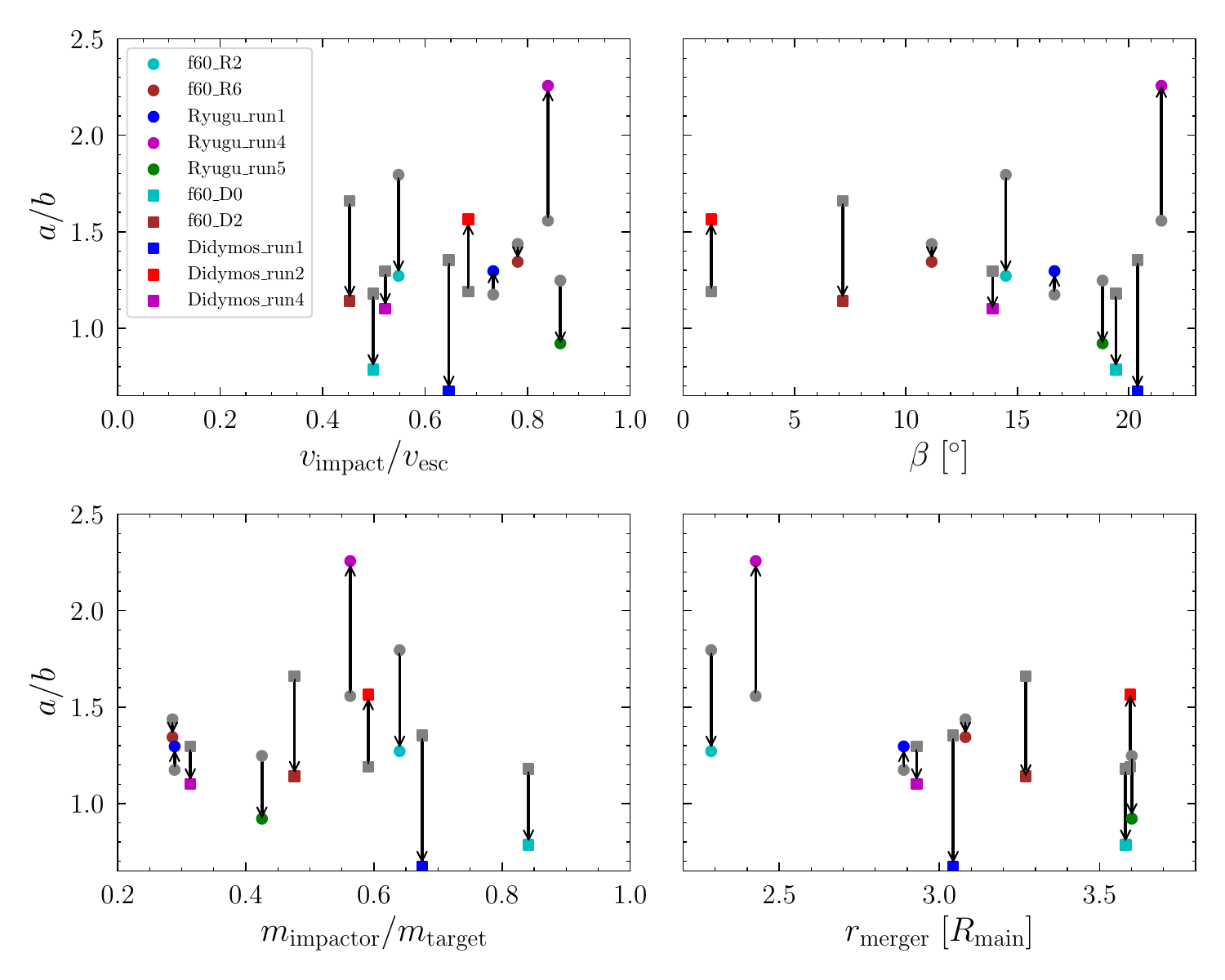}}
    \caption{The DEEVE axis ratio $a/b$ value before (grey) and after (coloured) the critical merger which determines the final shape of the considered secondaries at the final time step of our simulations, compared with four different impact properties. Arrows pointing upwards generally mean that the body became more prolate (elongated) after the merger, while an arrows pointing downwards indicates that a deformation to a more oblate shape. Note that we fix the definition of each axis as $a$ or $b$ before the merger, which is why the $a/b$ axis ratio can be below unity for some cases (e.g.\ Didymos run 1). \textbf{Upper left:} the $a/b$ axis ratios and the corresponding impact velocity divided by the mutual escape velocity. \textbf{Upper right:} same axis ratio change plotted against the impact angle between the two velocity vectors. \textbf{Lower left:} the axis ratio compared to the mass ratio of the impactor and target. \textbf{Lower right:} the distance of the merger from the barycenter of the primary in terms of main radii.}
    \label{fig:merger_ab_vs_impact_properties}
\end{figure*}

In figure \ref{fig:merger_ab_vs_impact_properties}, we have plotted several impact properties of a few select critical mergers from our simulations. The selection criterion was that the merger had to be the final significant reshaping event for the largest remnant during its formation history. In each plot, the grey marker represents the $a/b$ value before the merger, while the coloured marker is the same ratio afterwards by the final time step of the simulation when the resulting aggregates have settled. In the top left plot, we show the impact velocity, i.e.\ $|\Vec{v}_\mathrm{target} - \Vec{v}_\mathrm{impactor}|$, normalised by the mutual escape velocity given by 

\begin{equation}
    v_\mathrm{esc} = \sqrt{\frac{2G(m_\mathrm{target}+m_\mathrm{impactor})}{(R_\mathrm{target}+R_\mathrm{impactor})}}.
\end{equation}

Here, the target body is always referring to the largest of the two. Notably, the impact velocities we observe in our simulations are much lower than for previous studies exploring deformation through mergers such as \cite{Leleu_et_al_2018}, who investigated collisions between initially spheroidal moons in the rings of Saturn for impact velocities between $v_\mathrm{impact}/v_\mathrm{esc}$ values between 1 and 4, leading to growth in the pyramidal regime. We attribute this to the fact that our work explores a different regime entirely, where bodies on similar orbits near the primary undergo comparably low-energy gravitational encounters, leading to soft mergers rather than the violent, higher-velocity impacts that occur for the migrating moonlets in the rings of Saturn and cause significant deformation. This is quantified in the impact velocity being approximately proportional to $\mu^{-1/3}$, where $\mu = (m_\mathrm{impactor}+m_\mathrm{target})/M_\mathrm{main}$ \citep{Leleu_et_al_2018}. Furthermore, the dissipative nature of the dynamics in our dense disks with numerous inelastic particle--particle collisions will also contribute to the low velocities observed as it effectively dampens any orbits with larger eccentricities and inclinations, keeping the system from getting more dynamically excited. \citeauthor{Leleu_et_al_2018} found that collisions between gravitational aggregates at these velocities lead to the characteristic ridges and deformations that can be seen from images of e.g.\ Pan and Atlas. Hence, if Dimorphos formed via a series of higher-velocity mergers, we would expect to observe distinct ridges and lineaments on its surface, yet it seems largely homogeneous \citep{Barnouin_et_al_2023} with no spatial preference for boulder density and sizes \citep{Pajola_et_al_2023}, indicating a different formation pathway was involved\footnote{We note that any ridges or lineaments on the surface of Dimorphos could have been homogenised since the coalescence of the satellite due to exchange of material from Didymos to Dimorphos \citep{Yu_et_al_2019,Madeira_et_al_2024} or isotropic single-grain accretion during the later stages of its formation history.}. Moreover, for lower impact velocities less than $0.75v_\mathrm{esc}$, \citeauthor{Leleu_et_al_2018} get unstable bilobate objects that end up separating because of tidal or Coriolis forces due to the close proximity to Saturn. While we cannot speak for the long-term stability of our objects, it is clear that our lowest velocity mergers below this value lead to more oblate shapes rather than generation of more prolate bodies. Intuitively, a low-velocity merger between two spherical bodies at a low impact angle should produce bilobate aggregates, which also is seen in other studies such as \cite{Jutzi_&_Asphaug_2015}. The dissimilarity in outcomes must accordingly be related to the prolate pre-impact shapes with $a/b>1.2$, as well as the orientation of the longest axes of the two merging bodies.

Indeed, the differences between the pyramidal and our hierarchical regime are emphasised by the initial shape of the aggregates in our mergers, as well as their masses which can be seen in the lower left plot. The largest mass ratio $m_\mathrm{impactor}/m_\mathrm{target}$ is 0.84, while the mean value is 0.51 and we observe that lower mass ratios lead to smaller effective changes in $a/b$ despite having a significant spread in $v_\mathrm{impact}/v_\mathrm{esc}$. In the upper right plot, the change in axis ratio has instead been plotted against the absolute angle between the velocity vectors of the two objects, denoted $\beta$. Seeing this angle is smaller than $23^\circ$ for all our cases, we can deduce that the orbits of the target and impactor are widely similar prior to the merger, which also aligns with the low impact velocities we find. Furthermore, there is no trend as to which impact angles produce which shapes or as to how far from the primary centre-of-mass the merger occurs, $r_\mathrm{merger}$, which has been plotted in the bottom right plot. Note, however, that the far-in locations of the mergers for runs \texttt{f60\_R2} and Ryugu run 4 do not coincide with their orbits by the final time step as each of these mergers is soon followed by a migration outwards farther away from the primary.

\begin{figure}
    \resizebox{\hsize}{!}{\includegraphics{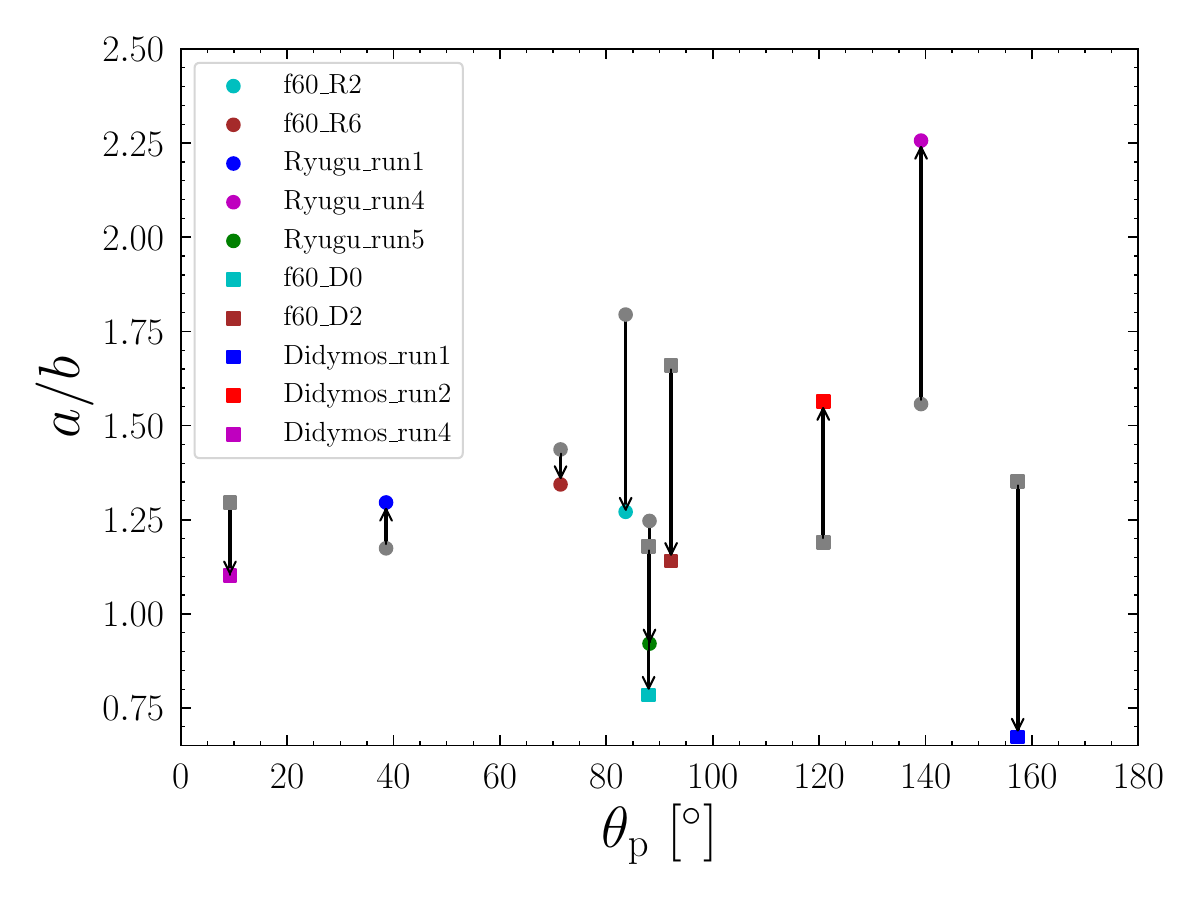}}
    \caption{The DEEVE axis ratio $a/b$ value before (grey) and after (coloured) the critical merger which determines the final shape of the considered secondaries at the final time step of our simulations, compared against $\theta_\mathrm{p}$ which is the angle between the longest principal axis and the relative velocity vector for the merger. A short-axis merger corresponds to $\theta_\mathrm{p}\sim 90^\circ$ while a long-axis merger would mean a value close to 0 or 180$^\circ$.}
    \label{fig:merger_ab_vs_prin_axis_angle}
\end{figure}

The reason why $\beta$ cannot directly predict the post-merger shape of the body is due to the fact that it tells us nothing of how the asymmetric target is rotated relative its trajectory and if it collides with the impactor at its short-axis or long-axis. To verify this idea, we define the angle $\theta_\mathrm{p}$ between the longest principal axis pre-merger, $a$, and the relative velocity vector. An angle of $\theta_\mathrm{p} = 0$ or $180^\circ$ then corresponds to a relative velocity vector completely aligned with $a$, while a value of $90^\circ$ tells us that the relative velocity vector is parallel to the shortest axis, $b$. We plot this value against the change in $a/b$ in figure \ref{fig:merger_ab_vs_prin_axis_angle}. There is a correlation between a reduction/increase in $a/b$ value of the largest remnant and the $\theta_\mathrm{p}$ value of the merger, indicating that short-axis/long-axis mergers indeed favour the creation of more oblate/prolate bodies. We study the two outlier cases, Didymos runs 2 and 4 which are both long-axis mergers per our definition but still show a reduction in $a/b$. The former does not become an oblate spheroid, but rather a prolate, bilobate body where the pre-merger axis $b$ instead is the longest axis. As for Didymos run 4, the impactor is a highly elongated but low-mass aggregate on an eccentric orbit that undergoes a close encounter with the primary, enters the Hill radius of the target and is subsequently slowed down before it is tidally disrupted and accreted, which makes it behave unlike any of the typical short- or long-axis mergers we observe.

\begin{figure}
    \resizebox{\hsize}{!}{\includegraphics{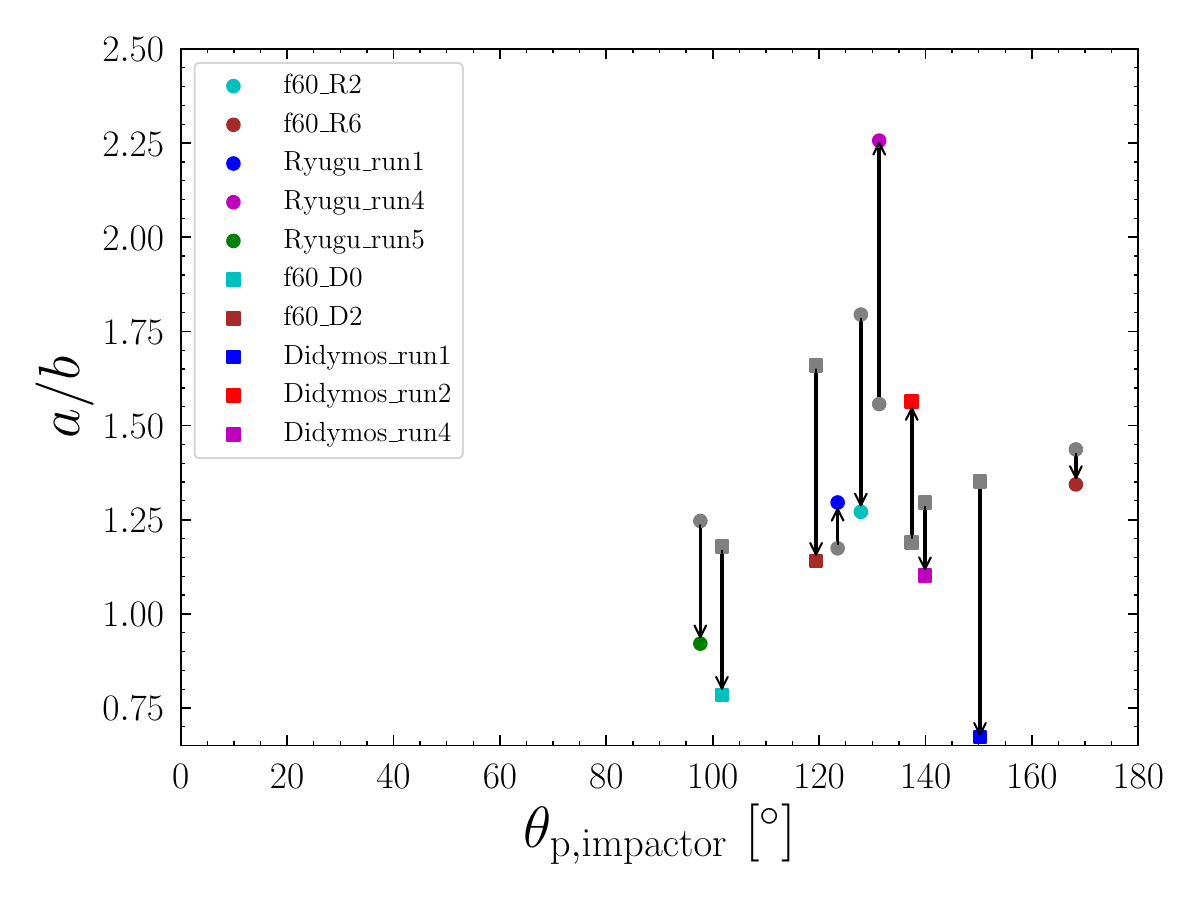}}
    \caption{Same as figure \ref{fig:merger_ab_vs_prin_axis_angle} but for the impactor in the merger.}
    \label{fig:merger_ab_vs_prin_axis_angle_impactor}
\end{figure}

While the orientation and shape of the impactor will also affect the final structure of the largest moon in our systems, the generally low mass ratio between the two objects renders the orientation and initial shape of the target far more important. This becomes evident when comparing figure \ref{fig:merger_ab_vs_prin_axis_angle} with figure \ref{fig:merger_ab_vs_prin_axis_angle_impactor}, where we instead have plotted the angle between the relative velocity and the largest principal axis of the impactor. Only two of the impactors have distinct short-axis mergers with $\theta_\mathrm{p,impactor}$ close to $90^\circ$ while five of the corresponding mergers generate more oblate bodies. Instead, their orbits have the largest effect on the outcome of the merger. Looking at the cases which generate more prolate bodies and contact binaries such as Ryugu run 4, as well as Didymos runs 1 and 2, this becomes evident when combining their distribution of impact angles with the corresponding $\theta_\mathrm{p}$ values. Here we see that the long-axis mergers Ryugu run 4 and Didymos run 1 have the largest impact angles as well, showing that the impactor has a dissimilar trajectory to the target. Indeed, investigating the merger in detail confirms that the impactor in Ryugu run 4 has a wider orbit than the target and approaches from the outside, while the opposite is true for Didymos run 1 where the impactor instead has an orbit closer to the primary and intersects the orbit of the target from inside.

From these results, there is an indication that a formation model more likely to generate low-velocity, short-axis mergers is efficient at forming oblate bodies. Tying this to the geometry of the disk, a wider initial radial extension similar to those generated by giant impacts onto planets \citep[e.g.][]{Ida_et_al_1997,Kokubo_et_al_2000} facilitate the fast formation of multiple aggregates. Due to the hierarchical nature of the process, one aggregate will quickly grow more massive than the other satellites in the system and often becomes prolate and tidally locked on a synchronous orbit. Undergoing one or more mergers with other satellites (depending on the mass ratios) under favourable circumstances such as low impact velocity, similar trajectories and $\theta_\mathrm{p}$ values close to 90$^\circ$ will then reduce the elongation of the body significantly and can in some cases generate oblate spheroids.

\subsubsection{Bilobate satellites}

Further recalling our analysis from the section \ref{section:discussion_geometry}, the IDH, MDH and CDH scenario outcomes are dependent on the position of the Roche limit relative to the outer edge of the disk and are as a result system-specific. Because the strength of the tidal forces largely affects the initial shape prior to mergers, forming most aggregates farther away from this limit will generate less prolate bodies that are less likely to have synchronous orbits, which is the case for our Didymos-like systems. For the corresponding IDH cases where growth is dominated by isotropic accretion of single grains and low-mass aggregates, the moonlet will become more oblate over longer timescales. However, for systems where there still are additional massive satellites, it follows that the long-axis mergers will be likelier, which in turn increases the probability of forming bilobate bodies. Hence, for our rotational failure model, primaries with smaller radii and higher densities with Roche limits closer to the surface should be more likely to have contact binary moons. 

This is an important topic for further research, especially given the recent discovery of the bilobate satellite Selam of the (152830) Dinkinesh system \citep{Levison_et_al_2024} by the Lucy mission during a flyby \citep{Levison_et_al_2021}. Dinkinesh has a distinct top-shape indicative of rotational failure and an effective diameter of $720\pm 24$ m making it smaller than both Didymos and Ryugu. The satellite Selam is orbiting the primary at a distance of $\sim 9R_\mathrm{main}$ and has a distinct bilobate structure where the components have diameters of $212\pm 21$ m and $234\pm 23$ m \citep{Levison_et_al_2024}. Despite the much greater separation between primary and secondary, the Lucy images show that the lobes of Selam align radially with the primary, indicative of a tidally locked orbit. In turn, there is a question of whether or not Selam formed closer to the primary and then migrated outwards via dynamical effects such as binary YORP \citep[BYORP,][]{Cuk_&_Burns_2005} or tidal interaction \citep{Jacobson_&_Scheeres_2011b} or if it formed farther out in the pyramidal regime according to the model of \cite{Madeira_&_Charnoz_2024}. Moreover, the analysis from \cite{Levison_et_al_2024} also shows that the mass ratio between the satellite and primary, assuming equal density for both objects of $\rho = 2400\pm350\ \mathrm{kg/m}^3$, is 0.06 which is significantly higher than any of the satellites we can form during the course of our simulations (see table \ref{tab:final_moon_properties}) as we assume a lower initial disk mass. Nevertheless, judging from our spin-up simulations of Didymos in table \ref{tab:didymos_spin_up} and the work of similar studies \citep[e.g.][]{Sugiura_et_al_2021,Hyodo_&_Sugiura_2022}, it is entirely possible for primaries with higher bulk densities to shed such large masses via rotational failure when introducing some weak cohesion to the primary. Hence, by performing follow-up SPH simulations with a primary based on Dinkinesh with higher resolution, we can get a better understanding of whether or not forming such a massive contact binary satellite is achievable with the circumasteroidal disk model. Given the low velocity of our impacts and the fact that our model can form weakly bilobate bodies even with low-resolution simulations, we believe this to be the case. 

With the pronounced equatorial ridge of Dinkinesh in mind, a history of deformation and mass shedding due to rotational spin-up is very likely. In turn, we expect that the existence of Selam cannot be explained with satellite formation via giant impacts \citep{Durda_et_al_2004,Durda_et_al_2007} alone. Yet, we can imagine a scenario where a combination of rotational failure and a grazing giant impact can lead to a similar type of debris disk investigated in this work and \cite{Agrusa_et_al_2024}, especially since Dinkinesh displays nonaxisymmetry and a significant concavity that could originate from a collision.

\subsection{Particle size distribution, shape and resolution}\label{section:discussion_particle_sfd}

One important caveat of the SFD in our model and the low resolution of our hand-off simulations is that it leads to unrealistically large boulders in the system and an absence of smaller grains. As shown in a series of recent publications related to the DART mission impact, the choice of boulder packing, cohesion and SFD highly affects the dynamics of rubble piles and whether or not they are disrupted or reshaped from high-energy collisions \citep{Raducan_et_al_2022,Raducan_et_al_2024a,Raducan_et_al_2024b}. Combining observations from the DART mission \citep{Li_et_al_2023,Dotto_et_al_2024} and SPH simulations of the DART impact, \cite{Raducan_et_al_2024b} found that Dimorphos is expected to be a cohesionless rubble pile with a boulder packing of less than 40\% on the surface and slightly below it. This accounts for boulders with $D_\mathrm{p}>2.5$ m, meaning most of the mass is constituted by smaller grains and boulders, which aligns with observations of Dimorphos surface \citep{Pajola_et_al_2023}. The volume of boulders in Dimorphos, their packing and SFD will be further constrained with the upcoming Hera mission \citep{Michel_et_al_2022}, which will revisit the binary system satellite once more post DART impact and determine how the body was reshaped. 

In our simulation, the presence of smaller particles has been omitted and their mass is simply accounted for by using larger clusters of particles represented by boulders with densities similar to the bulk density of the primary for computational purposes. Including particles with diameters less than 5 m would put a high computational strain on our simulations that cannot currently be handled with the resources at our disposal as the number of particles in each system would increase drastically, way above $10^4$. Nevertheless, results from \cite{Raducan_et_al_2024b} and previous studies such as \cite{Zhang_et_al_2017,Zhang_et_al_2021} show that large differences in particle sizes affect the stability and structural dynamics of rubble pile asteroids. For our purposes, excluding the small meter-sized particles from our simulations can generate undesired behaviour in the contact mechanics and artificially high mechanical strength for the inner structure from particle interlocking between our large, highly angular particles. This becomes particularly important when considering the high rate of tidal interactions and mergers we observe. While our SFD is polydisperse for the standalone \textit{N}-body simulations, it may be that smaller grains with diameters below 5 m could smooth out contacts between larger boulders, making the aggregates more fluid-like and thereby more susceptible to tidal disruption and deformation during mergers. Further, a more fluid-like behaviour would increase the time it takes for the newly formed aggregates to settle in a specific shape as particle interlocking effects would have less influence on the mechanical integrity of the body. With weaker mechanical strength, dynamical effects such as tidal forces, impacts, spin and self-gravity would have a more dominant role in the determination of the final configuration of the constituents. Keeping these potential caveats in mind, the specific shapes of the moons formed in our simulations may be unphysical, which is why we have opted to solely rely on DEEVE estimates for their geometrical properties. Moreover, another consequence of altering our SFD in this manner could be that it facilitates the creation of a distinct neck between two moonlets during the formation of a bilobate object. The low impact velocities in our model would allow for longer gravitational interactions prior to the contact and smaller loosely bound grains could flow towards the other body, creating the characteristic neck between them, seemingly present between the lobes of Selam \citep{Levison_et_al_2024}. However, if rubble piles get their mechanical strength from the presence of larger boulders beneath their surfaces, interlocking effects between granular particles could potentially also provide the cohesive effect needed to keep structural integrity for the smaller constituent in contact binaries \citep{Meyer_&_Scheeres_2024}. We aim to investigate the validity of our SFD and how it affects the stability of our aggregates in a follow-up study using higher-resolution simulations of select mergers from our simulations and tidal encounters with the primary.

With these model characteristics in mind, we also note that there are no evident dissimilarities in outcomes and dynamical behaviour of the hand-off and pure \textit{N}-body systems in this work, which use drastically different SFDs. 

\subsection{Length of simulations and system stability}

The heavy computational task of resolving contacts between polyhedra not only limits the number of particles we can use, but also how long we can realistically simulate a system, let alone several. Therefore, we had to settle for resolving each system a maximum of ten days post rotational failure. While we can capture and assess the main dynamical mechanisms involved in generating oblate and bilobate shapes, we cannot speak for the long-term stability of our aggregates and their orbits. There are many reasons a potential moon could end up being deformed or even disrupted. Since a majority of the debris disks in our systems have yet to be cleared of matter, further mergers with aggregates still remaining in the system that migrate over time and end up intercepting the orbit of the largest satellite can lead to further growth and reshaping. Furthermore, while unlikely, a lower mass object on an eccentric orbit that undergoes a close encounter with the primary body can be scattered, directly collide with the largest remnant and disrupt it or cause significant deformation due to the high-velocity impact. Orbit decay through long-term tidal interactions with the primary or scattering events with other aggregates could also lead to further mass shedding events and deformation due to tidal disruption. 

Despite the mass available in our systems and the number of aggregates that interact gravitationally, most systems remain dynamically cold in the sense that there the orbital inclinations are kept small, causing a large majority of the close encounters to lead to mergers with low impact velocities. In consequence, changes in orbits mainly occur through energy dissipation from impacts, tidal effects and gravitational scattering with the primary. The mergers also lead to breaking of the often initially synchronous orbits of the largest remnant, leaving only a few of the satellites in such a state. One possible result is that the perturbed aggregate then enters a state of non-principal axis rotation, limiting or breaking any potential for migration via BYORP \citep{Cuk_&_Burns_2005,Quillen_et_al_2022}. Depending on the inertia ratios of the satellite and its collisional history, it can also enter a state of tumbling \citep{Agrusa_et_al_2021}. For bodies that retain their synchronous orbit, BYORP becomes important as a mechanism for migration and if multiple satellites remain in the system, mean motion resonances can further affect the orbital evolution. Long-term dynamical effects of binary asteroid systems are complex and cannot be captured in a meaningful way over the course of just a few days of dynamical evolution and we refer the interested reader to \cite{Agrusa_et_al_2024} who analysed the orbital stability and rotational state of 96 simulations lasting 100 days for a similar, debris disk based binary asteroid formation model. 

\subsection{Comparison to \cite{Agrusa_et_al_2024}}\label{section:agrusa_comparison}

While there have been numerous previous studies on asteroid binary formation \citep{Walsh_et_al_2008,Walsh_et_al_2012,Jacobson_&_Scheeres_2011a,Tardivel_et_al_2018,Hyodo_&_Sugiura_2022,Madeira_et_al_2023,Madeira_&_Charnoz_2024}, the work most related to ours is \cite{Agrusa_et_al_2024} (henceforth referred to as A24). In this section, we summarise any differences and similarities between the work brought up throughout the text with the previous discussion in mind. The two models can consistently form oblate spheroid satellites in wide debris disks around similar primary bodies over timescales on the order of days, a few \% of the cases in A24 and 3 out of 14 simulations for this work. Judging from the distribution of DEEVE axis ratios for the satellites formed in A24 (see their figure 12a), the simulations also seem to have resulted in a few bilobate objects. The dynamical mechanisms that determine the final shape are highly similar in both models as oblate spheroids form in cases where the largest remnant in each system undergoes a catastrophic re-shaping event due to a tidal encounter and/or a critical merger. Therefore, the overall trends for the dynamical evolution of the disks appear similar, independent of differing SFDs, particle shapes and contact physics. However, while it is difficult to argue that our implementation, on average, forms more oblate spheroids because of the low number of simulations we could carry out, it would align with previous points made during the discussion. Given how mergers and tidal encounters with the primary body are the main mechanisms that determine the shape of a satellite in this regime, this would mainly be attributed to the number of aggregates that simultaneously can exist in the system throughout the evolution of the disk, especially during the later stages. This number is governed by how efficiently a given disk can create aggregates and their mechanical strength after formation. With this in mind, we proceed to identify the main differences between the models in our work and A24 that could cause our aggregates to form more easily and survive for longer.

As mentioned in sections \ref{section:method_nbody} and \ref{section:method_nbody_setup}, the methods used to perform the spin-up simulations are based on two distinct codes and setups, which accounts for the variation in initial conditions for the subsequent dynamical evolution of the resulting debris disks. Crucially, the disks simulated in A24 were symmetric, less radially extended (by up to $1R_\mathrm{main}$), thinner and less massive (max $0.04M_\mathrm{main})$. Hence, it is clear that the type of mass-shedding event considered in A24 was less chaotic and more confined to areas around the equator of their primary. As discussed in section \ref{section:discussion_geometry}, the geometry of the debris disk appears to affect the number of aggregates available in the system, as the ability of an aggregate to survive the tidal forces depend on where they form in relation to the fluid Roche limit. The thickness of the disk should also affect its efficiency at forming satellites, as a thicker disk would let an aggregate accrete isotropically for longer before it starts accreting only within the plane, which is less efficient (analogous to 2D and 3D accretion rates in planet formation \citep[e.g.][]{Johansen_&_Lambrechts_2017}). Thus, a satellite forming in a thick disk would be more efficient at clearing its vicinity of matter, reach higher masses and survive for longer before it ultimately merges with another aggregate. The longevity of a given aggregate is also dependent on its mechanical strength. Given that the use of non-spherical particles, as compared to using spherical particles with approximated granular behaviour, has been shown to increase the rigidity of rubble pile objects \citep{Ferrari_&_Tanga_2020} and ability to withstand tidal disruption \citep{Marohnic_et_al_2023}, it is likely that our use of angular particles with rich geometric diversity and non-linear contact mechanics also contribute to a higher mechanical strength. This allows aggregates to more easily survive under the influence of strong tidal forces in our model.

Another main difference between the two studies comes from the disk masses considered. The satellites in A24 with masses similar to that of Dimorphos all occur in cases where the disk mass is 0.03$M_\mathrm{main}$ or less, which offers an explanation for why our model generates more massive largest remnants. While the oblate spheroid satellites from our results all have lower masses than their more prolate counterparts ($<0.02M_\mathrm{main}$), the systems they form in are not fully resolved by the final time step as mentioned above. Therefore, it would be meaningful to run additional SPH spin-up simulations with a smaller effective friction angle to get initial conditions for less massive debris disks. Moreover, by running standalone \textit{N}-body simulations with lower initial disk masses and altering their disk geometry to be more similar to the disks in A24, we would be able to further narrow down the driving factors that increase the frequency of systems with atypically shaped satellites.

\section{Conclusions}\label{section:conclusion}

In light of recent close-up observations of satellites in binary asteroid systems that diverge from the typical elongated shape, such as the oblate spheroid Dimorphos targeted by the DART mission \citep{Rivkin_et_al_2021}, or the contact binary Selam of the Dinkinesh system detected by the Lucy mission \citep{Levison_et_al_2021,Levison_et_al_2024}, we have aimed to narrow down the mechanisms involved in the formation process of these objects in this work. In turn, we performed short-term dynamical evolution of circumasteroidal debris disks formed around rapidly spinning asteroids. Taking initial conditions obtained from SPH simulations of mass shedding through rotational failure, we used a newly developed tool based on geometrical algorithms to perform numerical hand-off to a granular \textit{N}-body code that captures complex contact mechanics and gravitational interactions between irregularly shaped particles. Investigating two different systems, based on the asteroids Ryugu and Didymos, we carried out two hand-off simulations and five standalone \textit{N}-body simulations for each case. Out of the 14 simulations, we end up with three systems with oblate spheroid secondaries and three systems with bilobate satellites by the final time steps.

We find that a radially extended disk wider than 2 primary radii favours mergers and deformation via tidal interactions with the primary body. Growth is rapid, hierarchical and primarily driven by mergers, allowing satellites between 0.014$M_\mathrm{main}$ and 0.032$M_\mathrm{main}$ to form in just a few days. As most moonlets initially end up in orbits near the fluid Roche limit, they are most often prolate and on synchronous orbits that are subsequently perturbed by mergers and close encounters with the main asteroid. In our model, only systems where the largest satellite undergoes mild or catastrophic disruption (losing less or more than half its mass, respectively) during close encounters with the primary, making it less prolate, are capable of forming satellites with Dimorphos-like shapes. 

Analysis of critical mergers during our simulations shows that the impact velocities in this formation regime are below one mutual escape velocity, which is significantly lower compared to previous studies which have mainly been focused on merger-driven moon formation in planetary rings \citep[e.g.][]{Leleu_et_al_2018}. This hints at the existence of a different regime for mergers during the formation of asteroid moons which results in homogeneous bodies without the significant ridges and deformations caused by high-velocity impacts. The low-energy mergers facilitate the formation of bilobate satellites with two lobes connected by a thin neck, which we expect to find when performing higher-resolution simulations of some impacts. Moreover, we find that the initial shape and the orientation of the longest principal axis of the most massive body in the merger highly influence the post-merger shape. Given the bodies are initially prolate before impact, mergers where the relative velocity vector is parallel with the shortest axis lead to more oblate shapes while mergers where the same vector is more closely aligned with the longest axis result in more prolate bodies and bilobate objects.

Due to the heavy computational task of resolving contacts between irregularly shaped particles, we had to settle for a shorter simulation time than desired of at most 10 days and an SFD not allowing for particles smaller than 5 m in diameter. In turn, we cannot draw any conclusions regarding the stability of our systems and the potential presence of higher-order dynamical effects such as BYORP, mean motion resonance between multiple satellites and tumbling. Nevertheless, our results provide valuable insight into the formation mechanisms that govern deformation of the initially prolate bodies asteroid satellites that make up a majority of the population \citep{Pravec_et_al_2016}. Judging from the chaotic formation pathways in our simulations, oblate spheroids and bilobate satellites may be more prevalent in asteroid binaries than indicated by lightcurve measurements, which are biased against detecting satellite shapes that are more oblate \citep{Pravec_et_al_2016,Agrusa_et_al_2024}.

While computationally expensive, the added internal mechanic strength from the contact mechanics in our \textit{N}-body simulations caused by particle--particle interlocking, dilatancy and complex random packing of the irregularly shaped grains better our understanding of how granular aggregate satellites behave during mergers and close encounters with their primaries. When complete tidal disruption is more difficult, more aggregates will survive the earliest stages of formation, allowing for more mergers and reshaping events. However, recent studies such as \cite{Raducan_et_al_2024b} have emphasised that the structural integrity of aggregates is highly dependent on the distribution between large boulders and smaller particles with diameters less than 2.5 m. Furthermore, the observed boulder diameters on the surface of Dimorphos go down to 1 m and up to 16 m, while we include much larger particles serving as clusters of smaller constituents. Given that the presence of smaller particles can smooth out contacts during deformation events such as mergers and tidal interactions, making the body behave more like a fluid, we recognise that our aggregates may be too rigid and will further investigate the effect of particle SFD choice for our implementation. With upcoming missions such as Hera \citep{Michel_et_al_2022}, which will send a spacecraft to revisit the Didymos system in 2027 after the DART impact and determine the resulting shape of Dimorphos and its internal structure, the boulder volume fraction and SFD can be further constrained, allowing us to significantly improve our formation model. 

\printcredits

\section*{Declaration of competing interest}
The authors declare that they have no known competing financial interests or personal relationships that could have appeared to influence the work reported in this paper.

\section*{Acknowledgments}

J.W. and F.F. acknowledge funding from the Swiss National Science Foundation (SNSF) Ambizione grant No.\ 193346. M.J. and S.D.R. acknowledge support from SNSF project No. 200021\_207359. G.M. acknowledges the support of the European Research Council, France (101001282, METAL). We thank the two anonymous reviewers for their constructive feedback, as well as Harrison F. Agrusa and Masatoshi Hirabayashi for helpful discussions and input during this project.

\bibliographystyle{model2-names.bst}

\bibliography{manuscript_wimarsson_et_al.bib}

\end{document}